\documentclass[english,notitlepage,nofootinbib]{revtex4-1}
\usepackage[T1]{fontenc}
\usepackage[latin9]{inputenc}
\usepackage{color}
\usepackage{babel}
\usepackage{amsmath}
\usepackage{amssymb}
\usepackage{graphicx}
\PassOptionsToPackage{normalem}{ulem}
\usepackage{ulem}
\usepackage[unicode=true,
 bookmarks=false,
 breaklinks=false,pdfborder={0 0 1},backref=false,colorlinks=false]
 {hyperref}

\makeatletter

\usepackage{tikz}
\usetikzlibrary{calc}

\usepackage{babel}

\usepackage[]{ulem}

\makeatother

\begin{document}
\title{Exclusive production of quarkonia pairs in collinear factorization
framework}
\author{Marat Siddikov, Iván Schmidt}
\affiliation{Departamento de Física, Universidad Técnica Federico Santa María,~~~~~~\\
 y Centro Científico - Tecnológico de Valparaíso, Casilla 110-V, Valparaíso,
Chile}
\begin{abstract}
In this paper we analyze the exclusive photoproduction of heavy quarkonia
pairs in the collinear factorization framework. We evaluate the amplitude
of the process for $J/\psi\,-\eta_{c}$ quarkonia pair in the leading
order of the strong coupling $\alpha_{s},$ and express it in terms
of generalized parton distributions (GPDs) of gluons in the proton.
We made numerical estimates in the kinematics of the Electron Ion
Collider, and found that in the photoproduction regime, when the virtuality
of the photon is much smaller than the quarkonia mass, the cross-section
of the process is sufficiently large for experimental studies. We
demonstrate that the study of this channel can complement existing
studies of gluon GPDs from other channels. 
\end{abstract}
\maketitle

\section{Introduction}

Understanding the proton structure presents one of the central problems
in high energy physics. Usually this structure is parametrized in
terms of partonic and multipartonic distributions of different flavors.
In view of the nonperturbative nature of strong interactions, it is
not possible to evaluate these distributions theoretically from first
principles, and thus we have to extract them from experimental data.
For exclusive processes, the amplitudes are usually controlled by
the Generalized Parton Distributions (GPDs) of the target~\cite{Diehl:2000xz,Goeke:2001tz,Diehl:2003ny,Guidal:2013rya,Boer:2011fh,Burkert:2022hjz}.
However, extraction of the GPDs from experimental data suffers from
a number of technical challenges, and at present inevitably requires
the use of model assumptions, even for Compton scattering and meson
production, which are considered as references in nucleon tomography~\cite{Kumericki:2016ehc}.
Many observables might obtain simultaneously contributions of GPDs
with different helicity and flavor states, albeit with different,
process-dependent weights. For this reason, the extraction of partonic
distributions of individual flavors inevitably requires analysis of
multiple channels, and thus the extension of the number of possible
channels for study of GPDs is strongly desired~\cite{Pire:2015iza,Pire:2017lfj,Pire:2017yge,Pire:2021dad}.
Recently a new class of $2\to3$ processes has been suggested in the
literature~\cite{GPD2x3:9,GPD2x3:8,GPD2x3:7,GPD2x3:6,GPD2x3:5,GPD2x3:4,GPD2x3:3,GPD2x3:2,GPD2x3:1,Duplancic:2022wqn,ElBeiyad:2010pji,Boussarie:2016qop},
as potential new probes, which should complement existing studies,
provide more stringent constraints on existing phenomenological models
and in this way diminish theoretical uncertainty. Most of these studies
focused on the production of light mesons and photons. Such processes
are dominated by quark GPDs (both in chiral odd and chiral even sectors).
A factorization for such processes has been proven in the kinematics
when the relative transverse momenta of the produced hadrons and photon
($\sim$pairwise invariant masses) are large enough to avoid soft
final-state interactions~\cite{GPD2x3:10,GPD2x3:11}.

In these analyses special attention should be paid to the extraction
of gluon GPDs. Since the gluons do not couple directly to photons,
they contribute to many processes only as higher order corrections,
which adversely affects the precision of the extracted gluon GPDs.
However, knowledge of the gluon GPDs is important for solving many
puzzles (see~~\cite{Diehl:2000xz,Goeke:2001tz,Diehl:2003ny,Guidal:2013rya,Boer:2011fh,Burkert:2022hjz}
for overview). The best channel for the study of gluon GPDs is the
production of heavy quarkonia. Due to the expected smallness of intrinsic
heavy parton densities, the process gets a dominant contribution from
gluon GPDs, which might therefore be studied in detail. The heavy
mass of quarkonia plays the role of a natural hard scale in the problem~\cite{Korner:1991kf,Neubert:1993mb},
relaxing the conditions on other kinematic variables and potentially
opening the possibility to use perturbative methods even in photoproduction
regime. A modern NRQCD framework allows to incorporate systematically
various perturbative corrections~\cite{Bodwin:1994jh,Maltoni:1997pt,Brambilla:2008zg,Feng:2015cba,Brambilla:2010cs,Cho:1995ce,Cho:1995vh,Baranov:2002cf,Baranov:2007dw,Baranov:2011ib,Baranov:2016clx,Baranov:2015laa}.
The use of single quarkonia production for constraining the gluon
GPDs has been discussed in detail in~\cite{DVMPcc1,DVMPcc2,DVMPcc3,DVMPcc4},
and the coefficient functions have been evaluated, taking into account
next-to-leading order and some higher twist corrections. However,
the amplitude of this process provides information only about GPDs
convoluted with process-dependent coefficient functions, and, as mentioned
earlier, an inversion of the procedure might be impossible, especially
when the complicated structure of higher-order corrections is taken
into account. For this reason it is important to complement the analysis
with data from other channels. A natural and straightforward extension
of these studies is the production of multiple quarkonia (\emph{e.g}.
heavy quarkonia pairs). Such processes have been the subject of theoretical
studies since the early days of QCD~\cite{Brodsky:1986ds,Lepage:1980fj,Berger:1986ii,Baek:1994kj},
and recently got renewed interest due to the forthcoming launch of
high-luminosity accelerator facilities, as well as being a potential
gateway for the study of all-heavy tetraquarks, which might be molecular
states of quarkonia pairs~\cite{Bai:2016int,Heupel:2012ua,Lloyd:2003yc,Vijande:2006vu,Vijande:2012jw,Chen:2019vrj,Esposito:2018cwh,Cardinale:2018zus,Aaij:2018zrb,Capriotti:2019huu,LHCb:2020bwg}.

Previously, the exclusive production of quarkonia pairs has been studied
for $J/\psi\,J/\psi$ production, which might proceed only via a two-photon
mechanism, $\gamma\gamma\to M_{1}M_{2}$ ~\cite{Goncalves:2015sfy,Goncalves:2019txs,Goncalves:2006hu,Baranov:2012vu,Yang:2020xkl,Goncalves:2016ybl}
due to $C$-parity constraints and thus cannot be used for studies
of gluon GPDs. Recently we analyzed the production of quarkonia pairs
with opposite $C$-parities, which proceeds via photon-pomeron fusion
and thus have larger cross-sections~\cite{Andrade:2022rbn}. However,
our study was realized in the framework of the Color Glass Condensate
approach and relied on an underlying eikonal picture, which is valid
in the small-$x$ domain. At smaller energies, as well as in the kinematics
of large photon virtuality $Q^{2}$, the assumptions of this picture
are not well-justified, and it makes sense to analyze this process
in the complementary collinear factorization approach, which is expected
to give reasonable predictions in this kinematics and give access
to the aforementioned gluon GPDs of the target. This kinematic regime
might be studied in low-energy electron-proton collisions at the forthcoming
Electron Ion Collider (EIC)~\cite{Accardi:2012qut,DOEPR,BNLPR,AbdulKhalek:2021gbh}.

The paper is structured as follows. Below, in Section~\ref{sec:Formalism},
we discuss in detail the kinematics of the process and the framework
for the evaluation of the amplitude of the process. In Section~\ref{sec:Numer}
we present our numerical estimates for the cross-sections, in EIC
kinematics. Finally, in Section~\ref{sec:Conclusions} we draw conclusions.

\section{Exclusive photoproduction of meson pairs}

\label{sec:Formalism}

Previously, the exclusive production of\emph{ light} meson pairs was
analyzed in Bjorken kinematics in~\cite{LehmannDronke:1999vvq,LehmannDronke:2000hlo,Clerbaux:2000hb,Diehl:1999cg,ZEUS:1998xpo},
with the additional constraint that the invariant mass of meson pair
should be large. There it was demonstrated that the amplitude of that
process might be represented as a convolution of the quark and gluon
GPDs of the target, with novel 2-meson distribution amplitudes. However,
the extension of those results to quarkonia pairs is not straightforward,
since quarkonia masses and the invariant mass $M_{12}$ are very large,
so the Bjorken regime ($Q\gg M_{12})$ is achieved in the kinematics
where the cross-section is negligibly small. For this reason, it makes
sense to analyze the quarkonia pair production by treating the heavy
mass of the quark and the photon virtuality $Q$ as two independent
hard scales, with the photoproduction ($Q\ll M_{12}$) and Bjorken
($Q\gg M_{12}$) regimes as limiting cases. In the following subsection~\ref{subsec:Kinematics}
we discuss in detail the kinematics of the process, and in subsection~\ref{subsec:Amplitudes}
we discuss the evaluation of the amplitudes in the collinear factorization
approach, and their relation to the target gluon GPDs.

\subsection{Kinematics of the process}

\label{subsec:Kinematics} In order to facilitate the comparison with
experimental data, in what follows we will present our results in
the frame whose axis $z$coincides with the photon-proton collision
axis, so the light-cone decomposition of the momenta is given by

\begin{align}
q & =\,\left(-\frac{Q^{2}}{2q^{-}},\,q^{-},\,\,\boldsymbol{0}_{\perp}\right),\quad q^{-}=E_{\gamma}+\sqrt{E_{\gamma}^{2}+Q^{2}}\label{eq:qPhoton-2}\\
P & =\left(P^{+},\,,\frac{m_{N}^{2}}{2P^{+}}\,\,\boldsymbol{0}_{\perp}\right),\quad P^{+}=E_{p}+\sqrt{E_{p}^{2}-m_{N}^{2}}\\
p_{a} & =\left(\frac{M_{a}^{\perp}}{2}\,e^{-y_{a}}\,,\,M_{a}^{\perp}e^{y_{a}},\,\,\boldsymbol{p}_{a}^{\perp}\right),\quad a=1,2,\label{eq:MesonLC-2}\\
 & M_{a}^{\perp}\equiv\sqrt{M_{a}^{2}+\left(\boldsymbol{p}_{a}^{\perp}\right)^{2}},
\end{align}
where $q$ is the (virtual) photon momentum, $P$ and $P'$ are the
momenta of the proton before and after the collision, and $p_{1},\,p_{2}$
are the 4-momenta of the produced heavy quarkonia; the latter are
expressed in terms of the rapidities and transverse momenta $\left(y_{a},\,\boldsymbol{p}_{a}^{\perp}\right)$
of these heavy mesons. This frame allows for straightforward analysis
down to the photoproduction limit $(Q\to0)$. The relation of this
frame to the so-called symmetric frame~\cite{Radyushkin:1996nd,Radyushkin:1997ki,Collins:1998be,Ji:1996nm,Ji:1998xh,Diehl:1999cg,Goeke:2001tz,Diehl:2003ny},
which is used for the analysis in Bjorken kinematics ($Q\to\infty$),
is discussed in detail in Appendix~\ref{sec:Relation-Symm}. In the
limit $Q\to0$, this frame, up to a trivial longitudinal boost, coincides
with the frame used in earlier studies of exclusive photoproduction
$\gamma p\to\gamma Mp$ ~\cite{GPD2x3:9,GPD2x3:8,GPD2x3:7,GPD2x3:6,GPD2x3:5,GPD2x3:4,GPD2x3:3,GPD2x3:2,GPD2x3:1,Duplancic:2022wqn}.
in this frame, the polarization vectors of the longitudinally and
transversely polarized photons are chosen respectively as~\footnote{We've chosen the longitudinal vector in the light-cone gauge, so the
contribution of the longitudinal photons in the $ep$ amplitude might
be reinterpreted as instantaneous part of the photon propagator. The
results will not change under any redefinition of polarization vectors
$\varepsilon_{\mu}(q)\to\varepsilon_{\mu}(q)+{\rm const}\,q_{\mu}$
in view of the Ward identity (in this problem it remains valid even
for offshell photons, since all amplitudes with an omitted photon
vertex vanish due to $C$-parity).} 
\begin{equation}
\varepsilon_{L}=\left(\frac{Q}{q^{-}},\,0,\,\boldsymbol{0}_{\perp}\right),\quad\varepsilon_{T}^{(\pm)}=\left(0,\,0,\frac{1}{\sqrt{2}},\pm\frac{i}{\sqrt{2}}\right).\label{eq:PolVector}
\end{equation}
We also will use the notations 
\begin{align}
\Delta & =P'-P=q-p_{1}-p_{2}=\left(\Delta^{+},\,\Delta^{-},\,\,\boldsymbol{\Delta}^{\perp}\right),
\end{align}
\begin{align}
 & \Delta^{+}=-\frac{Q^{2}}{2q^{-}}-\frac{M_{1}^{\perp}e^{-y_{1}}}{2}-\frac{M_{2}^{\perp}e^{-y_{2}}}{2},\quad\Delta^{-}=q^{-}-M_{1}^{\perp}\,e^{y_{1}}-M_{2}^{\perp}\,e^{y_{2}},\quad\boldsymbol{\Delta}_{\perp}=-\boldsymbol{p}_{1}^{\perp}-\boldsymbol{p}_{2}^{\perp}
\end{align}
for the 4-vector of momentum transfer to the proton and its components,
and the notation $t$ for its square, 
\begin{align}
t & =\Delta^{2}=-\left(q^{-}-M_{1}^{\perp}\,e^{y_{1}}-M_{2}^{\perp}\,e^{y_{2}}\right)\left(\frac{Q^{2}}{q^{-}}+M_{1}^{\perp}e^{-y_{1}}+M_{2}^{\perp}e^{-y_{2}}\right)-\left(\boldsymbol{p}_{1}^{\perp}+\boldsymbol{p}_{2}^{\perp}\right)^{2}\label{eq:tDef}\\
 & =-Q^{2}+M_{1}^{2}+M_{2}^{2}-q^{-}\left(M_{1}^{\perp}e^{-y_{1}}+M_{2}^{\perp}e^{-y_{2}}\right)+\frac{Q^{2}}{q^{-}}\left(M_{1}^{\perp}\,e^{y_{1}}+M_{2}^{\perp}\,e^{y_{2}}\right)\nonumber \\
 & +2\left(M_{1}^{\perp}M_{2}^{\perp}\cosh\Delta y-\boldsymbol{p}_{1}^{\perp}\cdot\boldsymbol{p}_{2}^{\perp}\right).\nonumber 
\end{align}
After the interaction, the 4-momentum of the proton is given by 
\begin{equation}
P'=P+\Delta=\left(q^{-}+\frac{m_{N}^{2}}{2P^{+}}-M_{1}^{\perp}\,e^{y_{1}}-M_{2}^{\perp}\,e^{y_{2}}\,,\,P^{+}-\frac{Q^{2}}{2q^{-}}-\frac{M_{1}^{\perp}e^{-y_{1}}+M_{2}^{\perp}e^{-y_{2}}}{2},-\boldsymbol{p}_{1}^{\perp}-\boldsymbol{p}_{2}^{\perp}\right),
\end{equation}
and the onshellness condition $\left(P+\Delta\right)^{2}=m_{N}^{2}$
allows to get an additional constraint 
\begin{align}
q^{-}P^{+} & =P^{+}\left(M_{1}^{\perp}\,e^{y_{1}}+M_{2}^{\perp}\,e^{y_{2}}\right)-\frac{m_{N}^{2}+t}{2}+\frac{m_{N}^{2}}{4P^{+}}\left(M_{1}^{\perp}e^{-y_{1}}+M_{2}^{\perp}e^{-y_{2}}+\frac{Q^{2}}{q^{+}}\right).\label{qPlus}
\end{align}
Solving the Equation~(\ref{qPlus}) with respect to $q^{-}$, we
get 
\begin{align}
q^{-} & =\frac{M_{1}^{\perp}\,e^{y_{1}}+M_{2}^{\perp}\,e^{y_{2}}-\frac{m_{N}^{2}+t}{2P^{+}}+\frac{m_{N}^{2}}{4\left(P^{+}\right)^{2}}\left(M_{1}^{\perp}e^{-y_{1}}+M_{2}^{\perp}e^{-y_{2}}\right)}{2}\pm\label{qPlus-1}\\
 & +\frac{1}{2}\,\sqrt{\left(M_{1}^{\perp}\,e^{y_{1}}+M_{2}^{\perp}\,e^{y_{2}}-\frac{m_{N}^{2}+t}{2P^{+}}+\frac{m_{N}^{2}}{4\left(P^{+}\right)^{2}}\left(M_{1}^{\perp}e^{-y_{1}}+M_{2}^{\perp}e^{-y_{2}}\right)\right)^{2}+\frac{Q^{2}m_{N}^{2}}{\left(P^{+}\right)^{2}}},\nonumber 
\end{align}
which allows to express the energy of the photon $E_{\gamma}\approx q^{-}/2$
in terms of the kinematic variables $\left(y_{a},\,\boldsymbol{p}_{a}^{\perp}\right)$
of the produced quarkonia. For asymptotically large energies $q^{-},P^{+}\gg\{Q,\,M_{a},\,m_{N},\,\sqrt{|t|}\}$,
the result~(\ref{qPlus-1}) reduces to 
\begin{align}
 & q^{-}\approx M_{1}^{\perp}\,e^{y_{1}}+M_{2}^{\perp}\,e^{y_{2}}\label{eq:qPlus}
\end{align}
and in this limit the variable $t$ merely reduces to 
\begin{equation}
t\approx-\left(\boldsymbol{p}_{1}^{\perp}+\boldsymbol{p}_{2}^{\perp}\right)^{2}.
\end{equation}
In the photoproduction regime, the expression for $q^{-}$simplifies
to 
\begin{align}
q^{-} & =M_{1}^{\perp}\,e^{y_{1}}+M_{2}^{\perp}\,e^{y_{2}}-\frac{m_{N}^{2}+t}{2P^{+}}+\frac{m_{N}^{2}}{4\left(P^{+}\right)^{2}}\left(M_{1}^{\perp}e^{-y_{1}}+M_{2}^{\perp}e^{-y_{2}}\right)
\end{align}

The invariant energy $W$ of the $\gamma p$ collision and the invariant
mass $M_{12}$ of the produced heavy quarkonia pair in terms of these
variables might be rewritten as 
\begin{equation}
W^{2}\equiv s_{\gamma p}=\left(q+P\right)^{2}=-Q^{2}+m_{N}^{2}+2q\cdot P,\label{eq:W2}
\end{equation}
and 
\begin{align}
M_{12}^{2} & =\left(p_{1}+p_{2}\right)^{2}=M_{1}^{2}+M_{2}^{2}+2\left(M_{1}^{\perp}M_{2}^{\perp}\cosh\Delta y-\boldsymbol{p}_{1}^{\perp}\cdot\boldsymbol{p}_{2}^{\perp}\right)=\label{eq:M12}\\
 & =t-Q^{2}+2M_{1}^{\perp}Q\cosh\left(y_{1}+\delta y_{q}\right)+2M_{2}^{\perp}Q\cosh\left(y_{2}+\delta y_{q}\right),\nonumber \\
 & \delta y_{q}=\ln\left(Q/q^{+}\right).
\end{align}
respectively. Finally, the Bjorken variable $x_{B}$ might be evaluated
using the relation 
\begin{align}
x_{B} & \approx\frac{Q^{2}+M_{{\rm 12}}^{2}}{Q^{2}+W_{\gamma p}^{2}-m_{N}^{2}}\approx\frac{Q^{2}}{2q^{-}P^{+}}+\frac{M_{1\perp}}{P^{+}}e^{-y_{1}}+\frac{M_{2\perp}}{P^{+}}e^{-y_{2}}\label{eq:xB-1}
\end{align}

The cross-section of electroproduction is dominated by single-photon
exchange between leptonic and hadronic parts, and for this reason
might be represented as 
\begin{equation}
\frac{d\sigma_{ep\to eM_{1}M_{2}p}}{d\ln x_{B}dQ^{2}\,d\Omega_{h}}=\frac{\alpha_{{\rm em}}}{\pi\,Q^{2}}\,\left[\left(1-y\right)\frac{d\sigma_{\gamma p\to M_{1}M_{2}p}^{(L)}}{d\Omega_{h}}+\left(1-y+\frac{y^{2}}{2}\right)\frac{d\sigma_{\gamma p\to M_{1}M_{2}p}^{(T)}}{d\Omega_{h}}\right],\label{eq:LTSep}
\end{equation}
where $y$ is the inelasticity (fraction of electron energy which
passes to the virtual photon, which should not be confused with the
rapidities $y_{1},\,y_{2}$ of produced quarkonia); $d\Omega_{h}$
represents the phase volume of the produced quarkonia pair and will
be specified below. In~(\ref{eq:LTSep}) we assumed that the incident
protons and electrons are not polarized, and $d\sigma^{(T)},\,d\sigma^{(L)}$
are the contributions of the transversely and longitudinally polarized
virtual photons. While the former is expected to dominate for longitudinal
photons, the latter might get pronounced contributions at large virtualities.

\begin{figure}
\includegraphics[height=8.5cm]{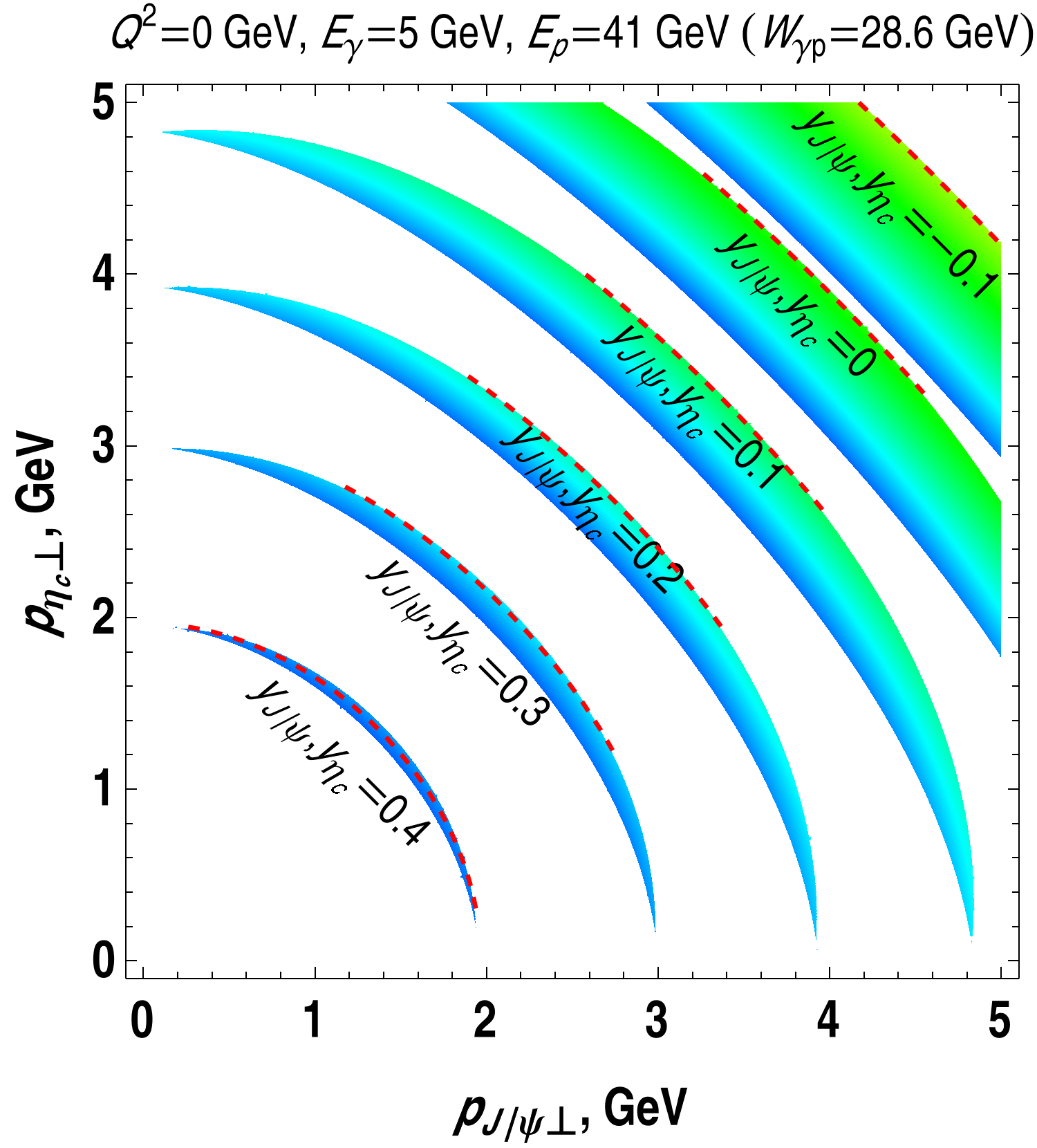}\includegraphics[height=8.5cm]{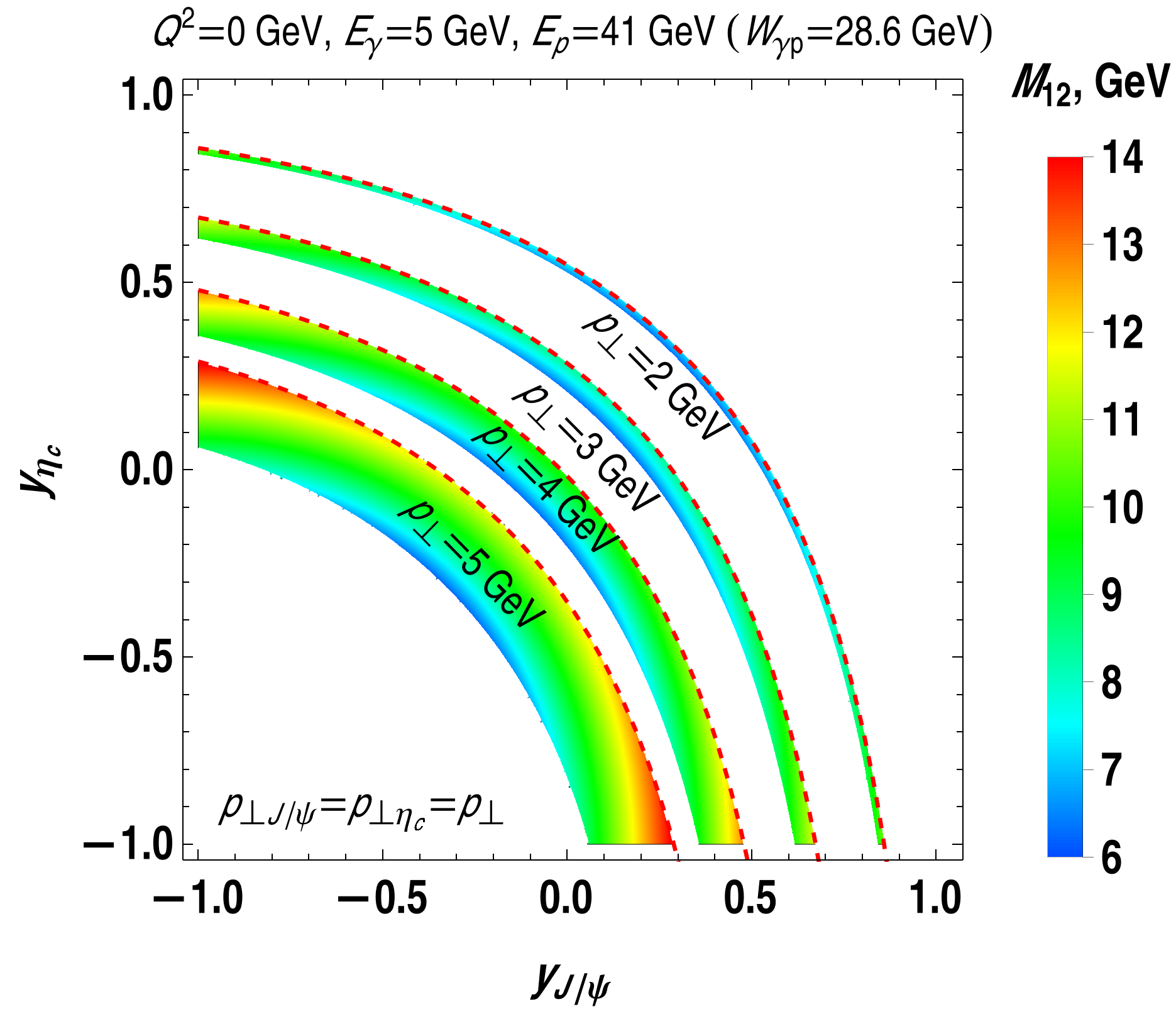}

\caption{\label{fig:Domain}(Color online) The colored bands represent kinematically
allowed regions for quarkonia pair production \uline{at fixed photon
energy} $E_{\gamma}$, virtuality $Q$ and proton energy $E_{p}$.
The left plot illustrates the allowed values of transverse momenta
for different fixed rapidities $y_{1.2}\equiv y_{J/\psi},y_{\eta_{c}}$
of both quarkonia. An increase of rapidities of both quarkonia leads
to higher longitudinal components of their momenta, and thus in view
of energy conservation leads to smaller transverse momenta of quarkonia.
The right plot illustrates the allowed values of rapidities at different
fixed transverse momenta$\left|\boldsymbol{p}_{1,2}\right|\equiv p_{J/\psi},\,p_{\eta_{c}}$.
Akin to the previous panel, in view of energy conservation, bands
with smaller $p_{\perp}$ require larger longitudinal components of
both quarkonia, which translates into higher quarkonia rapidities.
In both plots the color of each point encodes the value of the invariant
mass $M_{12}$ of the quarkonia pair, as given in the color bar legend
in the right panel. The red dashed line inside each band corresponds
to fixed momentum transfer to the proton $t=\Delta^{2}=-1\,{\rm GeV^{2}}$
(see the text for more explanation).}
\end{figure}

The photoproduction cross-section is related to the amplitude via
\begin{equation}
d\sigma_{\gamma p\to M_{1}M_{2}p}^{(L,T)}=\frac{dy_{1}dp_{1\perp}^{2}dy_{2}dp_{2\perp}^{2}d\phi_{12}\left|\mathcal{A}_{\gamma p\to M_{1}M_{2}p}^{(L,T)}\right|^{2}}{4\left(2\pi\right)^{4}\sqrt{\left(W^{2}+Q^{2}-m_{N}^{2}\right)^{2}+4Q^{2}m_{N}^{2}}}\delta\left(\left(q+P_{1}-p_{1}-p_{2}\right)^{2}-m_{N}^{2}\right)\label{eq:Photo}
\end{equation}
where the $\delta$-function guarantees onshellness of the recoil
proton. Taking into account that the vectors $q,P_{1}$ do not have
transverse momenta, we may rewrite the argument of $\delta$-function
as 
\begin{align}
 & \left(q+P_{1}-p_{1}-p_{2}\right)^{2}-m_{N}^{2}=\left(q+P_{1}-p_{1}^{||}-p_{2}^{||}\right)^{2}-\left(\boldsymbol{p}_{1}^{\perp}+\boldsymbol{p}_{2}^{\perp}\right)^{2}-m_{N}^{2}\label{eq:delta}\\
 & =\left(q+P_{1}-p_{1}^{||}-p_{2}^{||}\right)^{2}-\left(\left(p_{1}^{\perp}\right)^{2}+\left(p_{2}^{\perp}\right)^{2}+2p_{1}^{\perp}p_{2}^{\perp}\cos\phi_{12}\right)-m_{N}^{2}\nonumber 
\end{align}
where $\phi_{12}$ is the azimuthal angle between the transverse momenta
of quarkonia $\boldsymbol{p}_{1}^{\perp},\boldsymbol{p}_{2}^{\perp}$.
We may rewrite the $\delta$-function in~(\ref{eq:delta}) as 
\begin{align}
 & \delta\left(\left(q+P_{1}-p_{1}-p_{2}\right)^{2}-m_{N}^{2}\right)=\frac{\delta\left(\phi_{12}-\phi_{0}\right)}{2p_{1\perp}p_{2\perp}\left|\sin\phi_{0}\right|},\\
 & \phi_{0}=\arccos\left[\frac{\left(q+P_{1}-p_{1}^{||}-p_{2}^{||}\right)^{2}-\left(\left(p_{1}^{\perp}\right)^{2}+\left(p_{2}^{\perp}\right)^{2}+m_{N}^{2}\right)}{2p_{1}^{\perp}p_{2}^{\perp}}\right],
\end{align}
which allows to integrate out the dependence on $\phi_{12}$. The
restriction $\left|\cos\phi_{0}\right|\le1$ imposes an additional
constraint on possible $\left(y_{1},p_{1\perp}\right)$ and $\left(y_{2},p_{2\perp}\right)$
values, at \uline{fixed photon-proton energy}. In Figure~(\ref{fig:Domain})
we illustrate the typical kinematically allowed region for a fixed
choice of $E_{\gamma},\,E_{p}$, in EIC kinematics, as a function
of rapidities and transverse momenta of quarkonia. At very high energies
$P^{+},q^{-}\gg M_{1,2}$, the domain turns into a narrow strip surrounding
the curve~(\ref{qPlus}) and has a typical width $\sim1/P^{+}$.
In this regime the longitudinal momentum of the projectile remains
almost constant, so it corresponds to the kinematics $x_{B}\sim\xi\ll1$,
which is outside the scope of our study. The color of each point in
Figure~~(\ref{fig:Domain}) illustrates the value of the invariant
mass $M_{12}$ of the quarkonia pair. As we will show below, the dominant
contribution to the cross-section comes from the region $|t|_{{\rm min}}\lesssim|t|\lesssim1\,{\rm GeV}^{2}$,
for this reason we have also shown the line $t=-1\,{\rm GeV^{2}}$
(the line $t=t_{{\rm min}}$ corresponds to the upper border of each
colored domain). The observed anticorrelation between $|t|$ and $M_{12}$
might be understood, if we take into account that for fixed-energy
of the quarkonia pairs, the variable $|t|$ reaches its minimum (and
$M_{12}$ reaches its maximum) for quarkonia moving in \emph{opposite}
directions; vice versa, quarkonia moving in the \emph{same} direction,
will minimize $M_{12}$ but maximize $|t|$. In the experiment, due
to finite resolution in the measurement of the photon energy $W$
and the quarkonia kinematics ($y_{1,2},p_{1,2}$), the narrow domains
shown in Figure~\ref{fig:Domain} will get smeared. Due to this,
the values of $M_{12}^{2}$ and $t$ are not uniquely defined, but
rather are distributed in some interval. The size of this effect depends
crucially on the experimental setup, so we won't discuss it here with
more detail. However, for any reasonably narrow bins in rapidity $(\Delta y)$
or transverse momenta $(\Delta p_{\perp})$, the variables $y_{1,2},p_{1,2}$
remain restricted to some finite domain.

In electroproduction experiments, instead of conventional fixing the
photon energy, it might be easier to treat the quarkonia variables
($y_{1},p_{1\perp},y_{2},p_{2\perp},\phi$) as independent variables,
and express the photon energy in terms of these variables. The $\delta$-function
in~(\ref{eq:Photo}) can be rewritten as 
\begin{align}
 & \delta\left(\left(q+P_{1}-p_{1}-p_{2}\right)^{2}-m_{N}^{2}\right)=\delta\left(W^{2}+M_{12}^{2}-2\left(q+P_{1}\right)\cdot\left(p_{1}+p_{2}\right)-m_{N}^{2}\right)=\\
 & =\frac{\delta\left(W-W_{0}\right)+\delta\left(W+W_{0}\right)}{2W_{0}},\nonumber \\
 & W_{0}^{2}=2\left(q+P_{1}\right)\cdot\left(p_{1}+p_{2}\right)+m_{N}^{2}-M_{12}^{2}=\\
 & =\left(q^{-}+\frac{m_{N}^{2}}{2P^{+}}\right)\cdot\left(M_{1}^{\perp}e^{-y_{1}}+M_{2}^{\perp}e^{-y_{2}}\right)+2\left(P^{+}-\frac{Q^{2}}{2q^{-}}\right)\cdot\left(M_{1}^{\perp}e^{y_{1}}+M_{2}^{\perp}e^{y_{2}}\right)+m_{N}^{2}-M_{12}^{2}.\nonumber 
\end{align}
and $q^{-}$ can be fixed from~(\ref{qPlus-1}). After integration
over all possible energies $W$ (equivalent to integration over all
possible $x_{B}$), we get for the electroproduction cross-section

\begin{equation}
\frac{d\sigma_{ep\to eM_{1}M_{2}p}}{dQ^{2}\,d\Omega_{h}}=\frac{\alpha_{{\rm em}}}{4\pi\,Q^{2}}\,\left[\left(1-y\right)\frac{d\bar{\sigma}_{\gamma p\to M_{1}M_{2}p}^{(L)}}{d\Omega_{h}}+\left(1-y+\frac{y^{2}}{2}\right)\frac{d\bar{\sigma}_{\gamma p\to M_{1}M_{2}p}^{(T)}}{d\Omega_{h}}\right],\label{eq:LTSep-2}
\end{equation}
\begin{equation}
d\bar{\sigma}_{\gamma p\to M_{1}M_{2}p}^{(L,T)}=\frac{dy_{1}dp_{1\perp}^{2}dy_{2}dp_{2\perp}^{2}d\phi_{12}\left|\mathcal{A}_{\gamma p\to M_{1}M_{2}p}^{(L,T)}\right|^{2}}{4\left(2\pi\right)^{4}W_{0}^{2}\sqrt{\left(W_{0}^{2}+Q^{2}-m_{N}^{2}\right)^{2}+4Q^{2}m_{N}^{2}}}\label{eq:Photo-1}
\end{equation}
where now ($y_{1},p_{1\perp},y_{2},p_{2\perp},\phi$) are independent
variables, and $d\bar{\sigma}_{\gamma p\to M_{1}M_{2}p}^{(L,T)}$
corresponds to the photoproduction cross-section with photon's energy
evaluated using~(\ref{qPlus}).

\subsection{Amplitudes of the meson pair production process}

\label{subsec:Amplitudes}For the evaluation of the amplitudes $\mathcal{A}_{\gamma p\to M_{1}M_{2}p}^{(\mathfrak{a})}$
we will use the collinear factorization framework, which allows to
express the amplitude in terms of the target GPDs~\cite{Diehl:2000xz,Goeke:2001tz,Diehl:2003ny,Guidal:2013rya,Boer:2011fh,Burkert:2022hjz}.
We will assume that both the photon virtuality $Q^{2}$ and the quark
mass $m_{Q}$ are large parameters, and also disregard the transverse
momenta $\boldsymbol{\Delta}_{\perp},\boldsymbol{p}_{a\perp}$ in
the coefficient function. Furthermore, we will assume that the quarkonia
pairs are always produced with sufficiently large relative momentum
\begin{align}
p_{{\rm rel}} & \approx\frac{\left(2m_{Q}\right)\,v_{{\rm rel}}}{\sqrt{1-v_{{\rm rel}}^{2}}}\gtrsim\alpha_{s}\left(m_{Q}\right)m_{Q},\quad v_{{\rm rel}}=\sqrt{1-\frac{p_{1}^{2}p_{2}^{2}}{\left(p_{1}\cdot p_{2}\right)^{2}}}=\sqrt{1-\frac{4M_{1}^{2}M_{2}^{2}}{\left(M_{12}^{2}-M_{1}^{2}-M_{2}^{2}\right)^{2}}}
\end{align}
  both with respect to each other, as well as with respect to recoil
proton, to avoid potential factorization breaking by the exchange of soft
gluons in the final state. We expect that the factorization should
remain valid both in the Bjorken and in the photoproduction regimes.

The GPDs are conventionally defined in the symmetric frame specified
in Appendix~\ref{sec:Relation-Symm}, so for the coefficient functions
evaluation we will temporarily switch to that frame~\footnote{We need to mention that in early studies~\cite{Radyushkin:1996nd,Radyushkin:1997ki,Goeke:2001tz,Diehl:2003ny},
the GPDs were defined in an asymmetric frame, in which the momentum transfer
of the incident photon is zero. Up to a trivial longitudinal boost this
frame essentially coincides with the frame introduced in Section~\ref{subsec:Kinematics}.
It is possible to relate the GPDs defined in different frames using some
transformation of the arguments. However, since this frame is not
widely used in the recent literature dedicated to GPD properties,
we abstain from using it in what follows.}. In this frame the momenta of the active parton (gluon), before and
after interaction, are given explicitly by 
\begin{align}
k_{i} & =\left((x+\xi)\bar{P}^{+},\,0,\,-\frac{\boldsymbol{\Delta}_{\perp}}{2}\right),\quad k_{f}=\left((x-\xi)\bar{P}^{+},\,0,\,\frac{\boldsymbol{\Delta}_{\perp}}{2}\right)
\end{align}
where $x$ is the light-cone fraction of average momentum, $x=\left(k_{i}^{+}+k_{f}^{+}\right)\bigg/2\bar{P}^{+}$,
and the skewedness variable $\xi$ is related to $x_{B}$ defined
in~(\ref{eq:xB-1}) via the relations~\cite{Diehl:2003ny} 
\begin{equation}
\xi=-\frac{\Delta^{+}}{2\bar{P}^{+}}=\frac{x_{B}}{2-x_{B}},\quad x_{B}=\frac{2\xi}{1+\xi}.\label{eq:XiDef}
\end{equation}

\begin{figure}
\includegraphics[width=6cm]{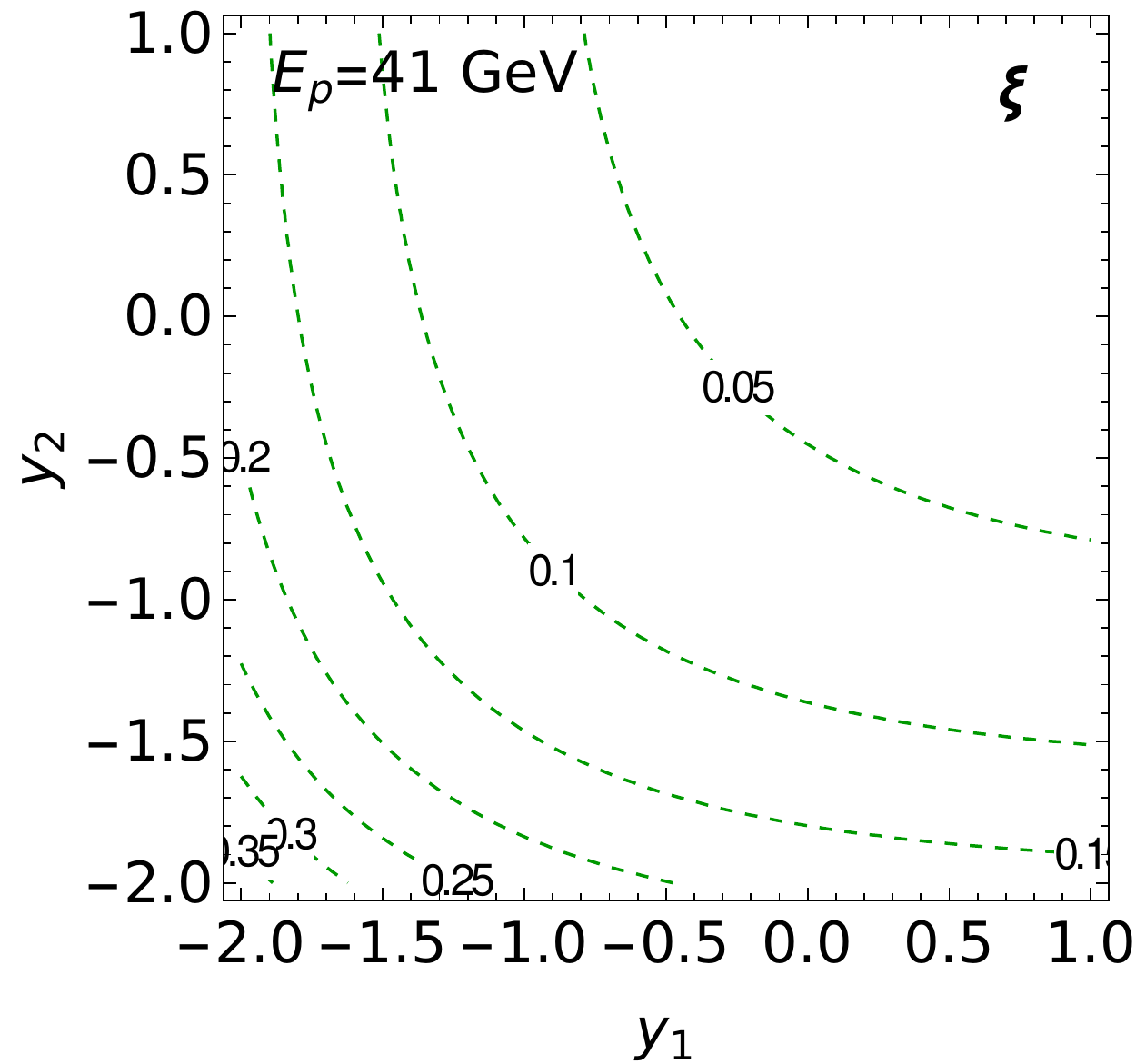}\includegraphics[width=6cm]{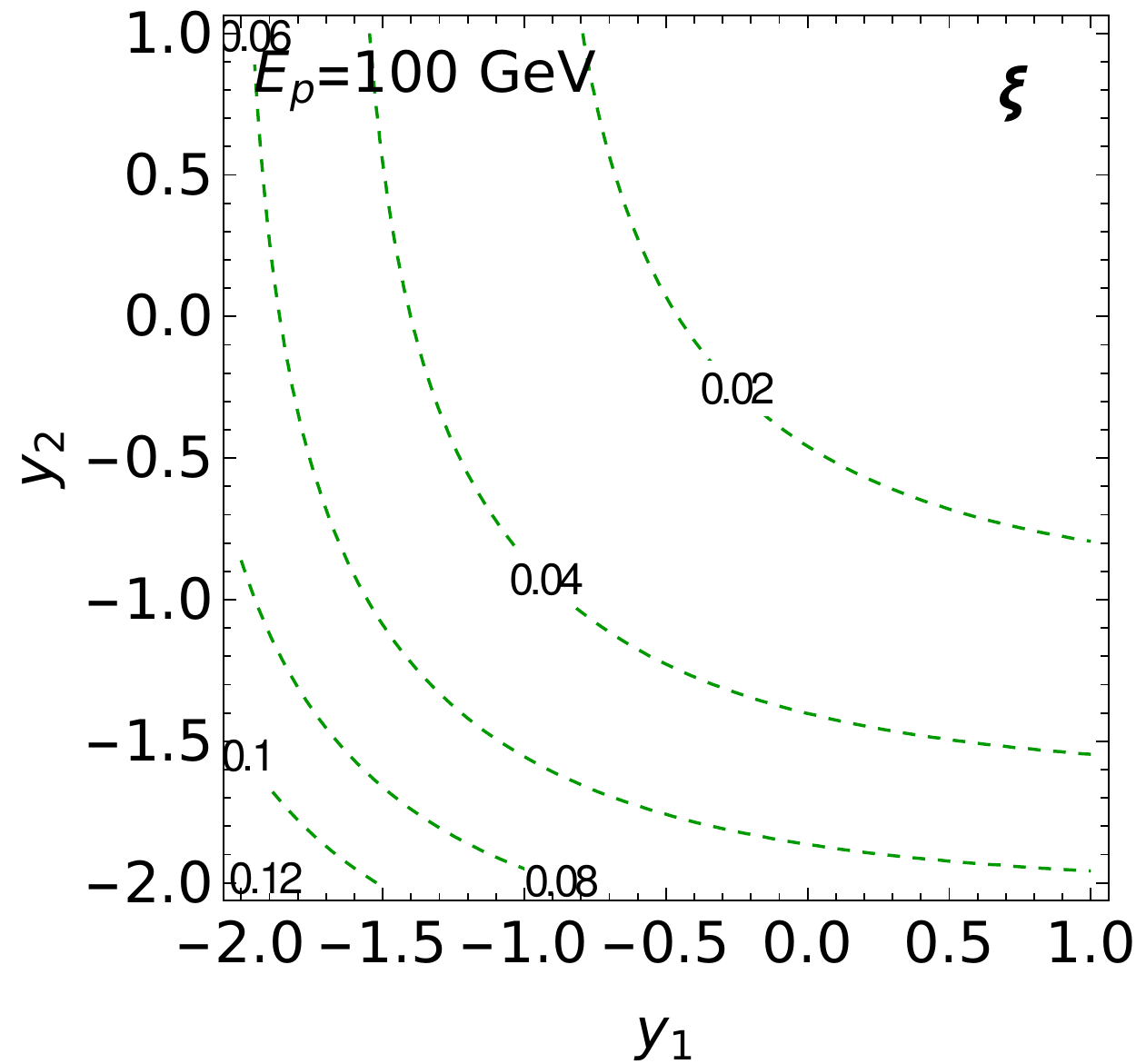}\includegraphics[width=6cm]{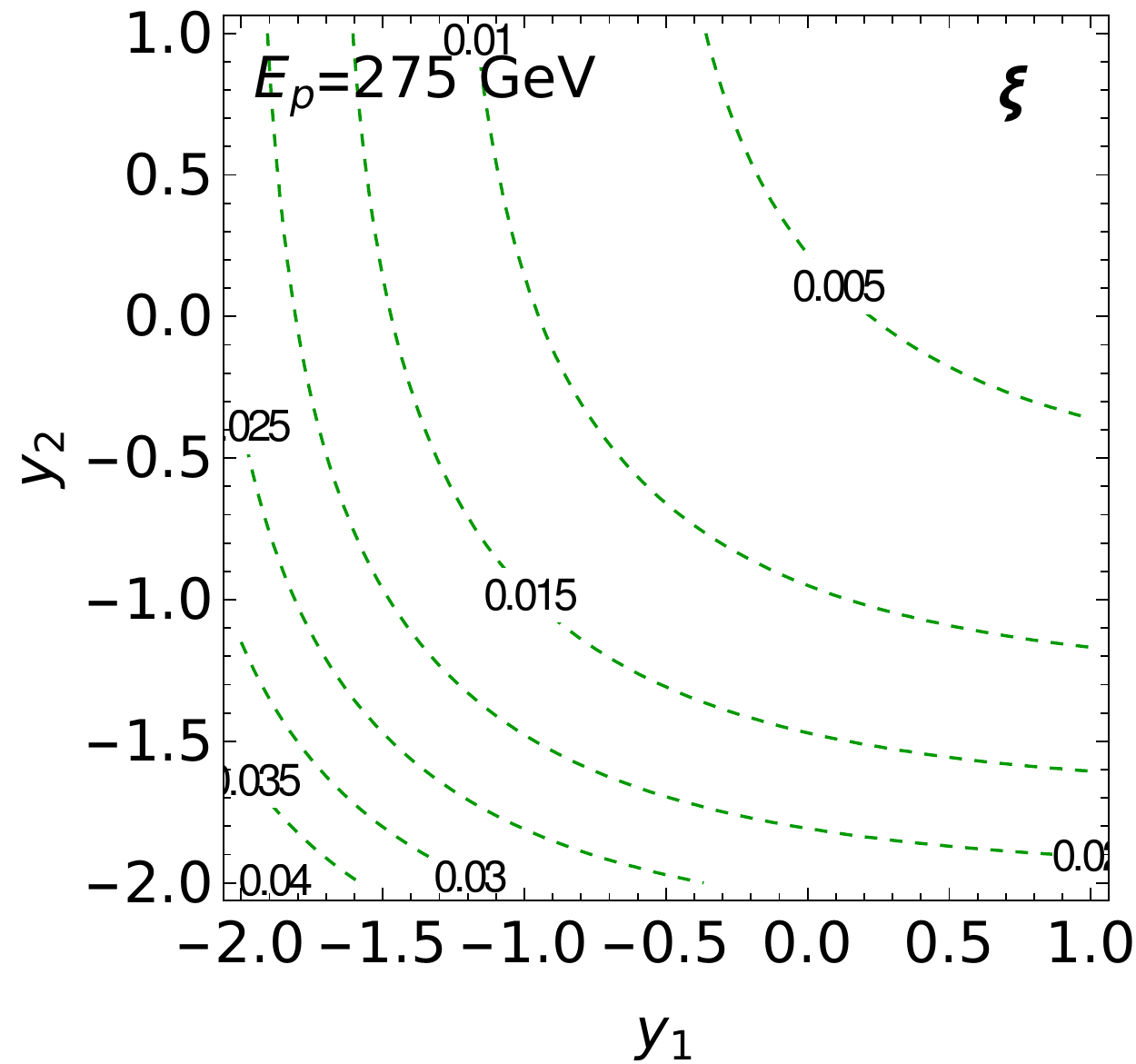}
\caption{\label{fig:xiRapidity}The contour plot illustrates the relation of
the skewedness variable $\xi=-\left(P_{f}^{+}-P_{i}^{+}\right)/\left(P_{f}^{+}+P_{i}^{+}\right)$
to the rest-frame quarkonia rapidities $y_{1},y_{2}$, for different
proton energies $E_{p}$, in EIC kinematics. For the sake of simplicity
we consider $J/\psi\,\eta_{c}$ production in the kinematics with
zero transverse momenta and zero photon virtuality $Q$, which gives
the dominant contribution to the total cross-section. Labels on contour
lines stand for the values of $\xi$.}
\end{figure}

In exclusive photoproduction, due to relation~(\ref{eq:xB-1}) it
is possible to express $\xi$ in terms of the produced quarkonia momenta.
In Figure~\ref{fig:xiRapidity} we illustrate the relation of the
variable $\xi$ to the rapidities $y_{1},y_{2}$ of the quarkonia
in the lab frame.

\begin{figure}
\includegraphics[scale=0.4]{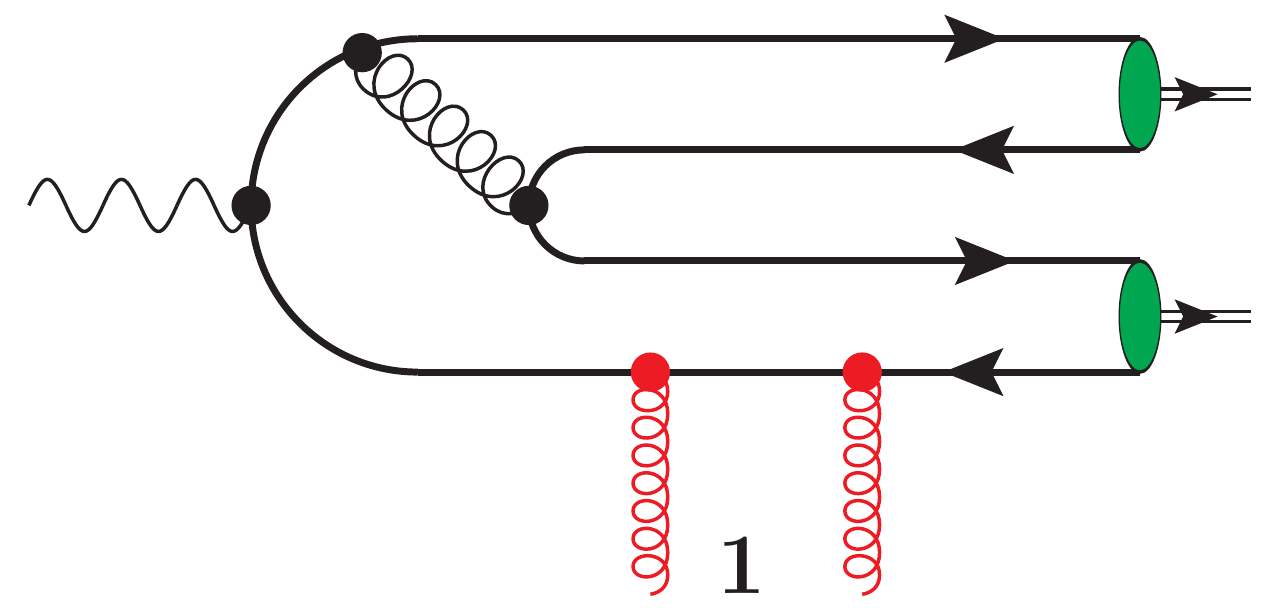}\includegraphics[scale=0.4]{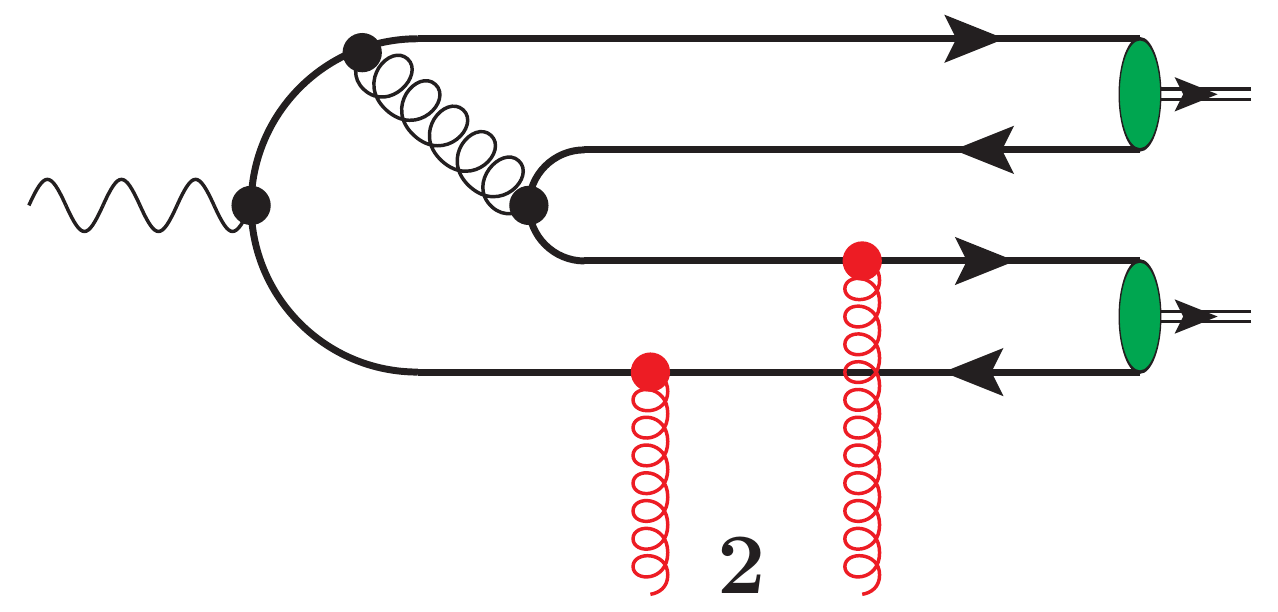}\includegraphics[scale=0.4]{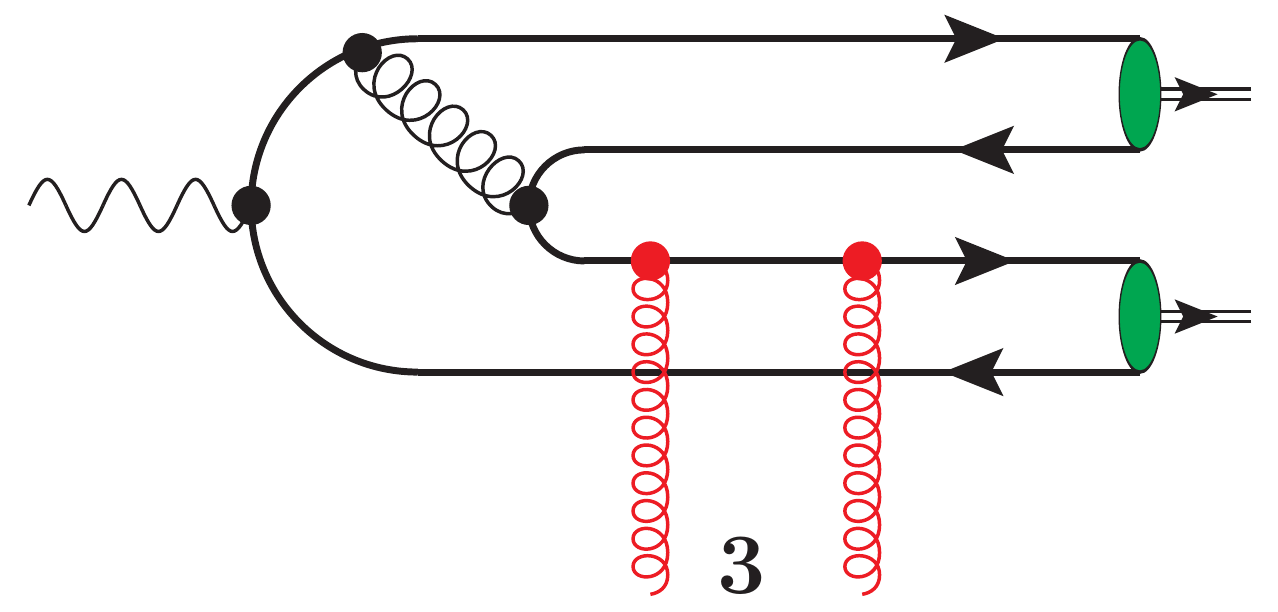}

\includegraphics[scale=0.4]{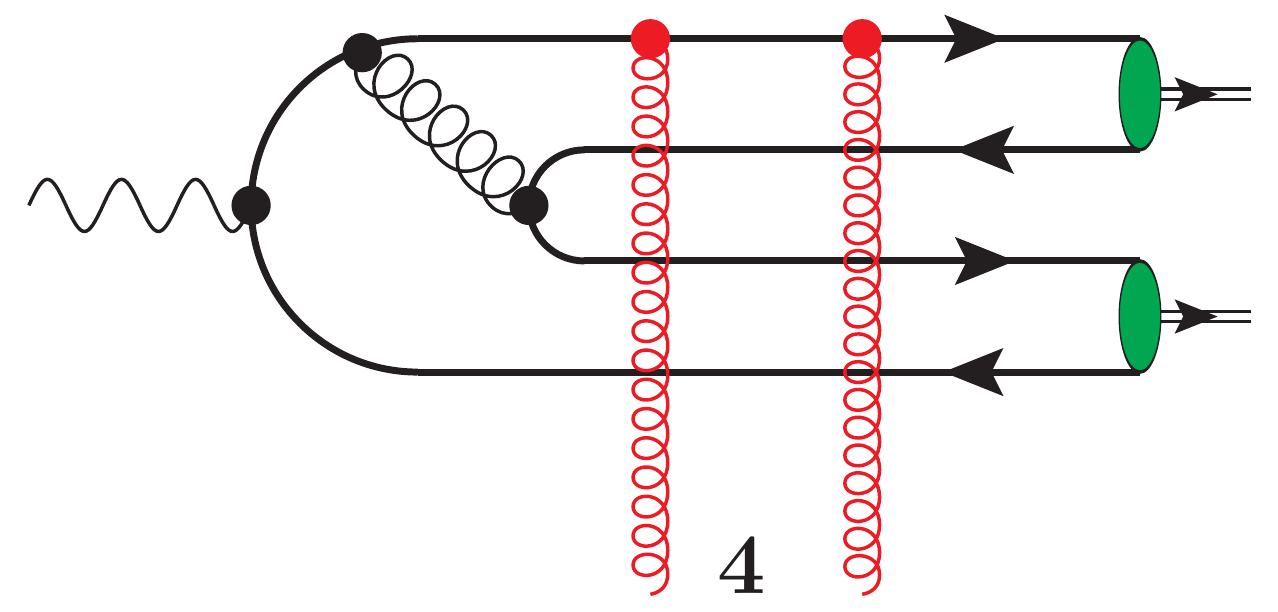}\includegraphics[scale=0.4]{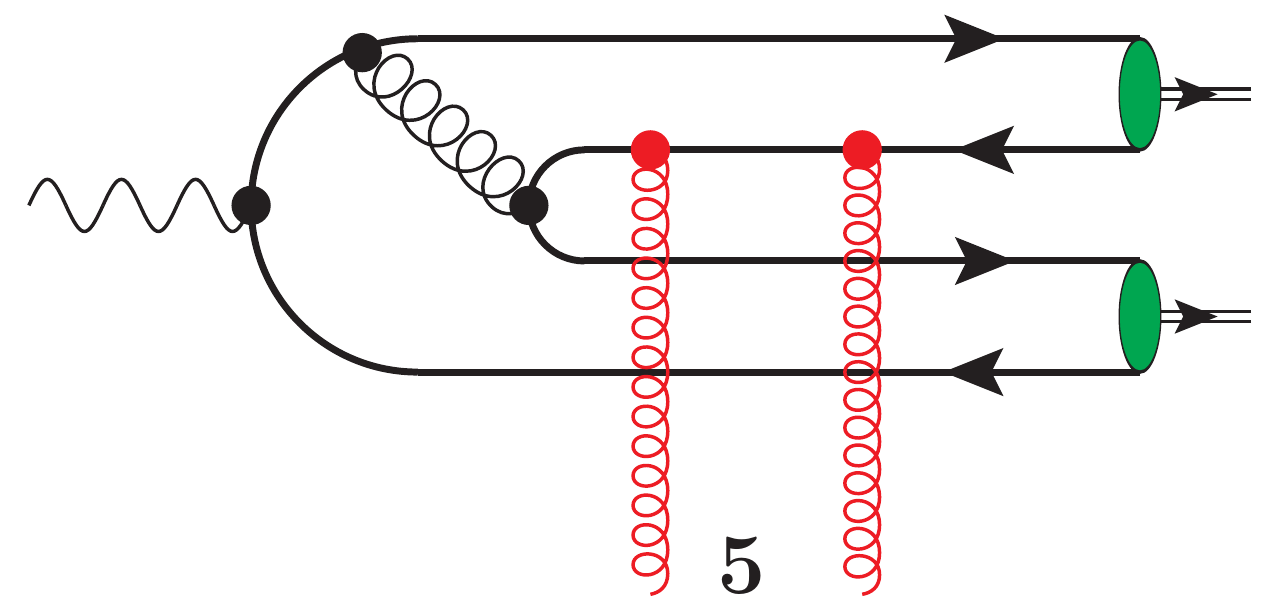}\includegraphics[scale=0.4]{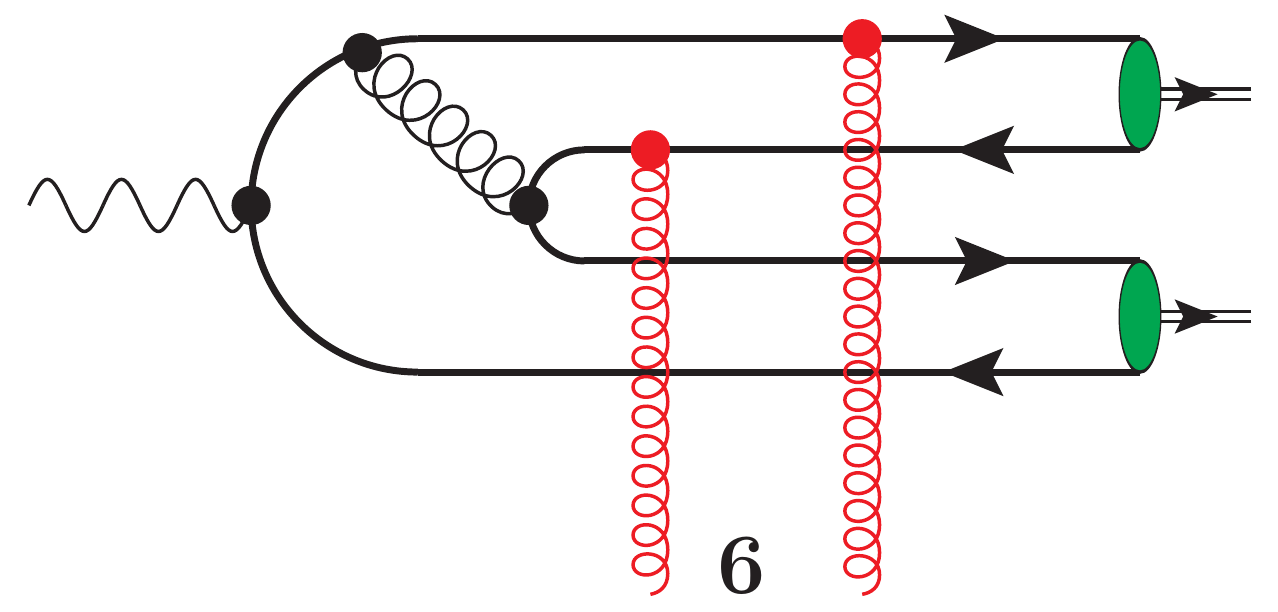}

\includegraphics[scale=0.4]{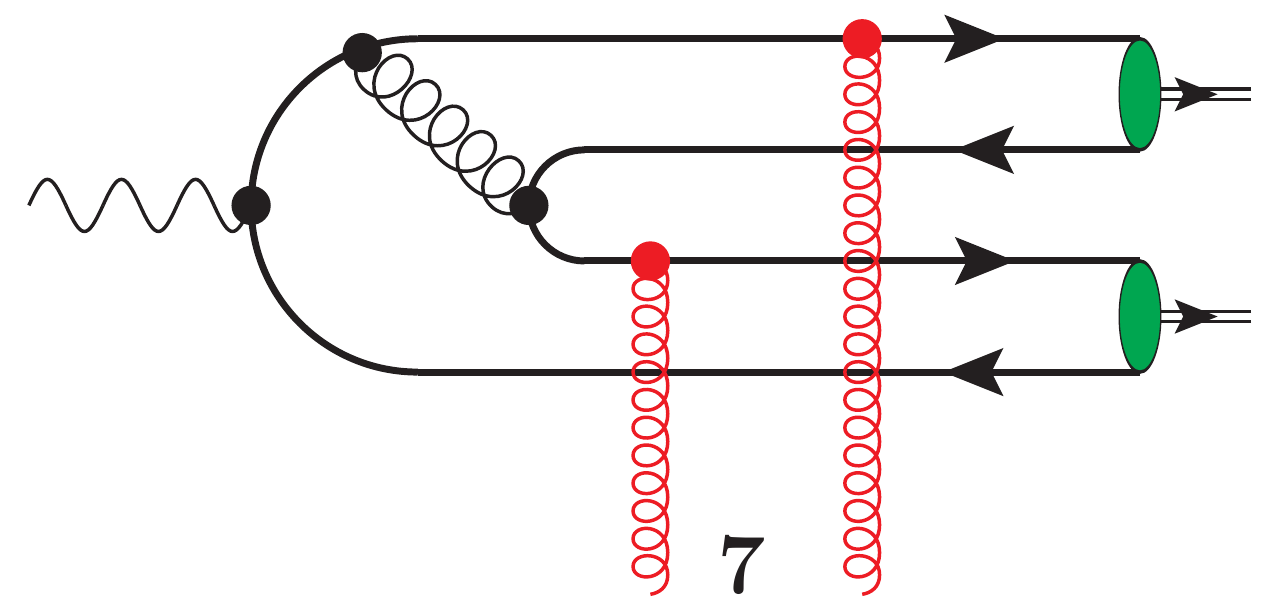}\includegraphics[scale=0.4]{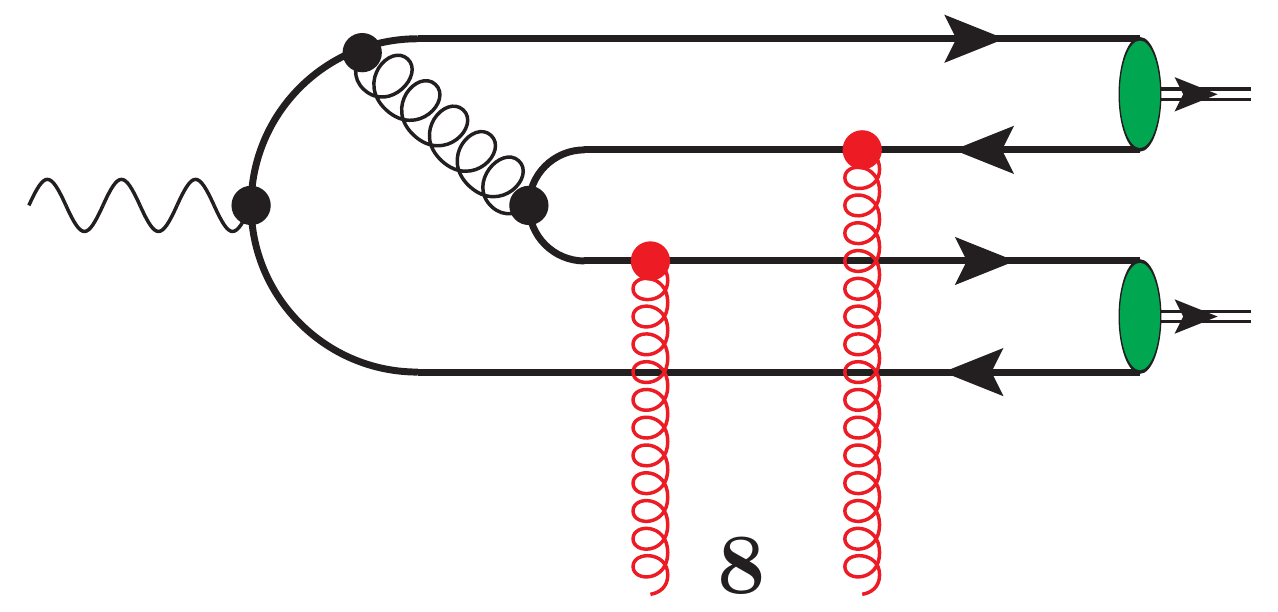}\includegraphics[scale=0.4]{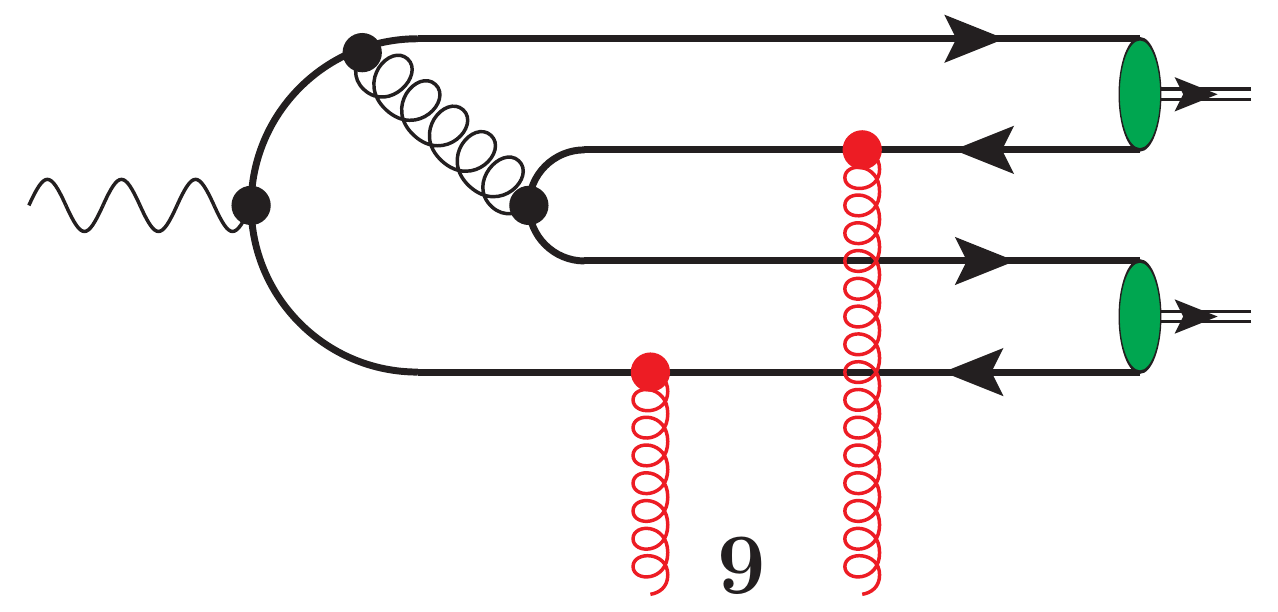}

\includegraphics[scale=0.4]{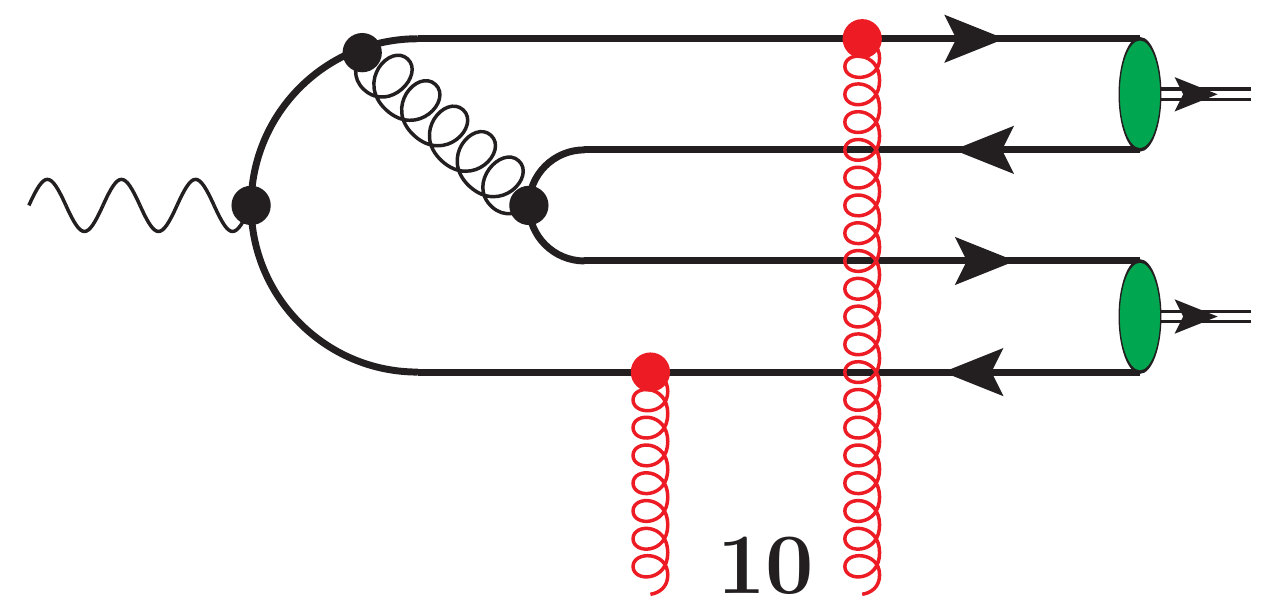}\includegraphics[scale=0.4]{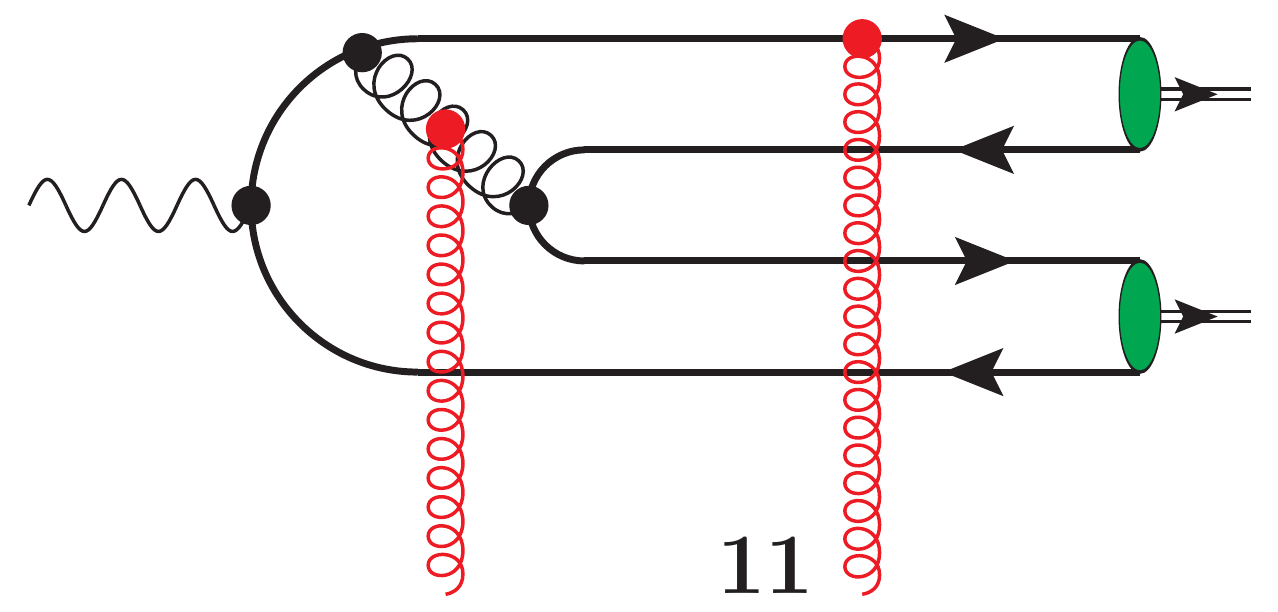}\includegraphics[scale=0.4]{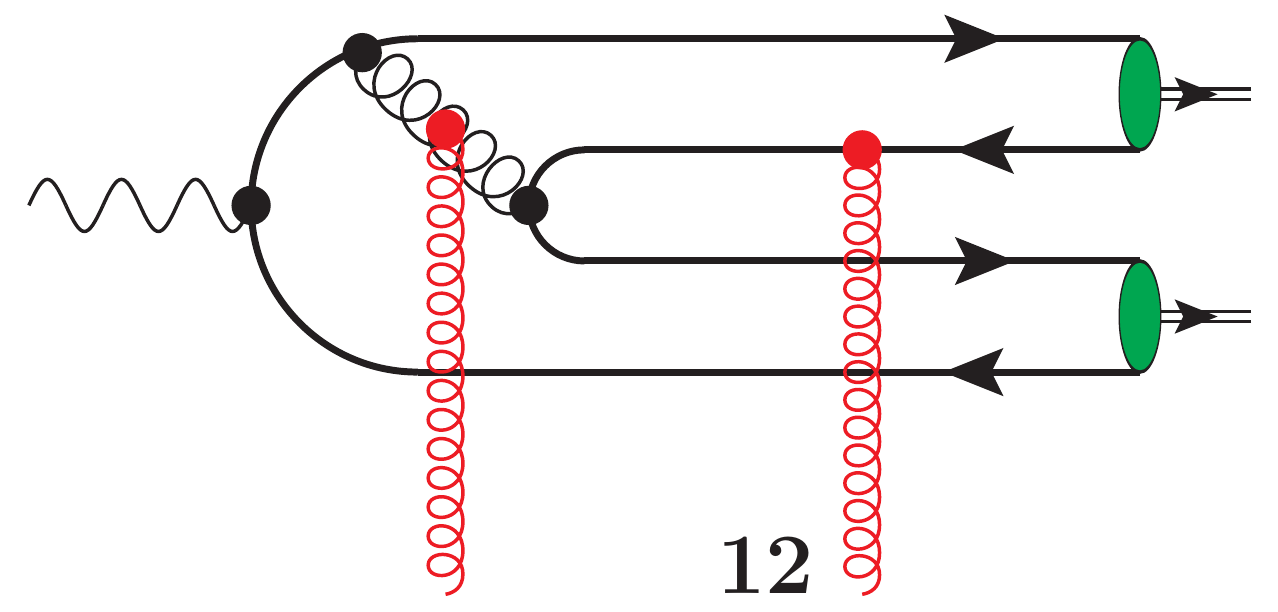}

\includegraphics[scale=0.4]{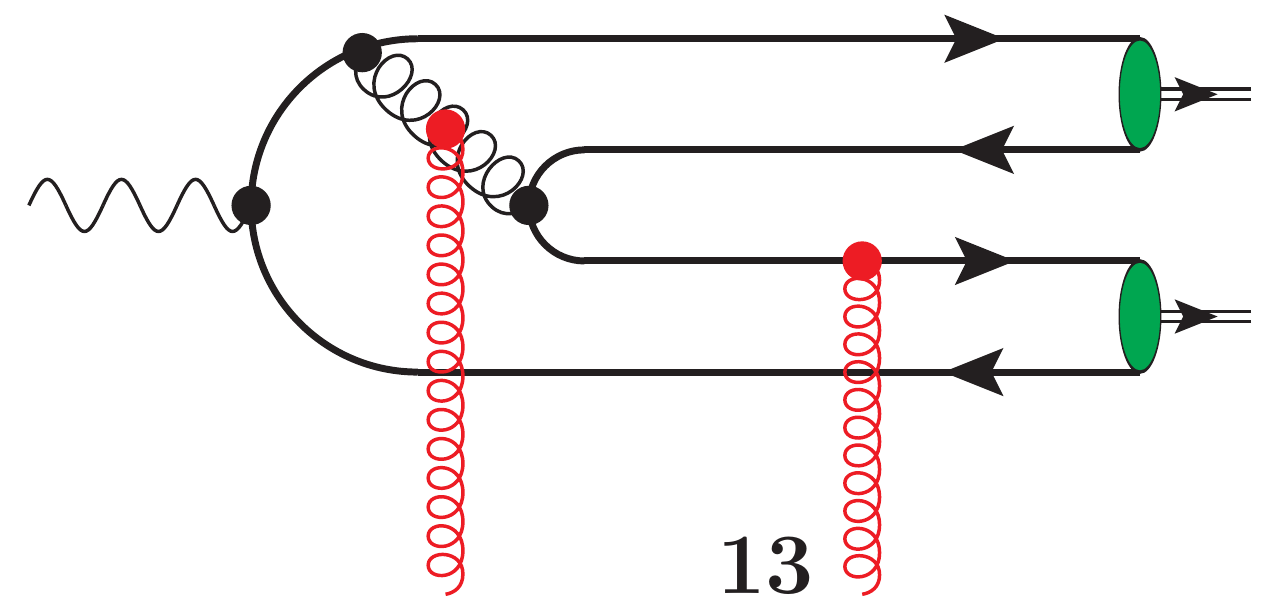}\includegraphics[scale=0.4]{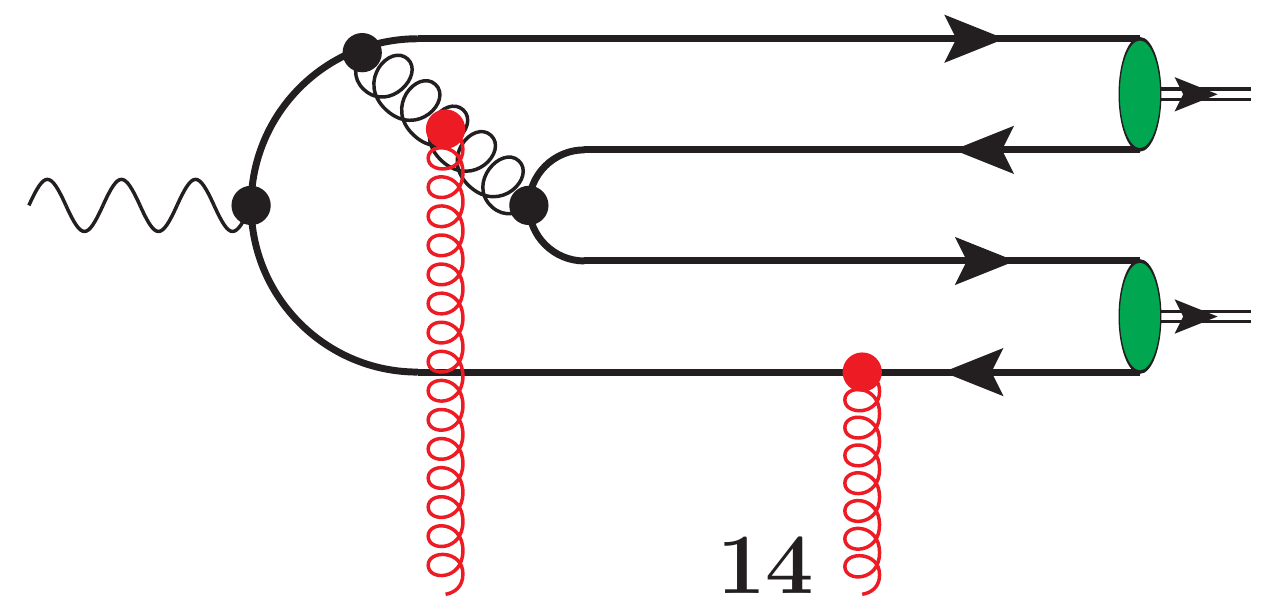}\includegraphics[scale=0.4]{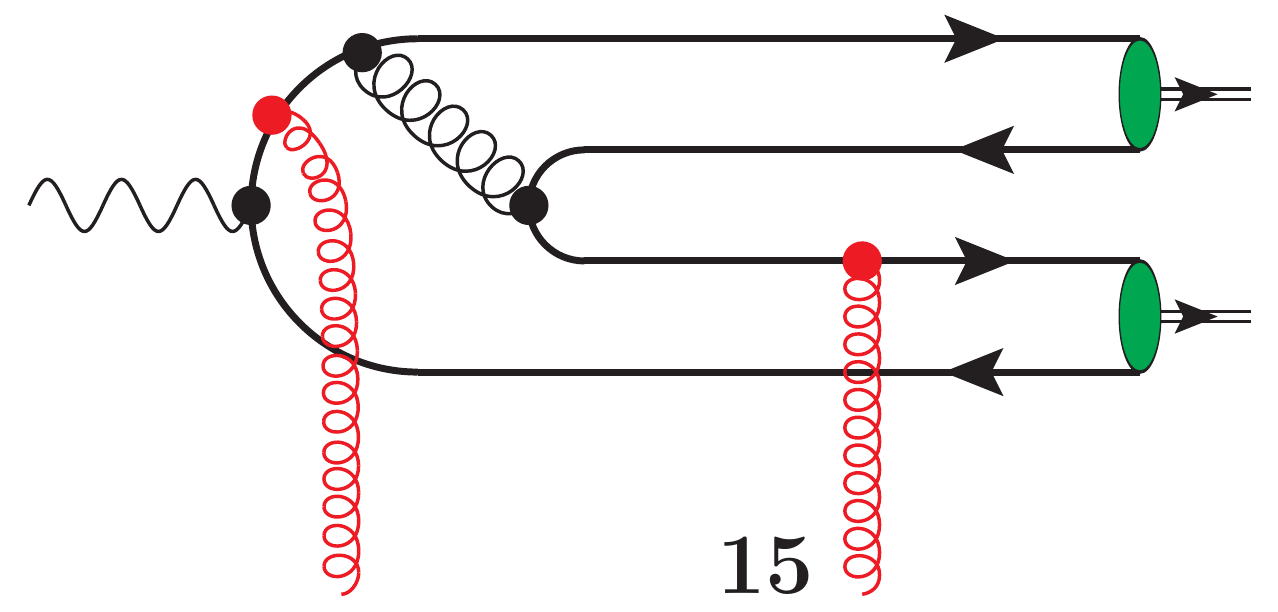}

\includegraphics[scale=0.4]{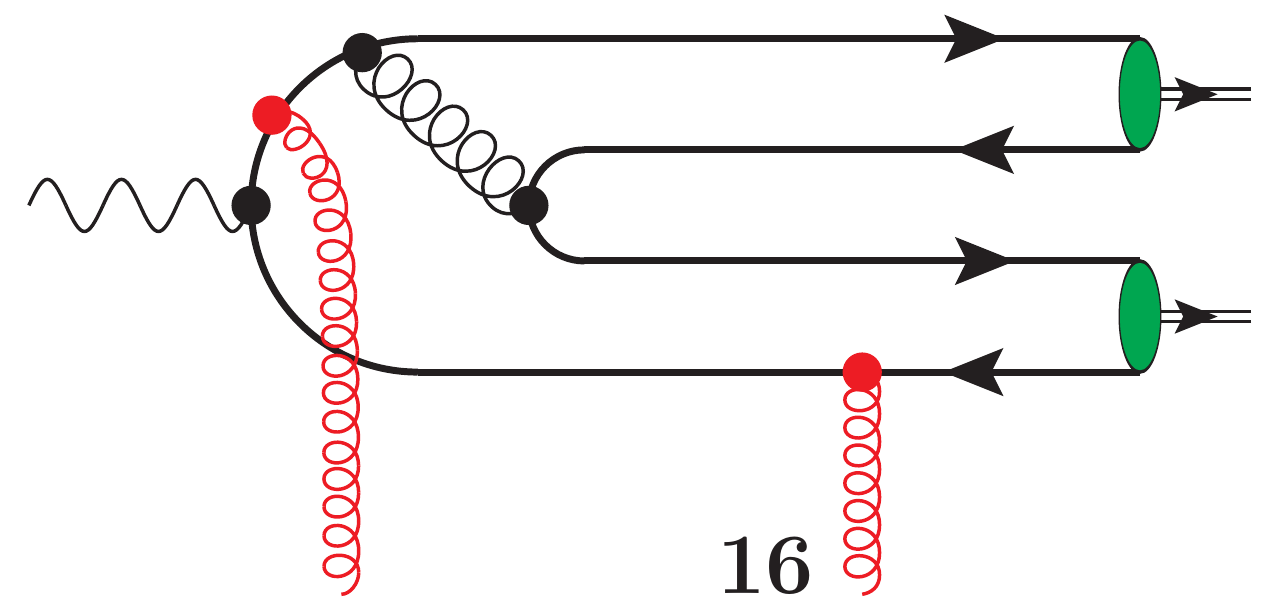}\includegraphics[scale=0.4]{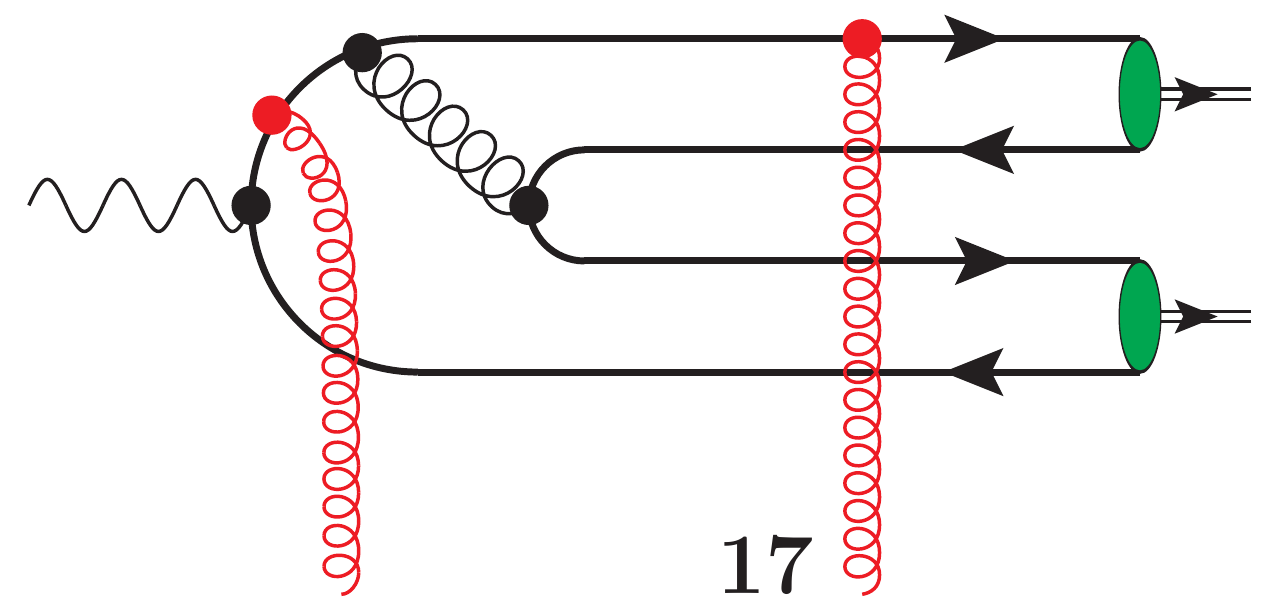}\includegraphics[scale=0.4]{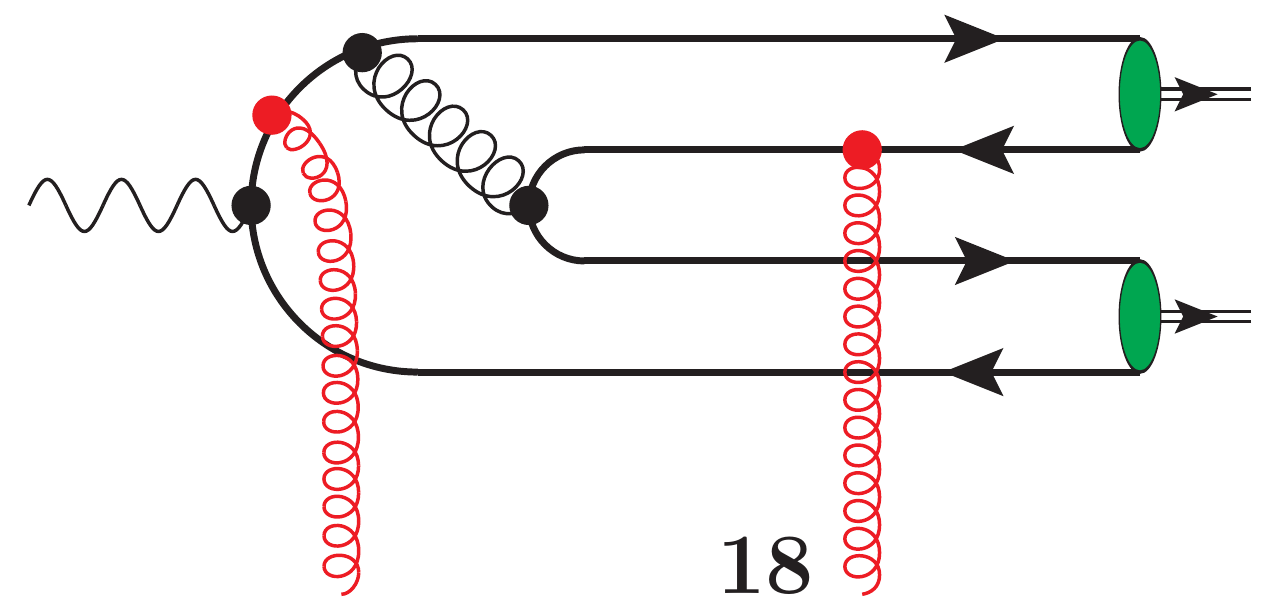}

\includegraphics[scale=0.4]{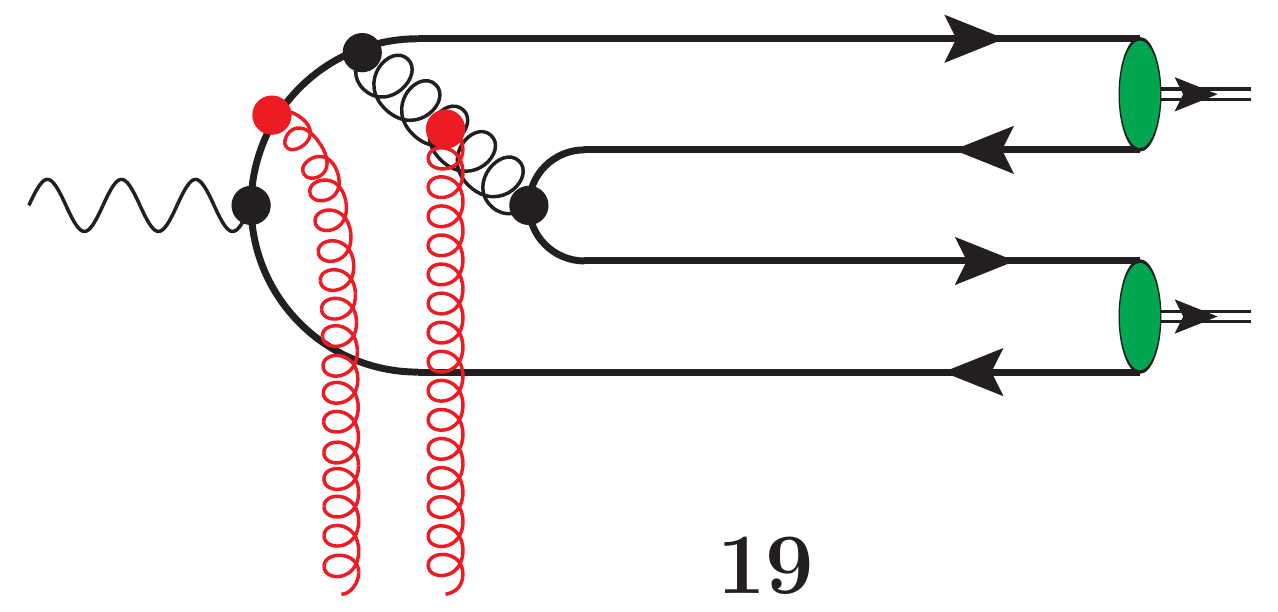}\includegraphics[scale=0.4]{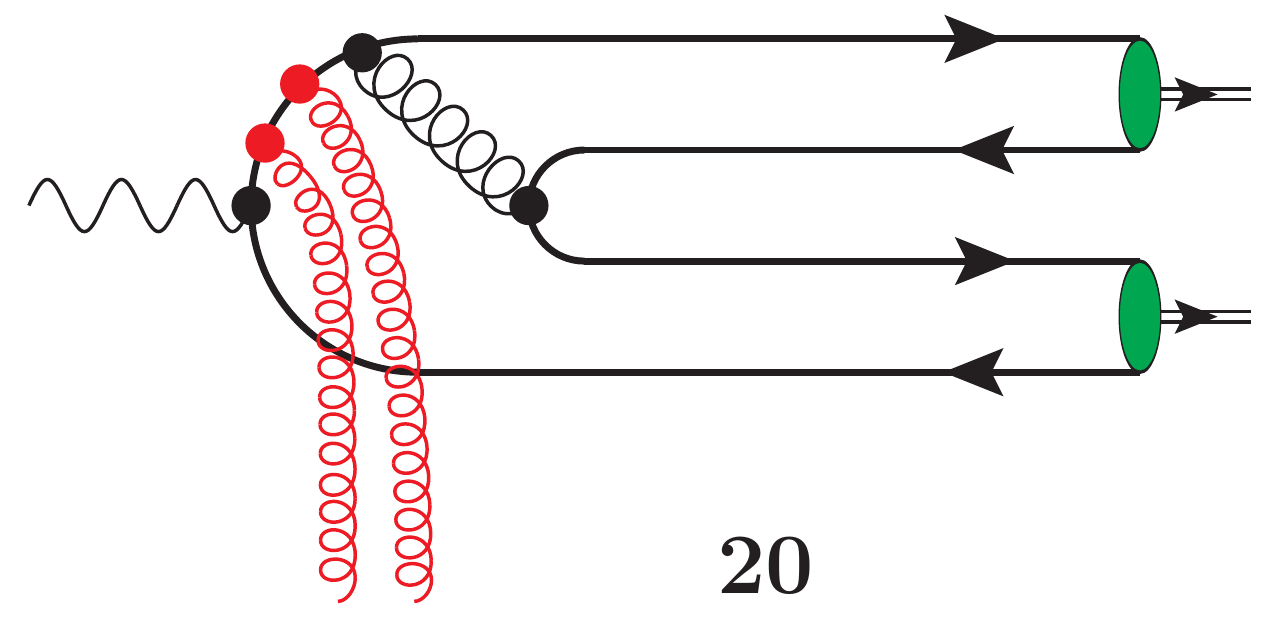}\includegraphics[scale=0.4]{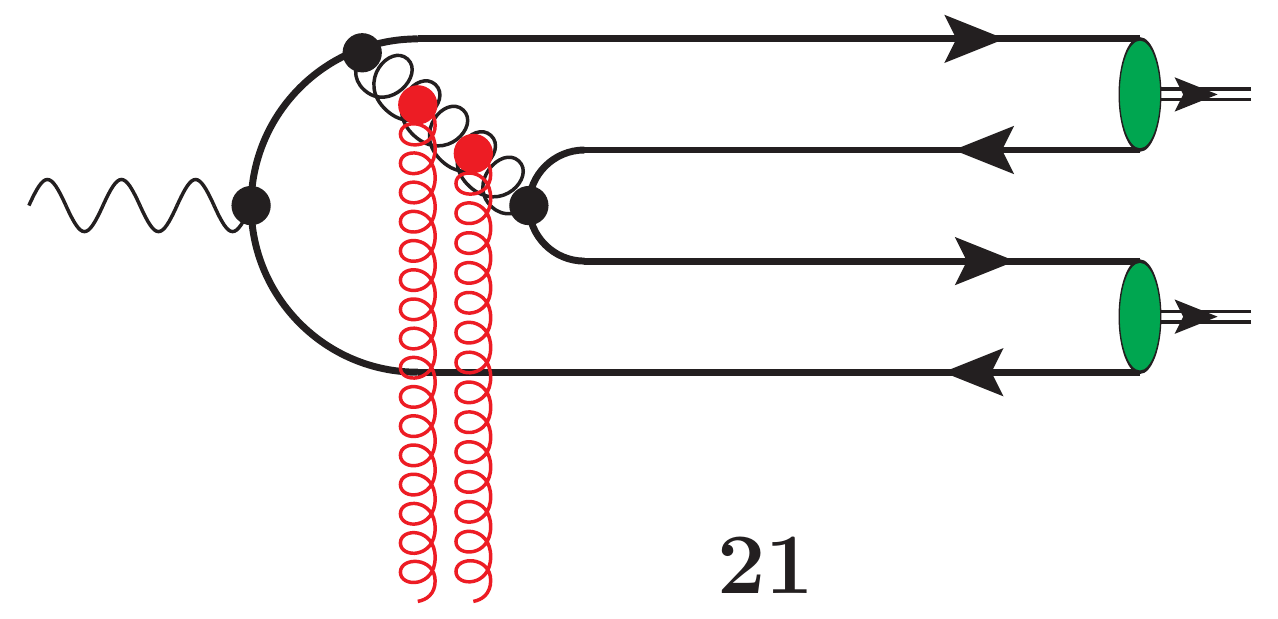}

\caption{\label{fig:Photoproduction-A}Schematic illustration of the single
quark loop (\textquotedblleft type-$A$\textquotedblright ) diagrams
which contribute to the meson pair production. In all plots it is
implied inclusion of diagrams which might be obtained by inversion
of heavy quark lines (\textquotedblleft charge conjugation\textquotedblright ).}
\end{figure}

\begin{figure}
\includegraphics[scale=0.4]{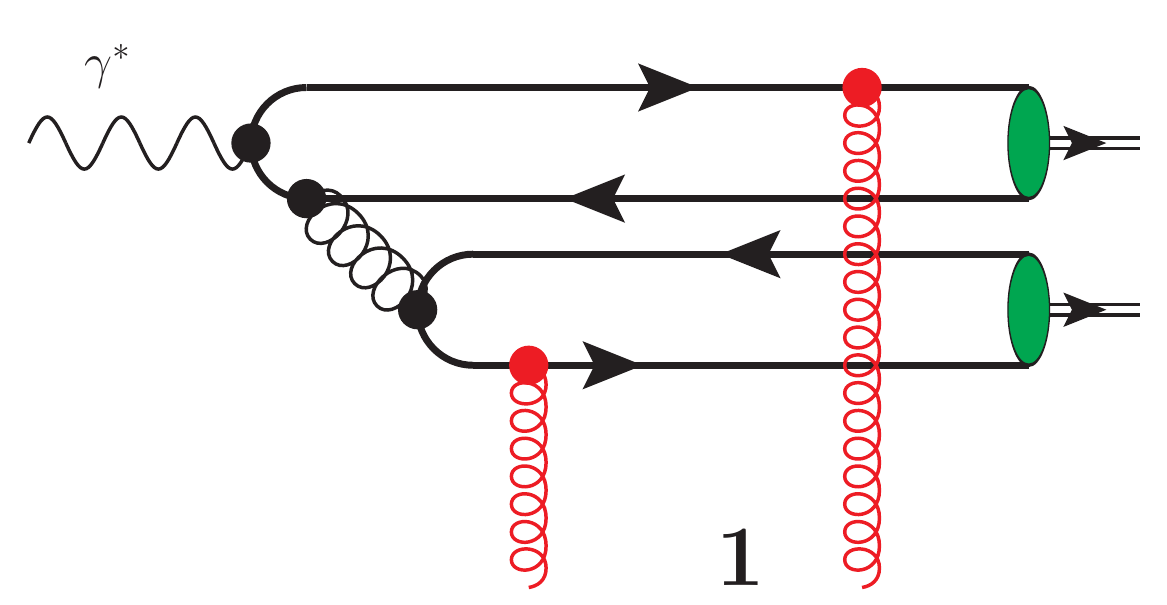}\includegraphics[scale=0.4]{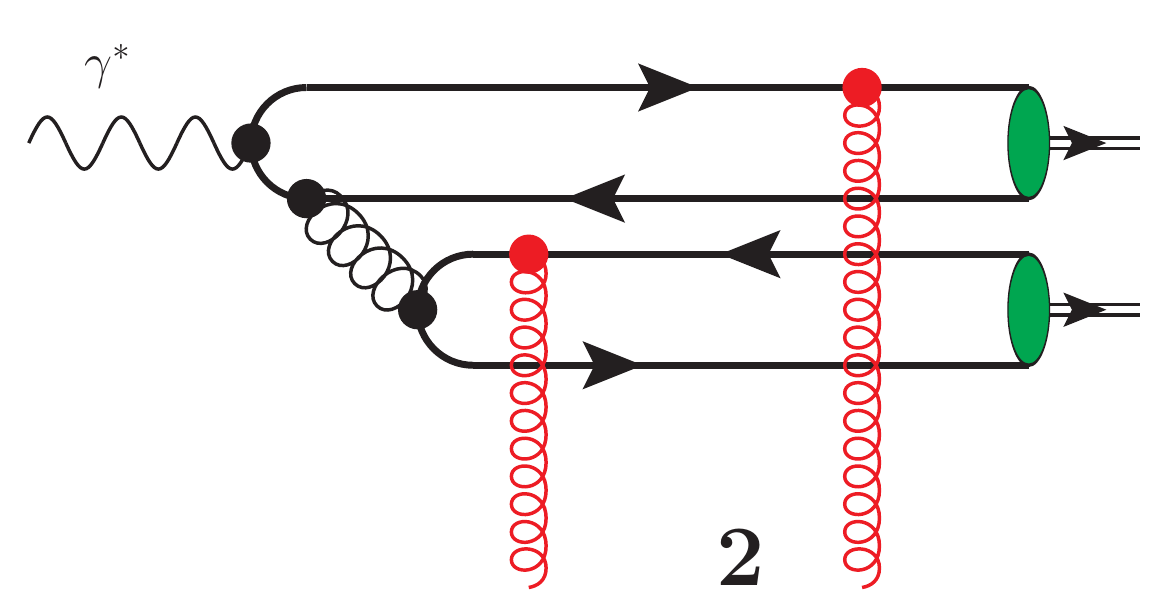}\includegraphics[scale=0.4]{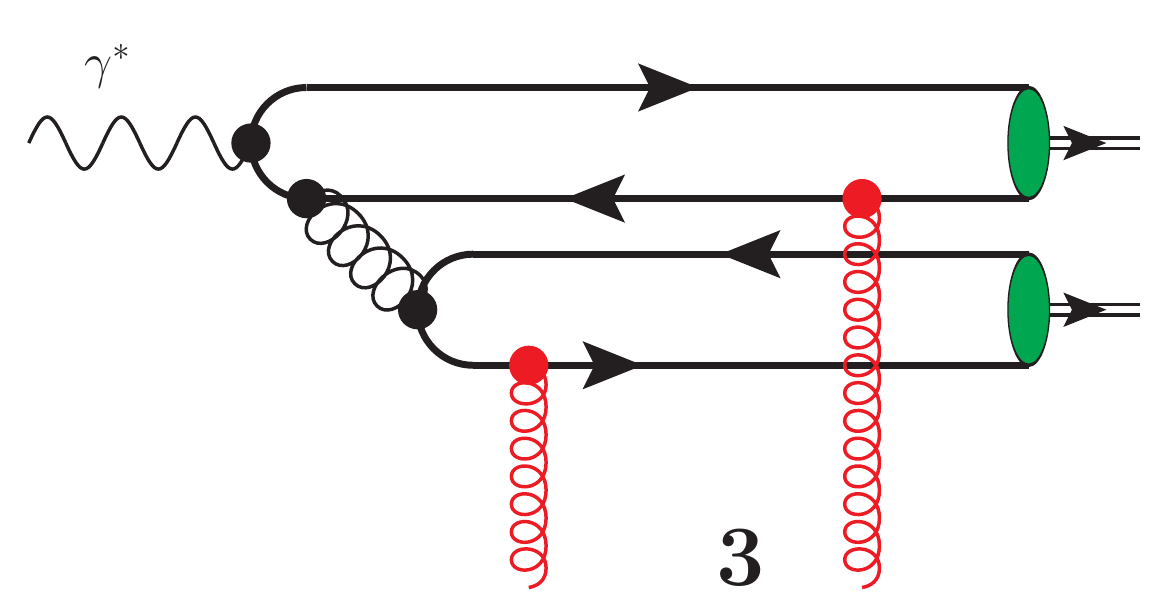}

\includegraphics[scale=0.4]{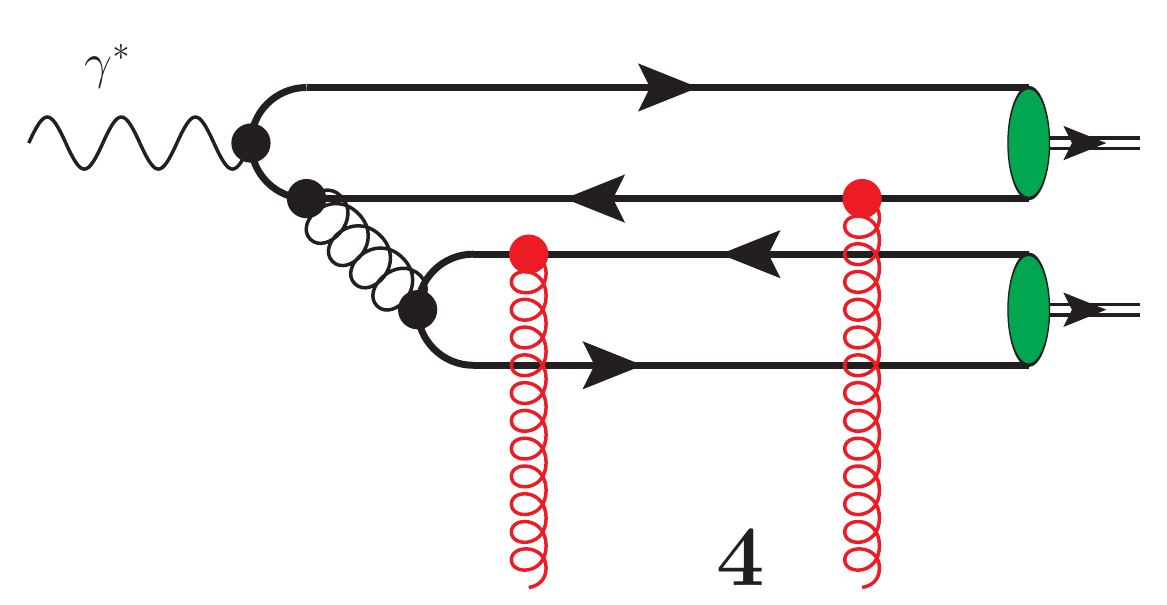}\includegraphics[scale=0.4]{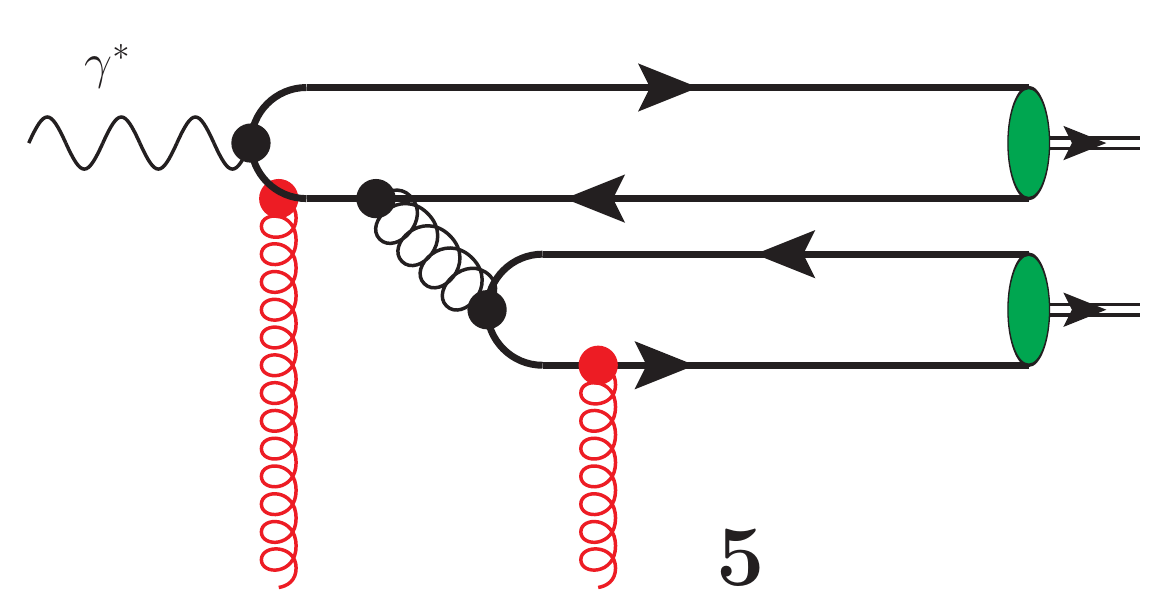}\includegraphics[scale=0.4]{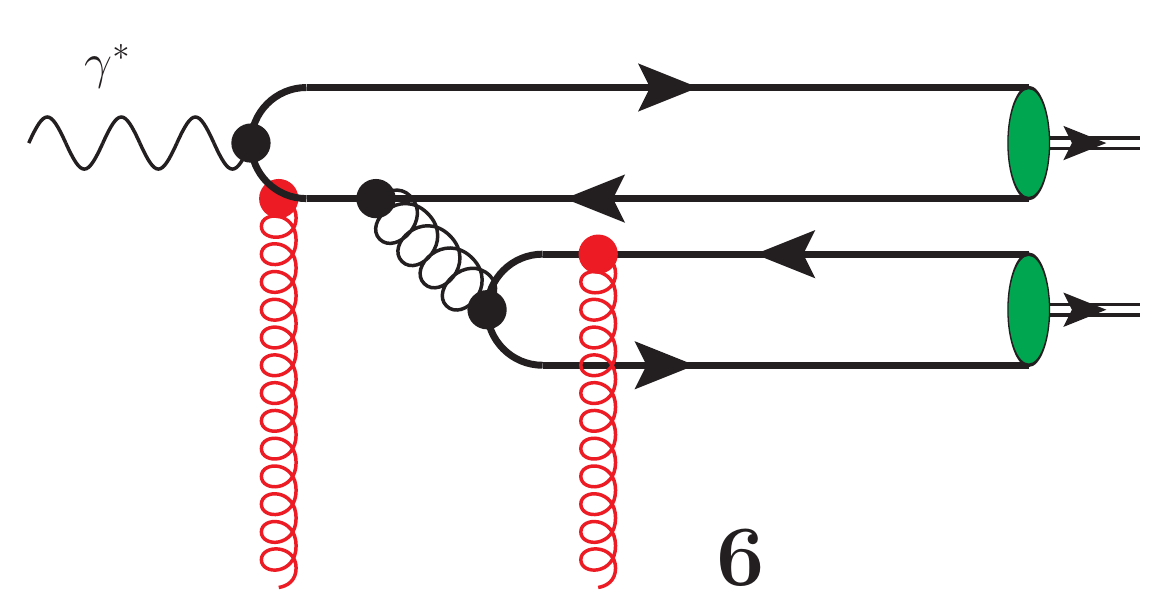}

\caption{\label{fig:Photoproduction-B}Schematic illustration of the double
quark loop ( \textquotedblleft type-$B$\textquotedblright ) diagrams
which contribute to the meson pair production. In all plots it is
implied inclusion of diagrams which might be obtained by inversion
of heavy quark lines (\textquotedblleft charge conjugation\textquotedblright )
in the first loop; diagrams 2,4,6 are related to diagrams 1,3,5 by
charge conjugation (symmetry $z_{2}\to1-z_{2}$).}
\end{figure}

In Bjorken kinematics, the leading order contribution to the amplitudes
of quarkonia production comes from the gluon GPDs. The contributions
of the light quark GPDs appear only via higher order loop corrections
and thus will be omitted in what follows. Furthermore, we will disregard
the contributions of the transversity gluon GPDs $H_{T}^{g},\,E_{T}^{g},\,\tilde{H}_{T}^{g},\,\tilde{E}_{T}^{g}$,
since at present there is no phenomenological parametrizations for
these GPDs, and existing experimental bounds suggest that they should
be negligibly small (see e.g. explanation in~\cite{Pire:2017yge,Goloskokov:2013mba}).
By their definition, the transversity GPDs appear in the amplitudes
multiplied by the momentum transfer to the proton $\Delta$, which
is small in the kinematics of interest, so we expect that their omission
should be numerically justified. The contribution of the chiral even
GPDs to the square of amplitude is given by

\begin{align}
\sum_{{\rm spins}}\left|\mathcal{A}_{\gamma p\to M_{1}M_{2}p}^{(\mathfrak{a})}\right|^{2} & =\frac{1}{\left(2-x_{B}\right)^{2}}\left[4\left(1-x_{B}\right)\left(\mathcal{H}_{\mathfrak{a}}\mathcal{H}_{\mathfrak{a}}^{*}+\tilde{\mathcal{H}}_{\mathfrak{a}}\tilde{\mathcal{H}}_{\mathfrak{a}}^{*}\right)-x_{B}^{2}\left(\mathcal{H}_{\mathfrak{a}}\mathcal{E}_{\mathfrak{a}}^{*}+\mathcal{E}_{\mathfrak{a}}\mathcal{H}_{\mathfrak{a}}^{*}+\tilde{\mathcal{H}}_{\mathfrak{a}}\tilde{\mathcal{E}}_{\mathfrak{a}}^{*}+\tilde{\mathcal{E}}_{\mathfrak{a}}\tilde{\mathcal{H}}_{\mathfrak{a}}^{*}\right)\right.\label{eq:AmpSq}\\
 & \qquad\left.-\left(x_{B}^{2}+\left(2-x_{B}\right)^{2}\frac{t}{4m_{N}^{2}}\right)\mathcal{E}_{\mathfrak{a}}\mathcal{E}_{\mathfrak{a}}^{*}-x_{B}^{2}\frac{t}{4m_{N}^{2}}\tilde{\mathcal{E}}_{\mathfrak{a}}\tilde{\mathcal{E}}_{\mathfrak{a}}^{*}\right],\qquad\mathfrak{a}=L,T\nonumber 
\end{align}
where the index $\mathfrak{a}$ refers to longitudinal or transverse
photons, and, inspired by similar analysis of Compton scattering and
single-meson deeply virtual production~\cite{Belitsky:2001ns,Belitsky:2005qn},
we introduced the double meson form factors

\begin{align}
\mathcal{H}_{\mathfrak{a}}\left(y_{1},y_{2},t\right) & =\int dx\,c_{\mathfrak{a}}\left(x,\,y_{1},\,y_{2}\right)H_{g}\left(x,\xi,t\right),\quad\mathcal{E}_{\mathfrak{a}}\left(y_{1},y_{2},t\right)=\int dx\,c_{\mathfrak{a}}\left(x,\,y_{1},\,y_{2}\right)E_{g}\left(x,\xi,t\right),\label{eq:Ha}\\
\tilde{\mathcal{H}}_{\mathfrak{a}}\left(y_{1},y_{2},t\right) & =\int dx\,\tilde{c}_{\mathfrak{a}}\left(x,\,y_{1},\,y_{2}\right)\tilde{H}_{g}\left(x,\xi,t\right),\quad\tilde{\mathcal{E}}_{\mathfrak{a}}\left(y_{1},y_{2},t\right)=\int dx\,\tilde{c}_{\mathfrak{a}}\left(x,\,y_{1},\,y_{2}\right)\tilde{E}_{g}\left(x,\xi,t\right),\label{eq:ETildeA}
\end{align}
where the variable $\xi$ should be understood as a function of $y_{1},y_{2}$,
as defined in~(\ref{eq:XiDef}). The corresponding partonic amplitudes
$c_{\mathfrak{a}},\,\tilde{c}_{\mathfrak{a}}$ might be evaluated
perturbatively, taking into account the diagrams shown in the Figures~\ref{fig:Photoproduction-A},~\ref{fig:Photoproduction-B}.
Since we assume that produced quarkonia are well-separated from each
other kinematically, the final Fock state of the system is a direct
product of Fock states of individual quarkonia, and thus it is possible
to express the amplitudes $c_{\mathfrak{a}}$,$\tilde{c}_{\mathfrak{a}}$
in terms of the wave functions or distribution amplitudes which encode
the nonperturbative structure of individual quarkonia. According to
NRQCD and potential models, the dominant Fock state in chamronium
is the color singlet $\bar{c}c$ pair in $^{3}S_{1}^{[1]}$ state
for $J/\psi$ , and $^{1}S_{0}^{[1]}$ state for $\eta_{c}$. The
distribution of the quarks over the light-cone momenta might be described
by the corresponding distribution amplitudes $\Phi_{M}\left(z_{a}\right)$,
where $z_{a}$ is the fraction of the quarkonium light-cone momentum
carried by the quark. The relative velocity of heavy quarks inside
the quarkonia in the heavy quark mass limit is suppressed as $\sim\alpha_{s}\left(m_{Q}\right)\ll1$,
and for this reason both heavy quarks inside each quarkonia carry
approximately half of its momentum. In this approximation, we may
replace both distribution amplitudes $\Phi_{\eta},\,\Phi_{J/\psi}$
with 
\begin{equation}
\Phi_{J/\psi}\left(z\right)\approx f_{J/\psi}\,\delta\left(z-1/2\right),\quad\Phi_{\eta_{c}}\left(z\right)\approx f_{\eta_{c}}\,\delta\left(z-1/2\right),
\end{equation}
where $f_{J/\psi},f_{\eta_{c}}$ are the (nonperturbative) decay constants
of the corresponding quarkonia states. In the language of NRQCD, $f_{J/\psi}^{2}$
and $f_{\eta_{c}}^{2}$ are proportional to the color singlet Long
Distance Matrix Elements (LDMEs) $\left\langle \mathcal{O}_{J/\psi}^{[1]}\left(^{3}S_{1}^{[a]}\right)\right\rangle $,
$\left\langle \mathcal{O}_{\eta_{c}}^{[1]}\left(^{1}S_{0}^{[a]}\right)\right\rangle $
respectively~\cite{DVMPcc1,Baranov:2012vu}. In this approach, the
functions $c_{\mathfrak{a}},\,\tilde{c}_{\mathfrak{a}}$ might be
related to partonic-level amplitudes $C_{\mathfrak{a}},\,\tilde{C}_{\mathfrak{a}}$
as 
\begin{align}
c_{\mathfrak{a}}\left(x,\,y_{1},\,y_{2}\right) & =\int\,dz_{1}\,dz_{2}C_{\mathfrak{a}}\left(x,\,z_{1},\,z_{2},\,y_{1},\,y_{2}\right)\Phi_{\eta}\left(z_{1}\right)\Phi_{J/\psi}\left(z_{2}\right)\approx f_{J/\psi}f_{\eta_{c}}C_{\mathfrak{a}}\left(x,\,\frac{1}{2},\,\frac{1}{2},\,y_{1},\,y_{2}\right)+\mathcal{O}\left(\alpha_{s}(m_{c})\right)\\
\tilde{c}_{\mathfrak{a}}\left(x,\,y_{1},\,y_{2}\right) & =\int\,dz_{1}\,dz_{2}\tilde{C}_{\mathfrak{a}}\left(x,\,z_{1},\,z_{2},\,y_{1},\,y_{2}\right)\Phi_{\eta}\left(z_{1}\right)\Phi_{J/\psi}\left(z_{2}\right)\approx f_{J/\psi}f_{\eta_{c}}\tilde{C}_{\mathfrak{a}}\left(x,\,\frac{1}{2},\,\frac{1}{2},\,y_{1},\,y_{2}\right)+\mathcal{O}\left(\alpha_{s}(m_{c})\right),
\end{align}
and $C_{\mathfrak{a}},\,\tilde{C}_{\mathfrak{a}}$ might be evaluated
in perturbative QCD. Assuming equal sharing of quarkonium momentum
between constituent quarks,  it is possible to show that the typical
virtuality of the gluon connecting different heavy lines is parametrically
of order $\sim M_{12}^{2}/4$ for the diagrams in Figures~\ref{fig:Photoproduction-A},
and of order $\sim{\rm min}\left(M_{1}^{2},M_{2}^{2}\right)$ for
the diagrams in Figure~\ref{fig:Photoproduction-B}. This justifies
the applicability of perturbation theory for evaluation of $C_{\mathfrak{a}},\,\tilde{C}_{\mathfrak{a}}$
, even for the diagrams which include 3-gluon vertices in Figure~\ref{fig:Photoproduction-A}.
The full expressions for the amplitudes are provided in Appendix~\ref{sec:CoefFunction}.

The contribution of longitudinal photons to $\mathcal{C}_{\mathfrak{a}}$
vanishes in the limit of small $\boldsymbol{p}_{a\perp}\ll M,Q$ in
view of combined Lorentz- and $P$-parity. The contributions of the
longitudinal photons to $\tilde{\mathcal{C}}_{\mathfrak{a}}$ do not
vanish in this limit, although in the cross-section it appears in
convolution with numerically small helicity flip gluon GPDs $\tilde{H}_{g},\,\tilde{E}_{g}.$
Since for quasireal photons the contribution of longitudinal photons
is suppressed by a factor $Q/m_{Q}$, we will disregard it altogether
in the total (unpolarized) cross-section.

\begin{figure}
\includegraphics[width=6cm]{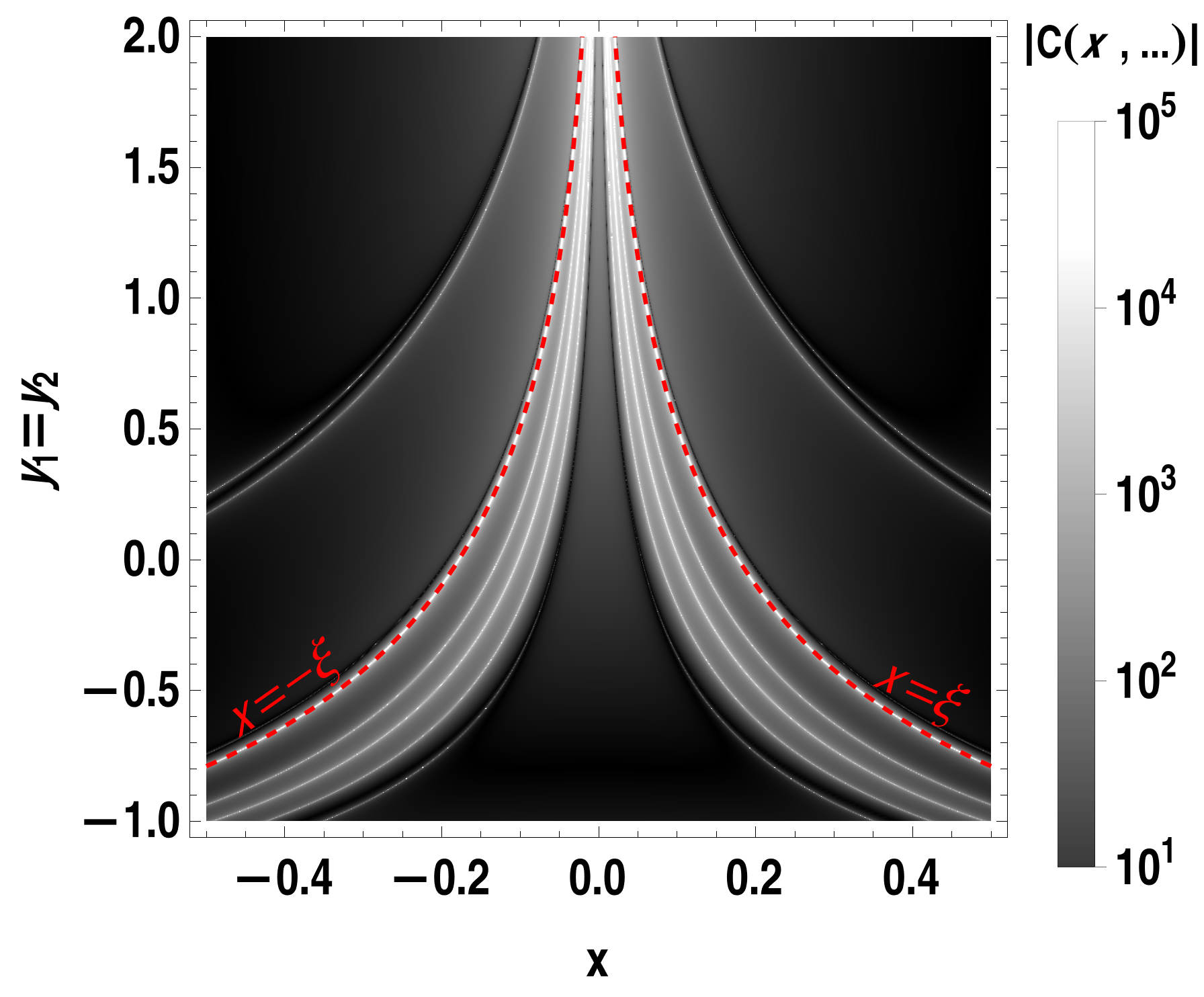}\includegraphics[width=6cm]{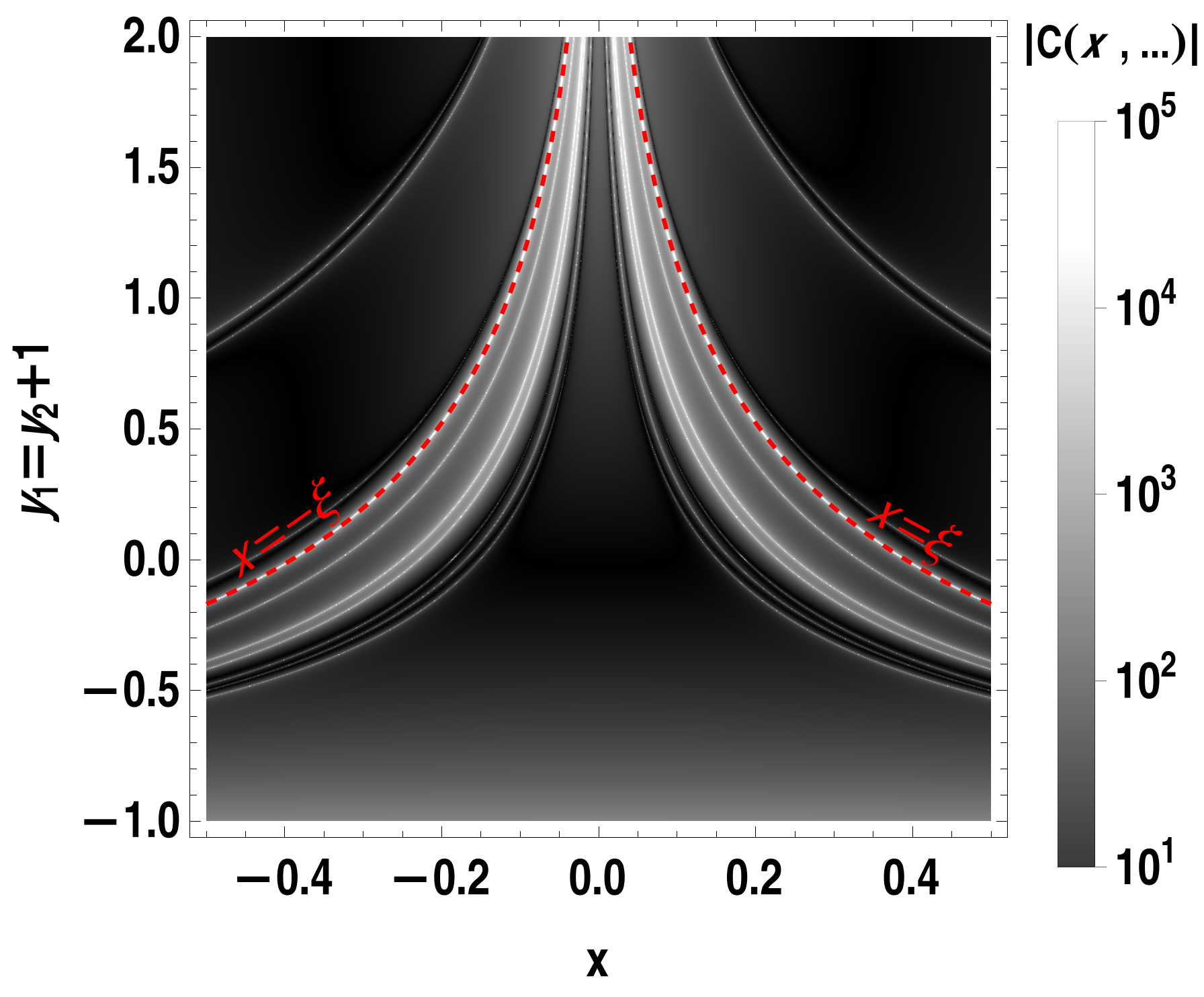}\includegraphics[width=6cm]{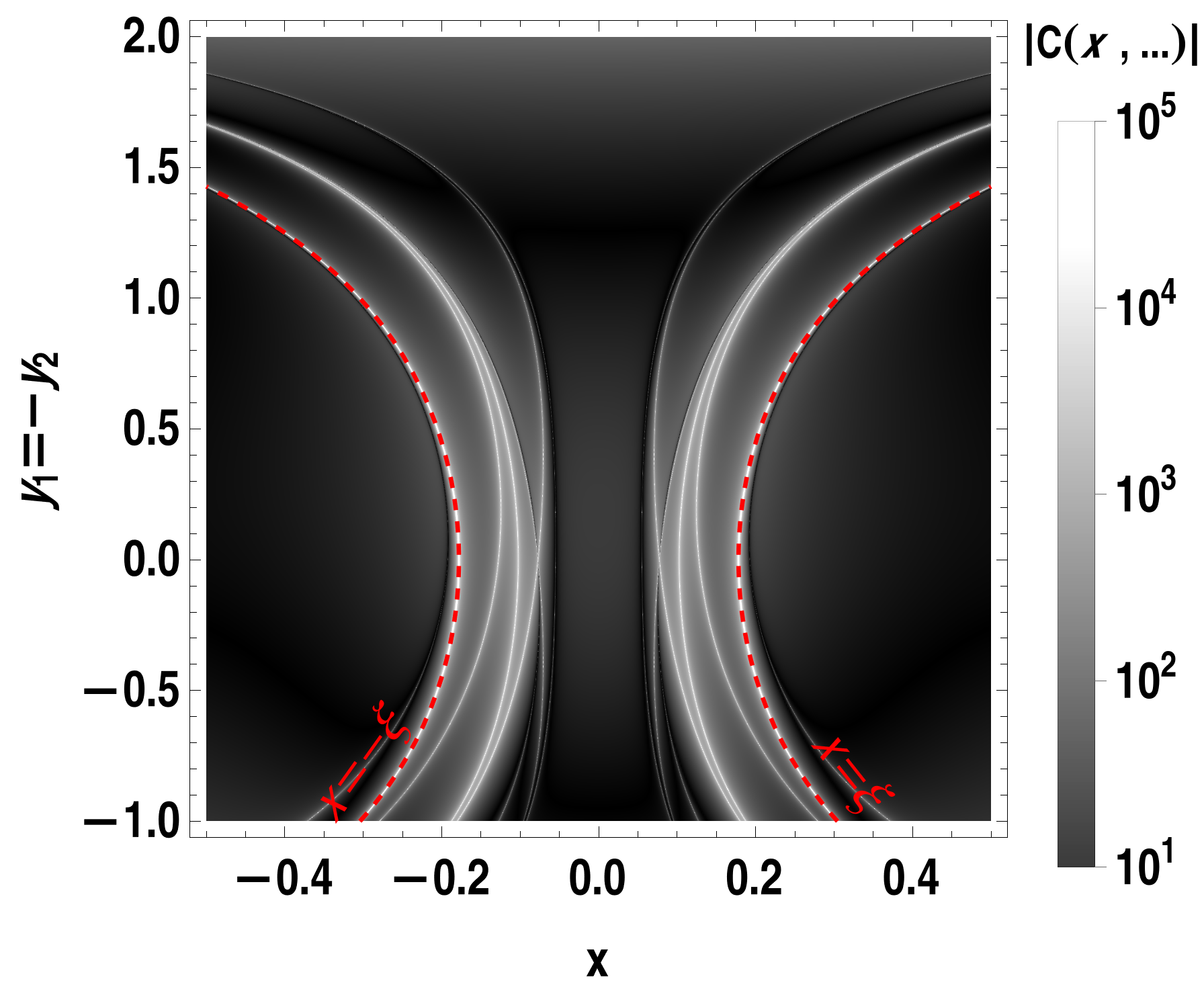}\caption{\label{fig:CoefFunction} Density plot which illustrates the coefficient
function $C_{T}$ (in relative units) as a function of the variables
$x$ and quarkonia rapidities $y_{1},y_{2}$. Left, central and right
plots correspond to $y_{1}=y_{2},\,y_{1}=y_{2}+1$ and $y_{1}=-y_{2}$
respectively. Rapidities are taken in the lab frame, for proton energy
$E_{p}^{(1)}={\rm 41}\,{\rm GeV}$; for other proton energies rapidities
should be shifted by $\text{\ensuremath{\Delta y=}}\ln\left(E_{p}/E_{p}^{(1)}\right)$.
For the sake of definiteness, we considered the photoproduction regime
($Q=0$) in all plots. White lines effectively demonstrate the position
of the poles $x_{k}^{\ell}$ of the coefficient function~(\ref{eq:Monome}).
For reference, we marked with red dashed lines the poles which correspond
to $x=\pm\xi$, where the skewedness $\xi=\xi\left(y_{1},y_{2}\right)$
was evaluated using~(\ref{eq:xB-1},\ref{eq:XiDef}). }
\end{figure}

The dependence on the variable $x$ in the coefficient functions might
be represented as a linear superposition of rational expressions
\begin{equation}
C_{\mathfrak{a}}\left(x,\,\frac{1}{2},\,\frac{1}{2},\,y_{1},\,y_{2}\right)\sim\sum_{\ell}\frac{\mathcal{P}_{\ell}\left(x\right)}{\prod_{k=1}^{n_{\ell}}\left(x-x_{k}^{(\ell)}+i0\right)}\label{eq:Monome}
\end{equation}
where $\mathcal{P}_{\ell}\left(x\right)$ is a smooth polynomial of
the variable $x$, and the denominator of each term in the sum~(\ref{eq:Monome})
might include a polynomial with up to $n_{\ell}=5$ nodes $x_{k}^{(\ell)}$
in the region of integration. The integral near the poles exists only
in the principal value sense and is evaluated using 
\begin{equation}
\frac{1}{x-x_{k}^{(\ell)}+i0}={\rm P.V.}\left(\frac{1}{x-x_{k}^{(\ell)}}\right)-i\pi\,\delta\left(x-x_{k}^{(\ell)}\right)
\end{equation}
The position of the poles $x_{k}^{(\ell)}$ depends on all kinematic
variables $y_{1},\,y_{2},\,Q$. In Figure~\ref{fig:CoefFunction}
we show the density plot which illustrates the behavior of the coefficient
function $C_{T}\left(x,\,\xi,\,z_{1}=z_{2}=1/2,\,y_{1},\,y_{2}\right)$
as a function of its arguments. While in the convolution integrals~(\ref{eq:Ha}-\ref{eq:ETildeA})
we need to take integral over all $x\in(-1,1)$, we expect that a
sizable contribution comes from the region near the poles of the coefficient
function. From the Figure~\ref{fig:CoefFunction} we can see that
in the coefficient function there are several poles, whose location
depends on the kinematics of produced quarkonia. For the special case
$Q=0$ and $y_{1}=y_{2}$ it is possible to express the position of
these poles in terms of the variable $\xi$ as 
\begin{equation}
\left|x_{k}\right|=\left\{ \xi,\,\xi\left(1-\frac{1}{1+\xi}\right),\,\xi\left(1-\frac{1}{2}\,\frac{1}{1+\xi}\right),\,\xi\left(1-\frac{2}{3}\,\frac{1}{1+\xi}\right),\,\xi\left(1-\frac{1}{3}\,\frac{1}{1+\xi}\right),\,3\xi\left(1+\frac{1}{6}\,\frac{1}{1+\xi}\right)\right\} .
\end{equation}
Varying the rapidities $y_{1},y_{2}$ of the observed quarkonia and
virtuality $Q^{2}$of the photon, it is possible to probe the gluon
GPDs in the full kinematic range $(x,\xi)$. For this reason, the
information about the gluon GPDs extracted from this process is complementary
to what could be extracted from single quarkonia production or DVCS,
which are mostly sensitive to gluon GPD near $x\approx\pm\xi$.

According to NRQCD~\cite{Bodwin:1994jh,Maltoni:1997pt,Brambilla:2008zg,Feng:2015cba,Brambilla:2010cs,Cho:1995ce,Cho:1995vh,Baranov:2002cf,Baranov:2007dw,Baranov:2011ib,Baranov:2016clx,Baranov:2015laa},
the color octet $\bar{Q}Q$ states might also contribute to quarkonia
production, so the expression~(\ref{eq:AmpSq}) should be generalized
as
\begin{equation}
\sum_{{\rm spins}}\left|\mathcal{A}_{\gamma p\to J/\psi\,\eta_{c}p}^{(\mathfrak{a})}\right|^{2}\approx\sum_{ij}\left\langle \mathcal{O}_{i}^{(J/\psi)}\right\rangle \left\langle \mathcal{O}_{j}^{(\eta_{c})}\right\rangle \left|\mathcal{A}_{\gamma_{T}p\to\left[\bar{Q}Q\right]_{i}\left[\bar{Q}Q\right]_{j}p}\right|^{2},\label{eq:Octets}
\end{equation}
where $\left\langle \mathcal{O}_{i}^{(M)}\right\rangle $ are the
nonperturbative color singlet and octet Long Distance Matrix Elements
(LDMEs) corresponding to a given state $i$ of the $\bar{Q}Q$. In
the heavy quark mass limit, the series~(\ref{eq:Octets}) is expected
to converge rapidly, so for numerical evaluations usually only the
first few terms are relevant. As mentioned earlier, the dominant color
singlet contribution is controlled by the LDMEs $\left\langle \mathcal{\mathcal{O}}_{J/\psi}\left(^{3}S_{1}^{[a]}\right)\right\rangle $,
$\left\langle \mathcal{\mathcal{O}}_{\eta_{c}}\left(^{1}S_{0}^{[a]}\right)\right\rangle $,
which according to phenomenological estimates have comparable values~\cite{Braaten:2002fi}
\begin{equation}
\left\langle \mathcal{\mathcal{O}}_{J/\psi}\left(^{3}S_{1}^{[a]}\right)\right\rangle \approx\left\langle \mathcal{\mathcal{O}}_{\eta_{c}}\left(^{1}S_{0}^{[a]}\right)\right\rangle \approx0.3\,{\rm GeV^{3}}.
\end{equation}
The evaluation of the color octet amplitudes $\mathcal{A}_{\gamma_{T}p\to\left[\bar{Q}Q\right]_{8}\left[\bar{Q}Q\right]_{8}p}$
is very similar to the color singlet case and differs only due to
different choice of the spin-color projections. However, according
to phenomenological estimates, the color octet LDMEs of $J/\psi$
mesons are very small~\cite{Baranov:2016clx}, 
\begin{align}
\left\langle \mathcal{O}^{J/\psi}\left(^{3}S_{1}^{[8]}\right)\right\rangle  & \approx2.32\times10^{-4}{\rm GeV}^{3},\\
\left\langle \mathcal{O}^{J/\psi}\left(^{1}S_{0}^{[8]}\right)\right\rangle  & \approx8.35\times10^{-3}{\rm GeV}^{3},\\
\left\langle \mathcal{O}^{J/\psi}\left(^{3}P_{0}^{[8]}\right)\right\rangle  & \approx0,
\end{align}
and the color octet LDMEs of the $\eta_{c}$ should be of the same
order in view of the heavy quark mass limit relations~\cite{Bodwin:1994jh}
\begin{align}
\left\langle \mathcal{O}^{\eta_{c}}\left(^{1}S_{0}^{[a]}\right)\right\rangle  & =\frac{1}{3}\left\langle \mathcal{O}^{J/\psi}\left(^{3}S_{1}^{[a]}\right)\right\rangle ,\quad a=1,8,\label{eq:CO1}\\
\left\langle \mathcal{O}^{\eta_{c}}\left(^{3}S_{1}^{[8]}\right)\right\rangle  & =\left\langle \mathcal{O}^{J/\psi}\left(^{1}S_{0}^{[8]}\right)\right\rangle ,\label{eq:CO2}\\
\left\langle \mathcal{O}^{\eta_{c}}\left(^{1}P_{1}^{[8]}\right)\right\rangle  & =3\left\langle \mathcal{O}^{J/\psi}\left(^{3}P_{0}^{[8]}\right)\right\rangle .\label{eq:CO3}
\end{align}

For this reason, in what follows we may safely omit the color octet
contributions~\footnote{We need to mention that at very large transverse momenta $p_{T}\gtrsim5-10\,{\rm GeV}$it
is known that color octet contributoins might give relevant contribution
to inclusive quarkonia production~\cite{Cho:1995ce,Cho:1995vh,DVMPcc1}.
However, in our evaluations we do not consider such large values of
$p_{T}$, since the exclusive cross-section is strongly suppressed
in that kinematics due to suppression of gluon GPDs at large $|t|\sim p_{\perp}^{2}$.}.

\section{Numerical results}

\label{sec:Numer} 
\begin{figure}
\includegraphics[width=12cm]{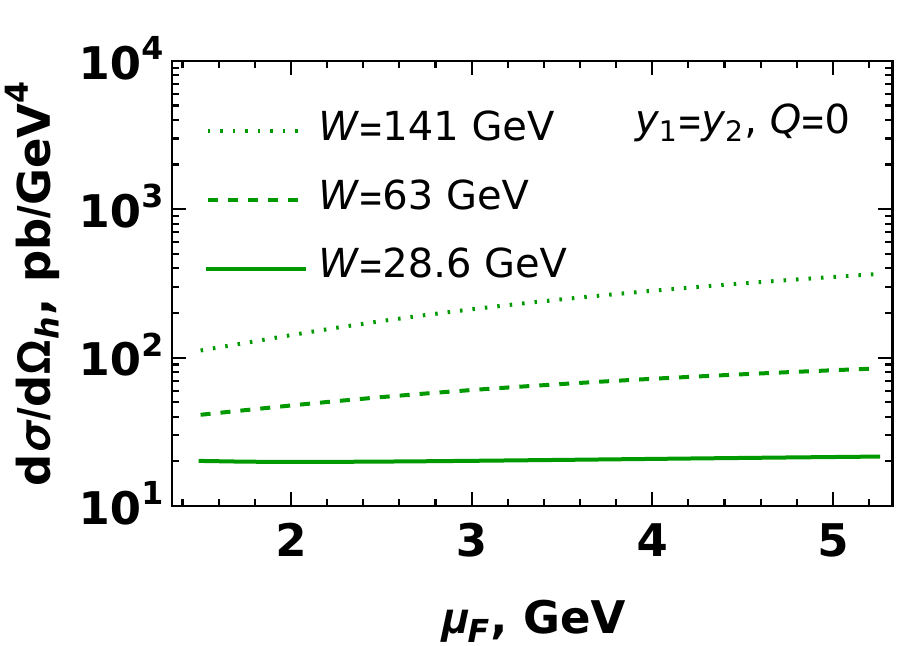} \caption{\label{fig:muFDependenceAll} Dependence of the cross-section on the
choice of factorization scale $\mu_{F}$. The frame label $d\sigma/d\Omega_{h}$
on the vertical axis is a shorthand notation for $d\sigma/dy_{1}dp_{1}^{2}dy_{2}dp_{2}^{2}d\phi$
. Chosen values of $W$ correspond to photon-proton energies $E_{\gamma}\times E_{p}=18\times275$
GeV, $100\times10$ GeV and $5\times41$ GeV respectively. In photoproduction
regime these values of $W$ correspond to values of Bjorken variable
$x_{B}\approx1.9\times10^{-3},$ $9.4\times10^{-3}$ and $4.5\times10^{-2}$
respectively. All frame-dependent variables are given in the laboratory
reference frame described in Section~\ref{subsec:Kinematics}.}
\end{figure}

For the sake of definiteness, for our predictions
we use the Kroll-Goloskokov parametrization of the gluon GPDs~\cite{Goloskokov:2006hr,Goloskokov:2007nt,Goloskokov:2008ib,Goloskokov:2009ia,Goloskokov:2011rd,Goloskokov:2013mba}.
This parametrization effectively takes into account the evolution
of the gluon distributions, introducing a mild dependence of the model
parameters on the factorization scale $\mu_{F}$. In what follows
for the sake of definiteness we will choose the scale $\mu_{F}=\mu_{R}\approx\sqrt{M_{J/\psi}^{2}+Q^{2}}$,
which interpolates smoothly between $\mu_{F}\approx M_{J/\psi}$ in
photoproduction regime, and $\mu_{F}\approx Q$ in Bjorken regime.
In Figure~\ref{fig:muFDependenceAll} we show the dependence
of the typical cross-section on the choice of this factorization scale.
We may observe that this dependence is mild at moderate energies,
but becomes very pronounced at very high energies (small $x_{B}$).
Such behaviour is not surprising: it is known from studies of \emph{single}
quarkonia photoproduction~\cite{DVMPcc1,DVMPcc2,DVMPcc3,DVMPcc4}
that this dependence is due to omitted loop corrections, and these
corrections become especially pronounced in the small-$x_{B}$ kinematics.

\begin{figure}
\includegraphics[width=9cm]{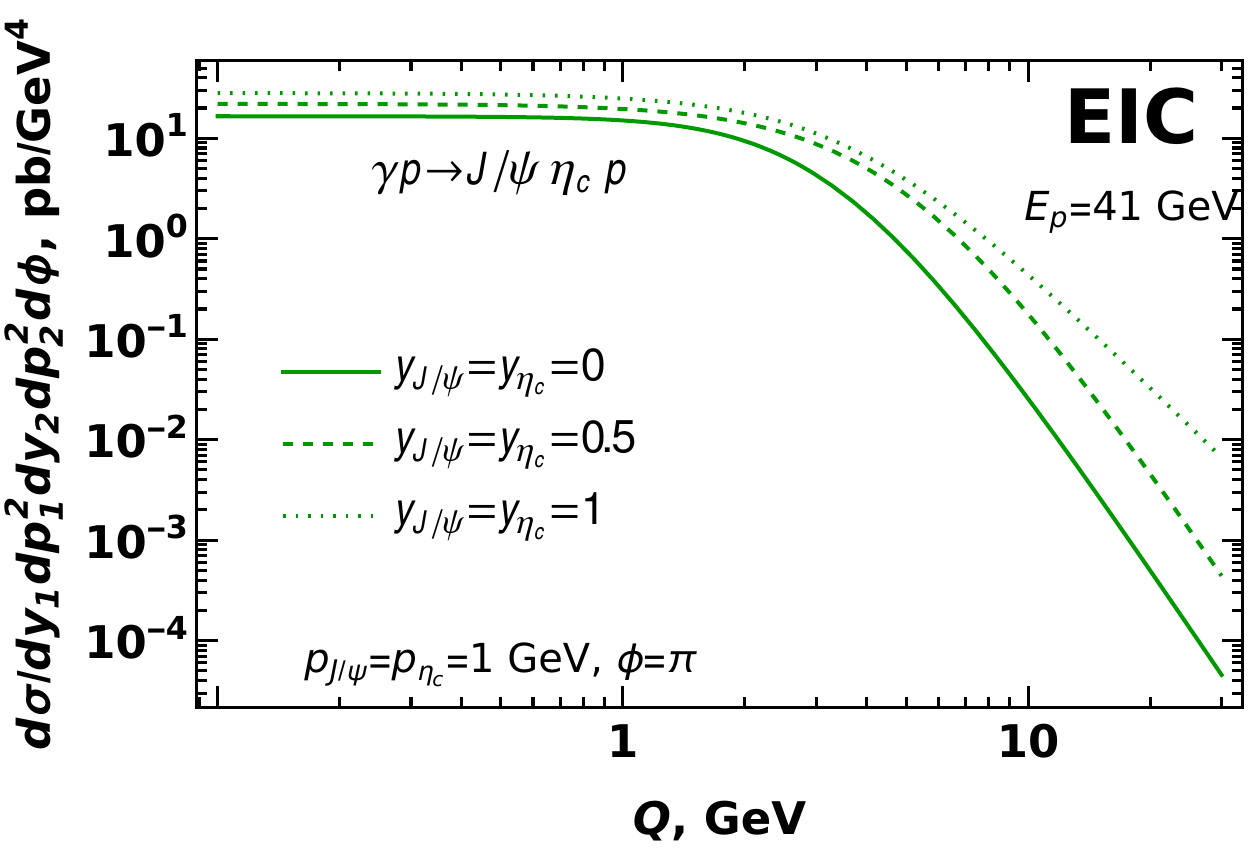}\includegraphics[width=9cm]{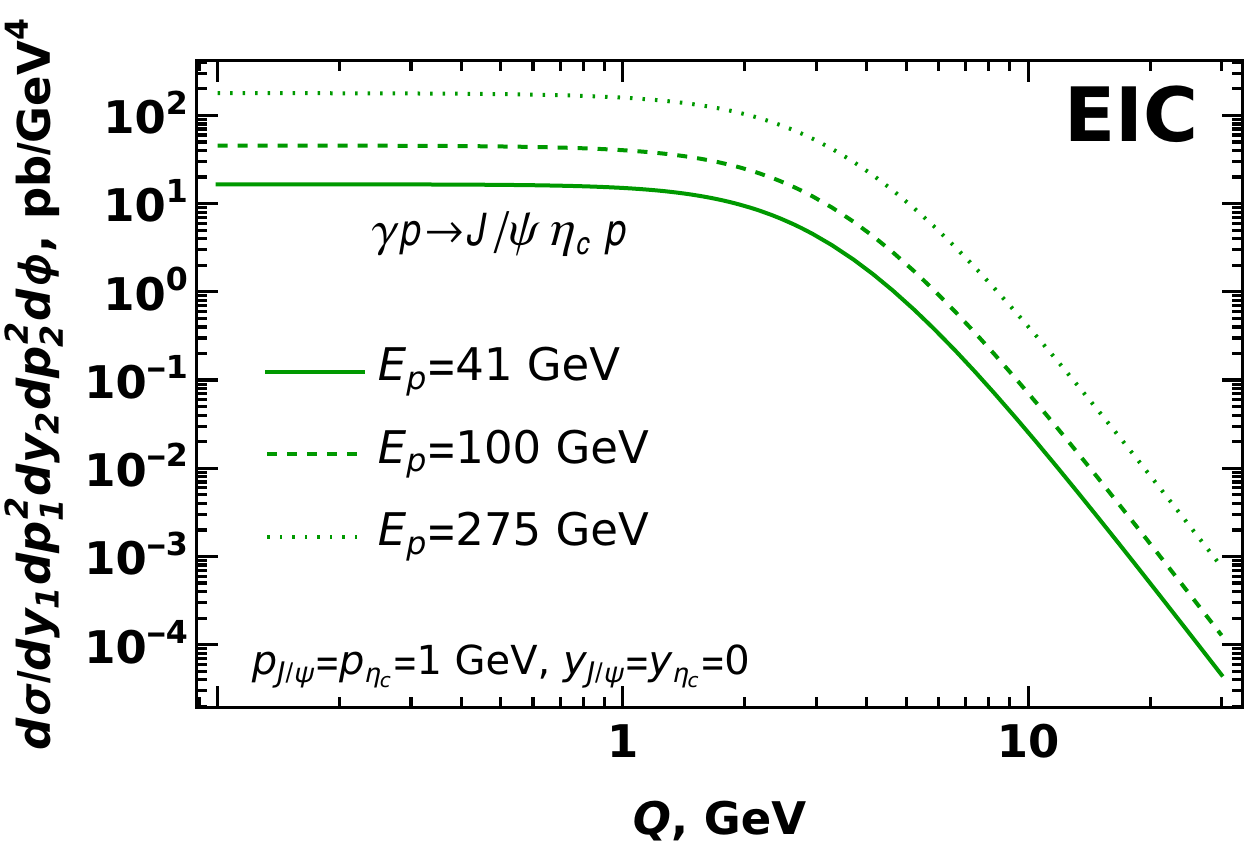}\caption{\label{fig:QDep}Dependence of the photoproduction cross-section~(\ref{eq:Photo-1})
on the virtuality $Q$ of the photon. In the left and right plots
we compare predictions for different rapidities $y_{J/\psi},y_{\eta_{c}}$
and different proton energies $E_{p}$. Both plots clearly illustrates
the transition from photoproduction to Bjorken regime in the region
$Q\sim1-2\,M_{J/\psi}$. In both plots the photon energy is evaluated
from~(\ref{eq:qPhoton-2},\ref{qPlus-1}). All frame-dependent variables
are given in the reference frame described in Section~\ref{subsec:Kinematics}.}
\end{figure}

We would like to start the presentation of results with a discussion
of the cross-section~(\ref{eq:Photo-1}) dependence on the virtuality
$Q$, shown schematically in Figure~\ref{fig:QDep}. This dependence
is very mild in the photoproduction regime ($Q\lesssim M_{J/\psi}$),
since the hard scale in this kinematics is controlled by the quarkonium
mass. In Bjorken regime ($Q\gg M_{J/\psi}$) the virtuality $Q$ plays
the role of the hard scale, which leads to a pronounced dependence
on $Q$. We can see that the cross-section is strongly suppressed,
so the experimental studies of this regime become very challenging.
For small $Q\lesssim M_{J/\psi}$, the cross-section is dominated
by the transversely polarized $J/\psi$ mesons, similar to single
$J/\psi$ production. This contribution is sensitive to the gluon
GPDs $H_{g},\,E_{g}$. The contribution of the longitudinally polarized
$J/\psi$ mesons is controlled by the helicity flip gluon GPDs $\tilde{H}_{g},\,\tilde{E}_{g}$,
which are less known phenomenologically, although they are clearly
significantly smaller than the unpolarized GPDs. We also observe that
the GPDs $H_{g},\,E_{g}$ might contribute to longitudinally polarized
photons via $\sim\mathcal{O}\left(\boldsymbol{p}_{\perp J/\psi}\right)$
corrections, although a systematic analysis of this contribution would
also require to take into account currently unknown twist-3 gluon
GPDs. In view of these uncertainties, we abstain from making predictions
for the longitudinal polarization. 

\begin{figure}
\includegraphics[width=6cm]{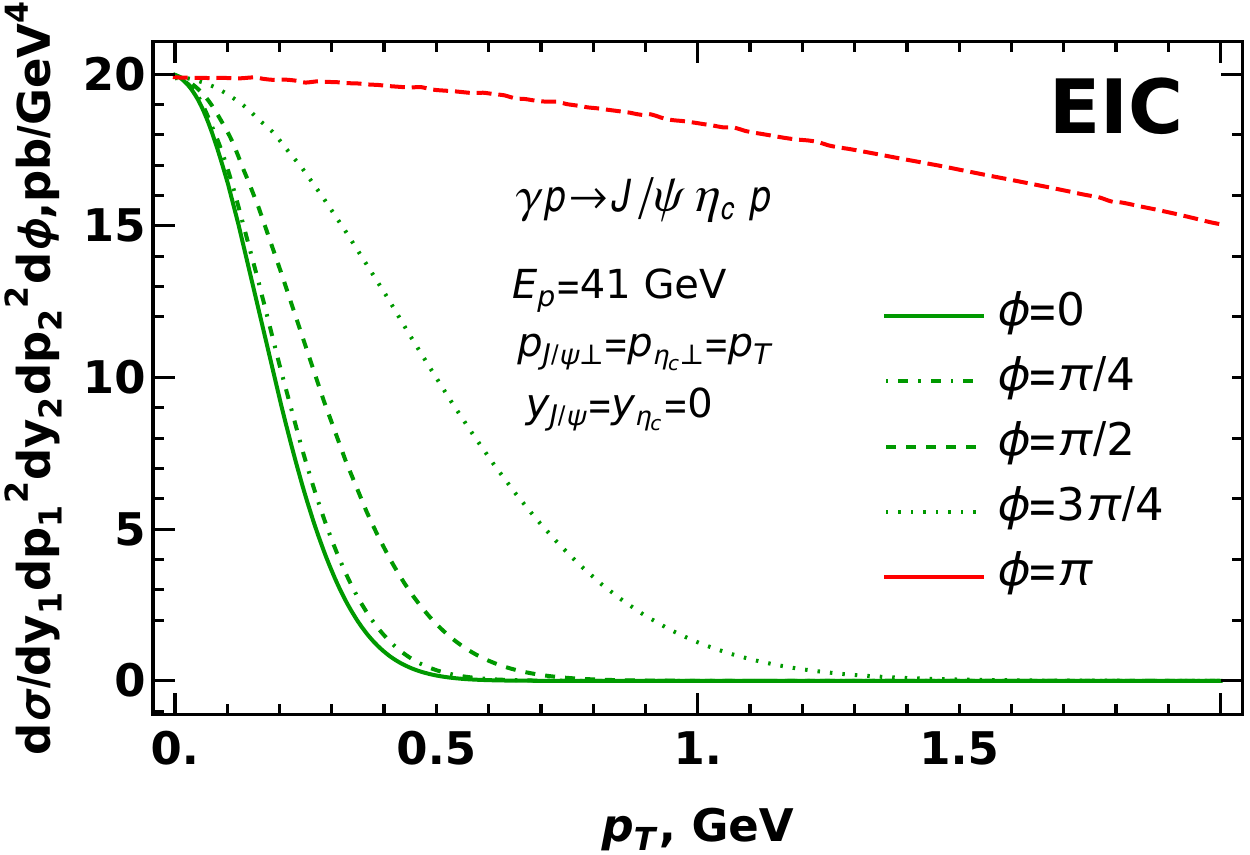}\includegraphics[width=6cm]{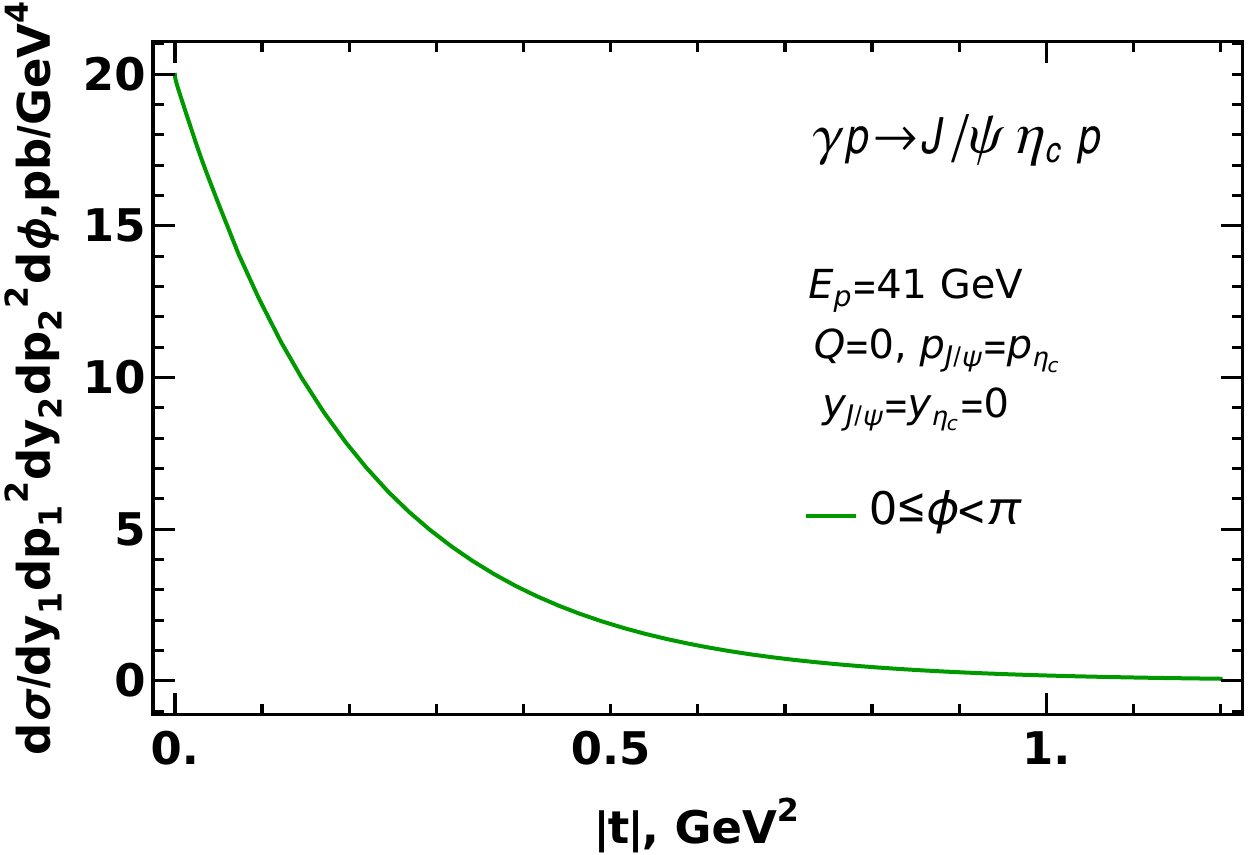}\includegraphics[width=6cm]{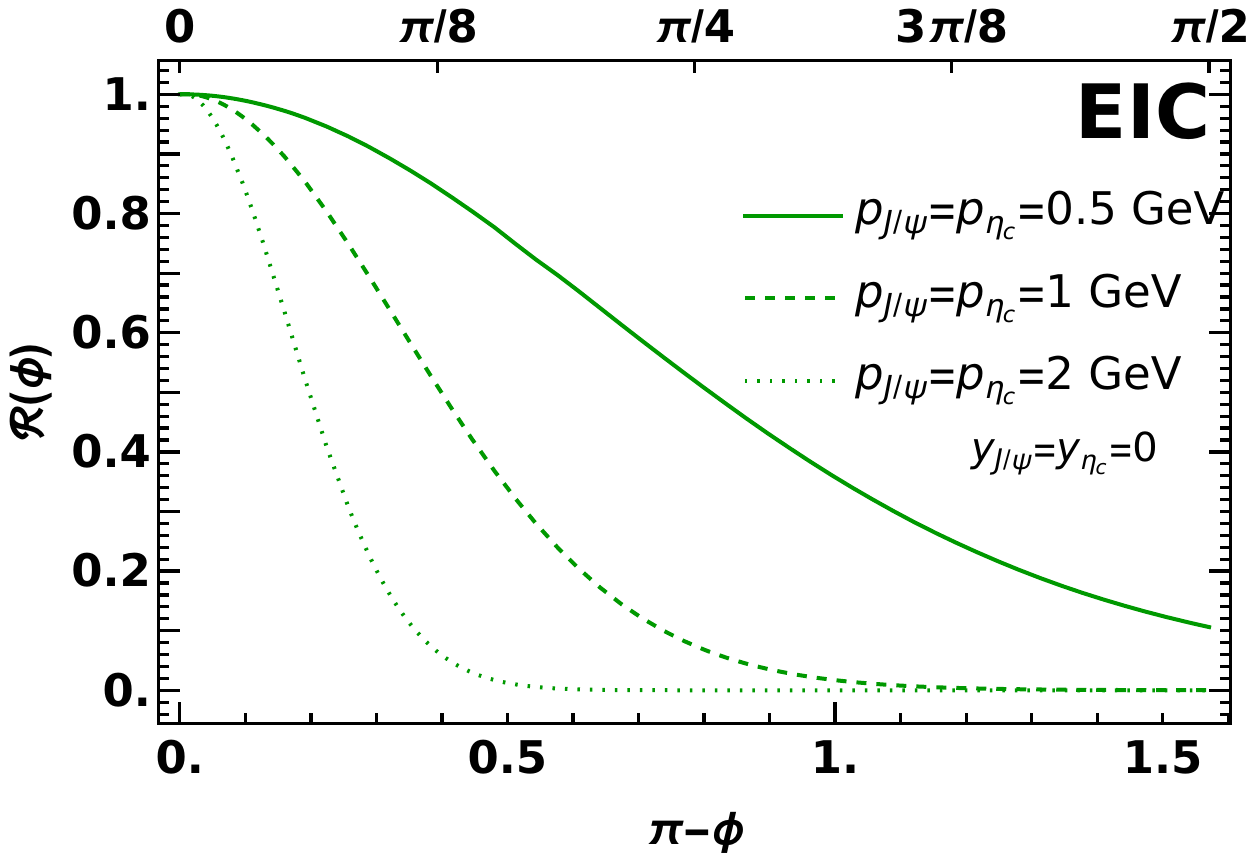}

\caption{\label{fig:tDep}Dependence of the photoproduction cross-section~(\ref{eq:Photo-1})
on the transverse momenta $p_{T}$of the quarkonia (left panel), invariant
momentum transfer $t$ to the proton (central panel) and the angle
$\phi$ between the quarkonia (right panel). Since the cross-sections
at different $p_{T}$ differ quite significantly,  in order to facilitate
the comparison of their $\phi$-dependence, in the right plot we normalized
them to unity in the maximum (angle $\phi=\pi$). For the sake of
definiteness, we considered the case photoproduction ($Q=0$) at central
rapidities ($y_{1}=y_{2}=0$) in the lab frame; for other virtualities
and rapidities the $p_{T}$- and $\phi$-dependence have very similar
shapes. All frame-dependent variables are given in the reference frame
described in Section~\ref{subsec:Kinematics}.}
\end{figure}

\begin{figure}
\includegraphics[width=6cm]{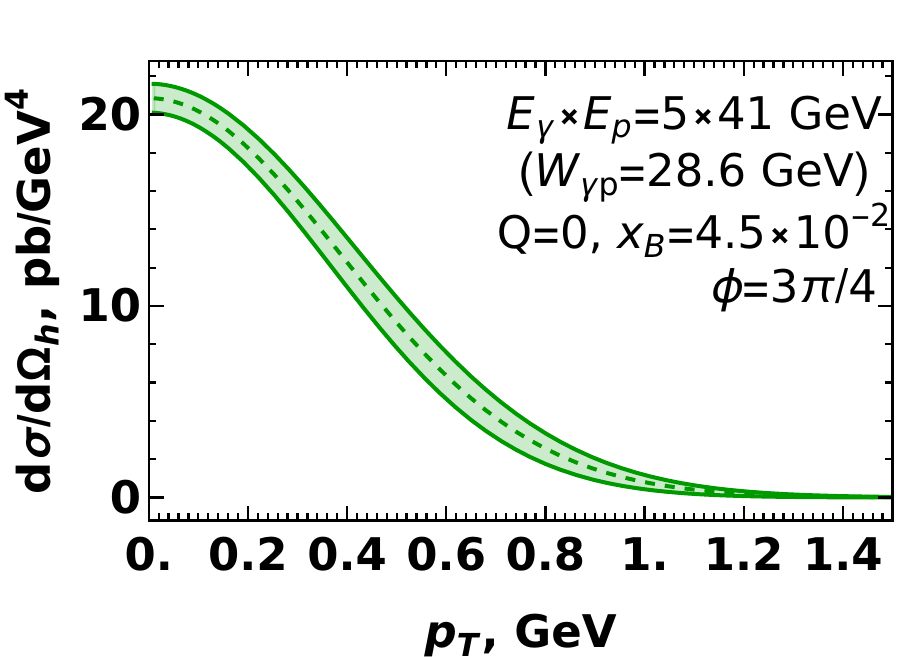}\includegraphics[width=6cm]{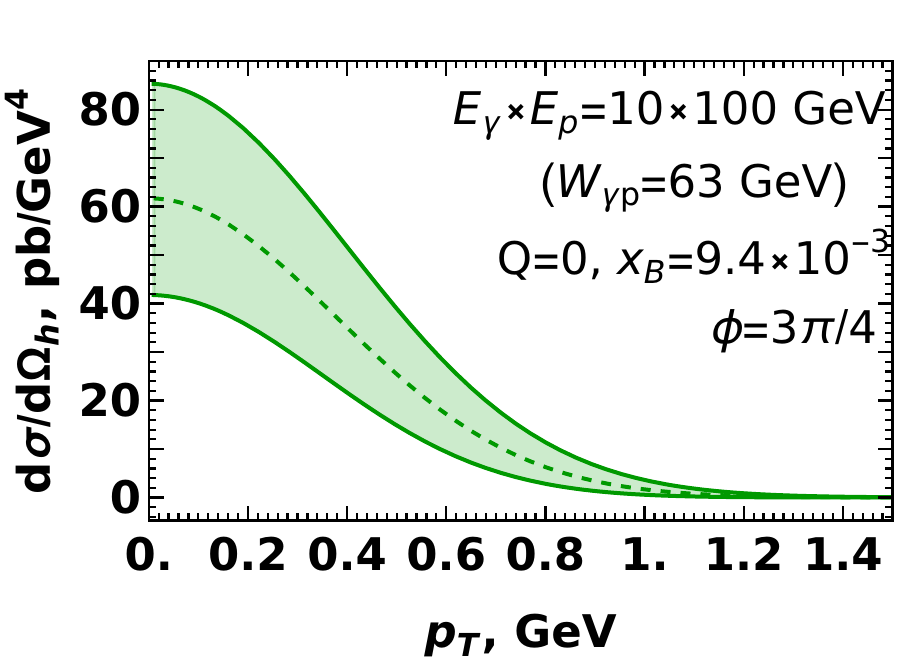}\includegraphics[width=6cm]{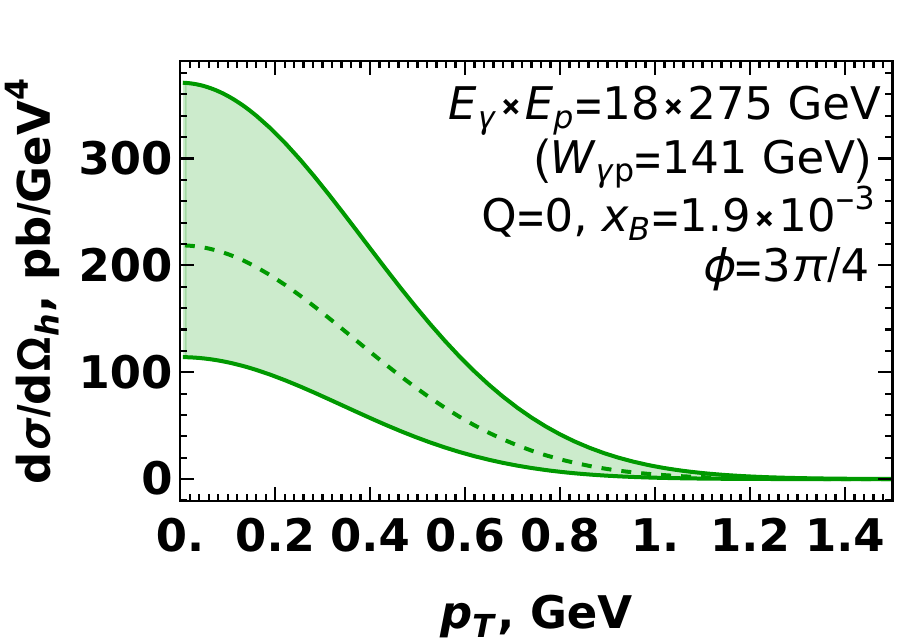}
\caption{\label{fig:muFDependence} Uncertainty of the cross-section due to
choice of factorization scale $\mu_{F}$. In all plots central dashed
line corresponds to $\mu_{F}=M_{J/\psi}$, whereas upper and lower
limits of the colored bands correspond to $\mu_{F}=2M_{J/\psi}$ and
$\mu_{F}=M_{J/\psi}/2$ respectively. For the sake of definiteness,
in all plots we considered that the angle between $J/\psi$ and $\eta_{c}$
is $\phi=3\pi/4$; for other angles the uncertanty due to choice
of $\mu_{F}$ has the same magnitude. }
\end{figure}

In Figure~\ref{fig:tDep} we show the dependence of the cross-section~(\ref{eq:Photo-1})
on the transverse momenta $\boldsymbol{p}_{1\perp},\,\boldsymbol{p}_{2\perp}$.
In the collinear factorization approach this dependence is largely
due to the gluon GPD dependence on the invariant momentum transfer
$t$~(\ref{eq:tDef}): most of the phenomenological models implement
a pronounced (nearly exponential) behavior at small $t$. At large
angles $\phi\approx\pi$ between transverse momenta of quarkonia (back-to-back
kinematics) the cross-section has a  sharp peak, which might be understood
from the definition~(\ref{eq:tDef}): this point minimizes $|t|$
at fixed $\left|p_{1}\right|,\left|p_{2}\right|$. As discussed in
Section~\ref{subsec:Kinematics}, the transverse momenta $\boldsymbol{p}_{1\perp},\,\boldsymbol{p}_{2\perp}$
also appear in other observables (e.g. via kinematic constraints,
``transverse'' masses $M_{1,2}^{\perp}$) and thus a mild $p_{T}$-dependence
exists even for $\boldsymbol{p}_{1\perp}=-\boldsymbol{p}_{2\perp}$,
as could be seen from the red long-dashed line in the left panel of
the Figure~\ref{fig:tDep}. Since in the collinear approach we neglected
the $p_{T}$-dependence in the coefficient functions, the results
are valid only for small $p_{T}\ll{\rm max}\left(Q,m_{Q}\right)$;
in the opposite limit (wide angle scattering kinematics) the cross-section
will be strongly suppressed as a function of the variable $p_{T}$
even for $p_{T}=\boldsymbol{p}_{1\perp}=-\boldsymbol{p}_{2\perp}$.
The central panel in the Figure~~\ref{fig:tDep} clearly demonstrates
that for any fixed $\phi\not=\pi$, the cross-section has the same
dependence on invariant momentum transfer $t$. This happens because
in collinear approach we disregard the transverse momenta in evaluation
of the coefficient function, so $\phi$-dependence exists only due
to $t$-dependence of the gluon GPDs. In the Figure~\ref{fig:muFDependence}
we illustrate the uncertainty of these cross-sections due to choice
of the scale $\mu_{F}$, varying it in the range $\mu_{F}\in\left(M_{J/\psi}/2,\,2M_{J/\psi}\right)$.
As discussed earlier, this uncertanty is very moderate at low energies,
yet becomes very pronounced (up to a factor of two) at high energies.
This indicates that loop corrections might give pronounced contribution
in that kinematics.

\begin{figure}
\includegraphics[width=9cm]{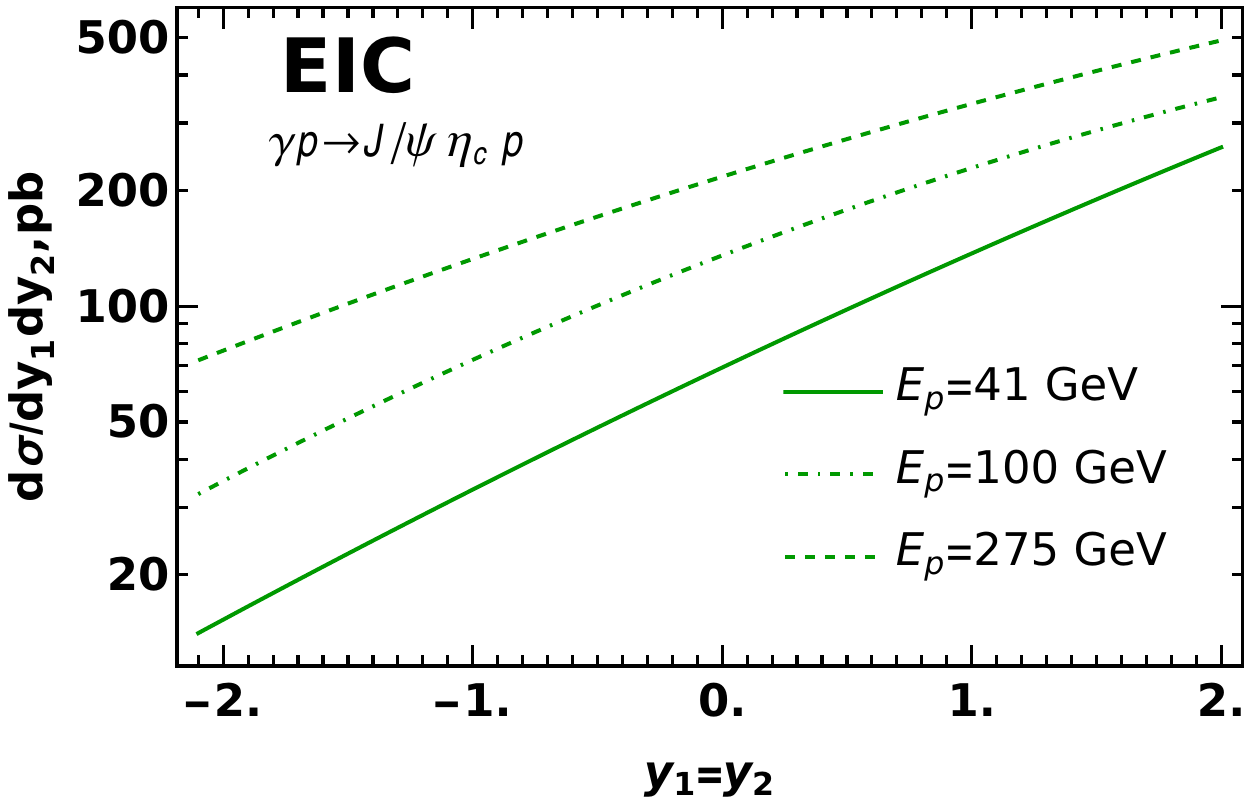}\includegraphics[width=9cm]{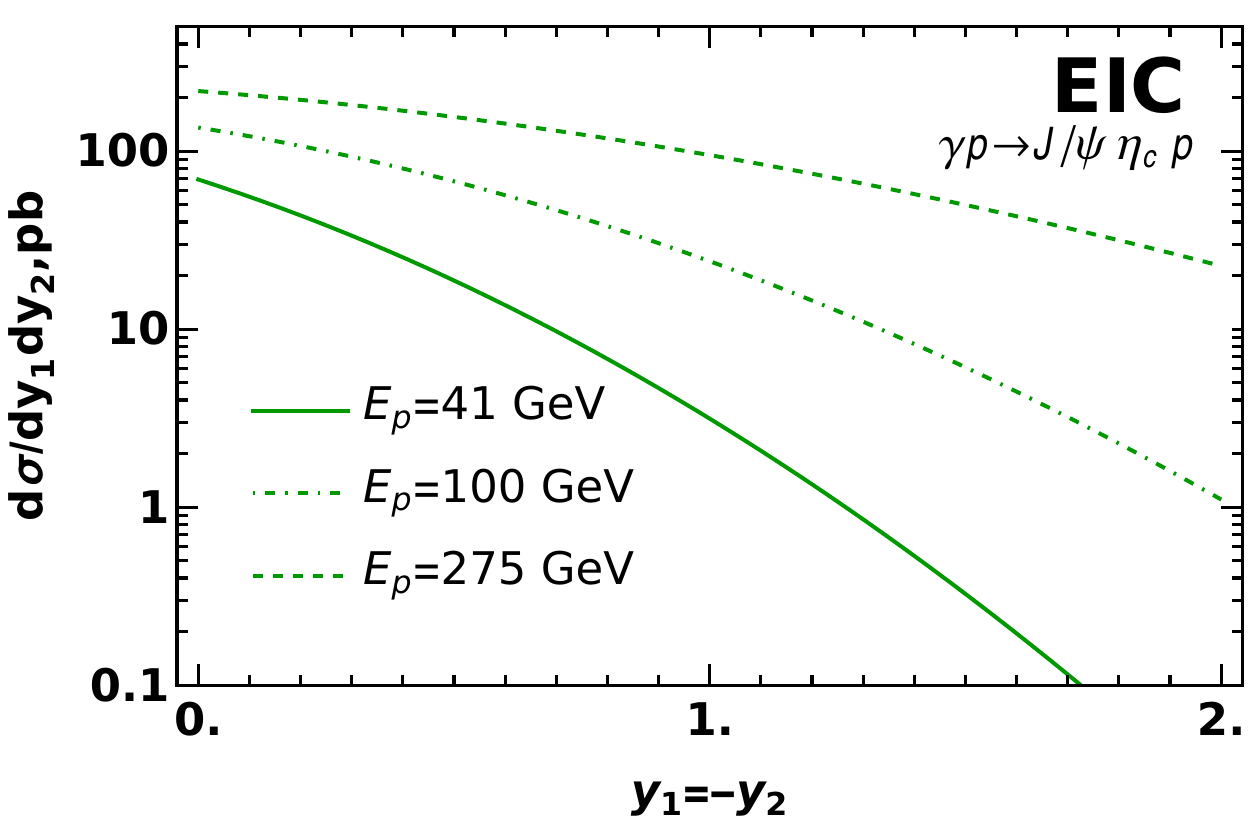}\caption{\label{fig:yDep} Dependence of the cross-section on the rapidities
$y_{1},y_{2}$ of the two quarkonia for several proton energies in
EIC kinematics. In the left plot we illustrate the dependence on the
average rapidity ($y_{1}=y_{2}$), and in the right plot we consider
the dependence on the rapidity difference at central rapidities ($y_{1}=-y_{2}=\Delta y/2)$. }
\end{figure}

\begin{figure}
\includegraphics[width=12cm]{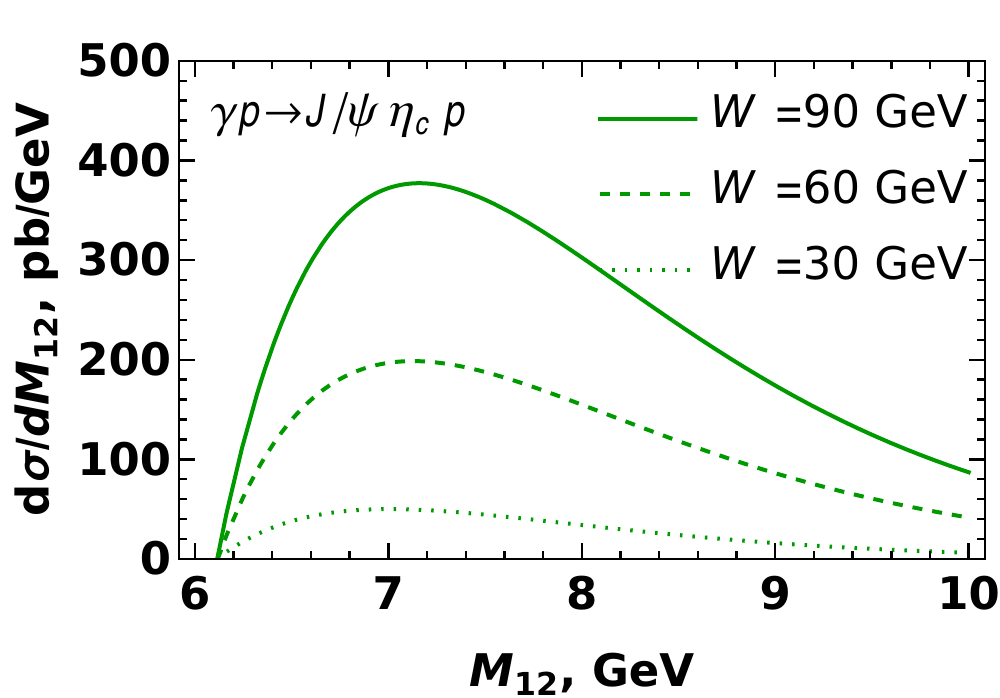}\caption{\label{fig:M12Dep} Distribution of the produced quarkonia pairs over
their invariant mass $M_{12}$, for several fixed invariant energies
$W$ of the $\gamma p$ collision.}
\end{figure}

In Figure~\ref{fig:yDep} we show the dependence of the $p_{T}$-integrated
cross-section on the rapidities of the produced quarkonia. In the
left panel, we show the dependence of the cross-section on the average
rapidity $y_{1}=y_{2}$. As expected, the cross-section grows with
$y$ due to the increase of photon energy, $W^{2}$, the corresponding
decrease of $x_{B},\xi$ and the growth of the gluon GPDs in that
kinematics. In the right panel we show the dependence on the rapidity
difference $\Delta y$ at central rapidities. The cross-section decreases
as a function of $\Delta y$, because the variables $x_{B},\xi$,
the longitudinal recoil to the proton, and the longitudinal momentum
transfer $\left|t_{{\rm min}}\right|$ grow as a function of $\Delta y$
at fixed $Y$, and the amplitude decreases due to suppression of gluon
GPDs with $|t|$. Finally, in Figure~\ref{fig:M12Dep} we show the
distribution of the produced $J/\psi\,\eta_{c}$ pairs over their
invariant mass $M_{12}$. The distribution has a pronounced peak near
$M_{12}\approx7\,{\rm GeV}$, which demonstrates that the quarkonia
pairs predominantly are produced with a small relative momentum $\sim2-3$
GeV.

\section{Conclusions}

\label{sec:Conclusions}In this paper we studied, in the collinear
factorization approach, the\emph{ }exclusive photoproduction of heavy
charmonia pairs with opposite $C$-parities ($J/\psi\,\eta_{c}$).
In our analysis we focused on the kinematics of moderate values of
$x_{B}$, achievable with low-energy $ep$ beams at the Electron Ion
Collider. This regime corresponds to values of Bjorken variable $x_{B}\in\left(10^{-3},\,10^{-1}\right)$.
We performed evaluations in leading order, assuming that higher order
corrections are suppressed at least as $\alpha_{s}\left(m_{Q}\right)$.
We focused on the photoproduction regime ($Q^{2}\approx0$) and found
that the dependence of the photoproduction cross-section on the virtuality
$Q$ is quite mild up to $Q\lesssim m_{Q}\approx1.2-1.5\,{\rm GeV}$.
The cross-section has a pronounced dependence on the invariant momentum
transfer $t$, and vanishes for $|t|\gtrsim1\,{\rm GeV^{2}}.$This
implies that the quarkonia pairs are produced predominantly in back-to-back
kinematics (with oppositely directed transverse momenta), which minimzes
$|t|$. The produced $J/\psi$ mesons are predominantly transversely
polarized, and the amplitude of the process obtains the dominant contribution
from the unpolarized gluon GPD $H_{g}$. The coefficient function
(partonic amplitude) has several poles (in addition to the classical
$x=\pm\xi$), whose positions depend on the kinematics of the produced
quarkonia. In view of the complexity of the coefficient function,
the deconvolution (direct extraction of GPDs from amplitudes) is apparently
not possible. Nevertheless, we believe that the process might be useful
to constrain existing models of phenomenological GPDs, especially
outside the $x=\pm\xi$ line.

The results presented here complement our earlier analysis~\cite{Andrade:2022rbn}
done in the color dipole framework in the kinematics $x_{B}\ll1$,
and agrees with it by an order of magnitude if extended to the region
of common validity (largest energy $ep$ beams at EIC, small $x_{B}\ll1$).
However, the collinear factorization approach might be not reliable
there due to large NLO corrections and onset of saturation effects.

Numerically, the evaluated cross-sections are on par with similar
estimates for $2\to3$ processes ($\gamma^{*}p\to\gamma Mp$, $M=\pi,\,\rho$)
suggested recently in the literature~\cite{GPD2x3:9,GPD2x3:8,GPD2x3:7,GPD2x3:6,GPD2x3:5,GPD2x3:4,GPD2x3:3,GPD2x3:2,GPD2x3:1,Duplancic:2022wqn}.
This happens because the emission of a photon in the final state leads
to a suppression by the fine-structure constant $\alpha_{{\rm em}}$,
on par with the suppression due to heavy quark mass in the production
of heavy quarkonia pairs. For this reason both $\gamma^{*}p\to\gamma Mp$
and heavy quarkonia production could be used as complementary tools
for the study of both quark and gluon GPDs.

\section*{Acknowldgements}

We thank our colleagues at UTFSM university for encouraging discussions.
This research was partially supported by Proyecto ANID PIA/APOYO AFB180002
(Chile) and Fondecyt (Chile) grants 1180232 and 1220242. \textquotedbl Powered@NLHPC:
This research was partially supported by the supercomputing infrastructure
of the NLHPC (ECM-02)\textquotedbl .

\appendix

\section{Symmetric frame}

\label{sec:Relation-Symm} In the collinear factorization framework
the evaluations in Bjorken kinematics are frequently performed in
the so-called symmetric frame~~\cite{Radyushkin:1996nd,Radyushkin:1997ki,Collins:1998be,Ji:1996nm,Ji:1998xh,Diehl:1999cg,Goeke:2001tz,Diehl:2003ny},
in which the vectors of photon momentum $q$ and $\bar{P}=(P_{i}+P_{f})/2$
(the average momentum of the target before and after collision) do
not have transverse momenta. This frame differs from the lab-frame
introduced in Section~\ref{subsec:Kinematics} by a transverse boost,
supplemented by a rotation in the transverse plane~\cite{Diehl:2003ny}.
In this paper we focus on the kinematics of small transverse momenta
$\Delta_{\perp}$, which eventually will be disregarded during evaluations
of the coefficient functions, so the parameters of the boost and rotation
are also small, $\sim\Delta_{\perp}/P^{+}$, and will give only $\mathcal{O}\left(\Delta_{\perp}^{2}\right)$
corrections to $\pm$ components of light-cone vectors. For this reason,
in what follows we will abuse notations and disregard possible differences
of $\pm$ components in lab- and symmetric frames.

Explicitly, the light-cone decomposition of photon and proton momenta
is given by 
\begin{align}
q & =\,\left(Z\bar{P}^{+},\,-\frac{Q^{2}}{2\,Z\bar{P}^{+}},\,\,\boldsymbol{0}_{\perp}\right),\label{eq:qPhoton-2-1}\\
\bar{P} & =\frac{P_{f}+P_{i}}{2}=\left(\bar{P}^{+},\,\frac{\overline{m}_{N}^{2}}{2\bar{P}^{+}},\,\,\boldsymbol{0}_{\perp}\right),\qquad\bar{m}_{N}^{2}=m_{N}^{2}-\frac{t}{4}\\
\Delta & =P_{f}-P_{i}=\left(-2\xi\,\bar{P}^{+},\,\frac{\xi\overline{m}_{N}^{2}}{\bar{P}^{+}},\,\,\boldsymbol{\Delta}_{\perp}\right)
\end{align}
so the momenta of proton before collision $(P_{i})$ and after collision
$(P_{f})$ are given explicitly by 
\begin{equation}
P_{f,i}=P\pm\frac{\Delta}{2}=\left((1\mp\xi)\bar{P}^{+},(1\pm\xi)\,\frac{\overline{m}_{N}^{2}}{2\bar{P}^{+}},\,\,\pm\frac{\boldsymbol{\Delta}_{\perp}}{2}\right)
\end{equation}
and the invariant momentum transfer to the proton is 
\begin{align}
t & =\Delta^{2}=-4\xi^{2}\left(m_{N}^{2}-\frac{t}{4}\right)-\Delta_{\perp}^{2}=-\frac{4\xi^{2}m_{N}^{2}+\Delta_{\perp}^{2}}{1-\xi^{2}}.
\end{align}
The variable $\bar{P}^{+}$ might be related to variables defined
in Section~~\ref{subsec:Kinematics} as 
\begin{align}
\bar{P}^{+} & =P^{+}+\frac{q^{+}-M_{1}^{\perp}\,e^{-y_{1}}-M_{2}^{\perp}\,e^{-y_{2}}}{2}=\frac{m_{N}^{2}}{2P^{-}}+\frac{q^{+}-M_{1}^{\perp}\,e^{-y_{1}}-M_{2}^{\perp}\,e^{-y_{2}}}{2}
\end{align}
The variable $Z$ might be fixed from conservation of plus-components
of momenta as 
\begin{align}
Z & =\frac{q^{+}}{\bar{P}^{+}}=-2\xi+\frac{M_{1\perp}}{2\bar{P}^{+}}e^{-y_{1}}+\frac{M_{2\perp}}{2\bar{P}^{+}}e^{-y_{2}}.\label{eq:Z}
\end{align}

\section{Evaluation of the coefficient functions}

\label{sec:CoefFunction} The evaluation of the coefficient functions
 (partonic amplitudes) relies on standard light--cone rules formulated
in~\cite{Lepage:1980fj,Brodsky:1997de,Diehl:2000xz,Diehl:2003ny,Diehl:1999cg,Ji:1998pc}.
We assume that both photon virtuality $Q$ and the quark mass $m_{Q}$
are large parameters, $Q\sim m_{Q}\sim\sqrt{s_{\gamma p}}$, tacitly
disregarding the proton mass and momentum transfer to the proton $t$.
As we discussed in Section~\ref{subsec:Amplitudes}, in the heavy
quark mass limit it is possible to disregard internal motion of the
quarks inside quarkonia, assuming that the momentum of the quarkonium
is shared equally between the quarks, and disregard the difference
of $J/\psi$ and $\eta_{c}$ masses, assuming $M_{J/\psi}\approx M_{\eta}\approx2m_{Q}$.
The evaluation of the partonic amplitudes requires computation of
the Feynman diagrams shown in Figures~\ref{fig:Photoproduction-A},~\ref{fig:Photoproduction-B},
and was done using FeynCalc package for \emph{Mathematica}~\cite{FeynCalc1,FeynCalc2}.
This evaluation resembles similar studies of the single quarkonia
photoproduction well-known from the literature~\cite{DVMPcc1,DVMPcc2,DVMPcc3,DVMPcc4}.
Below we provide some technical details which might help to understand
the main steps and assumtpions needed for derivation of the final
result.   

Since GPDs are conventionally defined in the symmetric frame, we perform
evaluation of the coefficient function in that frame, assuming that
all momenta might be related using the transformations described in
Section~\ref{sec:Relation-Symm}. The momenta of partons (gluons)
in this frame, before and after interaction, are given respectively
by 
\begin{equation}
k_{i,f}=\left((x\pm\xi)\bar{P}^{+},\,0,\,\boldsymbol{k}_{\perp}\mp\frac{\boldsymbol{\Delta}_{\perp}}{2}\right).
\end{equation}

Furthermore, to simplify further notations, we will shift the rapidities
of quarkonia and rewrite their momenta as 
\begin{align}
p_{a} & =\left(\,e^{\tilde{y}_{a}}\bar{P}^{+}\,,\,\frac{\left(M_{a}^{\perp}\right)^{2}e^{-\tilde{y}_{a}}}{2\bar{P}^{+}},\,\,\boldsymbol{p}_{a}^{\perp}\right),\quad a=1,2,\label{eq:MesonLC-2-1}\\
\tilde{y}_{a} & =-y_{a}+\ln\left(M_{a}^{\perp}/2\bar{P}^{+}\right).
\end{align}
This modification allows to suppress numerous factors $\sim M_{a}^{\perp}/\bar{P}^{+}$,
so the coefficient functions will depend only on 2 dimensional variables,
$m_{Q}^{2}$ and $Q^{2}$. For example, the variable $Z$ defined
in~(\ref{eq:Z}) will turn into a simple expression 
\begin{align}
Z & =-2\xi+e^{\tilde{y}_{1}}+e^{\tilde{y}_{2}}.\label{eq:Z-1}
\end{align}
Since we consider that formally both $M_{a}$ and $\bar{P}^{+}$ are
large parameters of the same order, the variables $y_{a}$ and $\tilde{y}_{a}$
are also of the same order, and thus switching from $y_{a}$ to $\tilde{y}_{a}$
does not require modification of the underlying counting rules.

The chiral even gluon GPDs, which are expected to give the dominant
contributions, are defined as~\cite{Diehl:2003ny,DVMPcc1}
\begin{align}
F^{g}\left(x,\xi,t\right) & =\frac{1}{\bar{P}^{+}}\int\frac{dz}{2\pi}\,e^{ix\bar{P}^{+}}\left\langle P'\left|G^{+\mu\,a}\left(-\frac{z}{2}n\right)\mathcal{L}\left(-\frac{z}{2},\,\frac{z}{2}\right)G_{\,\,\mu}^{+\,a}\left(\frac{z}{2}n\right)\right|P\right\rangle =\label{eq:defF}\\
 & =\left(\bar{U}\left(P'\right)\gamma_{+}U\left(P\right)H^{g}\left(x,\xi,t\right)+\bar{U}\left(P'\right)\frac{i\sigma^{+\alpha}\Delta_{\alpha}}{2m_{N}}U\left(P\right)E^{g}\left(x,\xi,t\right)\right),\nonumber \\
\tilde{F}^{g}\left(x,\xi,t\right) & =\frac{-i}{\bar{P}^{+}}\int\frac{dz}{2\pi}\,e^{ix\bar{P}^{+}}\left\langle P'\left|G^{+\mu\,a}\left(-\frac{z}{2}n\right)\mathcal{L}\left(-\frac{z}{2},\,\frac{z}{2}\right)\tilde{G}_{\,\,\mu}^{+\,a}\left(\frac{z}{2}n\right)\right|P\right\rangle =\label{eq:defFTilde}\\
 & =\left(\bar{U}\left(P'\right)\gamma_{+}\gamma_{5}U\left(P\right)\tilde{H}^{g}\left(x,\xi,t\right)+\bar{U}\left(P'\right)\frac{\Delta^{+}\gamma_{5}}{2m_{N}}U\left(P\right)\tilde{E}^{g}\left(x,\xi,t\right)\right).\nonumber \\
 & \tilde{G}^{\mu\nu,\,a}\equiv\frac{1}{2}\varepsilon^{\mu\nu\alpha\beta}G_{\alpha\beta}^{a},\quad\mathcal{L}\left(-\frac{z}{2},\,\frac{z}{2}\right)\equiv{\rm exp}\left(i\int_{-z/2}^{z/2}d\zeta\,A^{+}\left(\zeta\right)\right).
\end{align}
where the skewedness variable $\xi$ was defined in~(\ref{eq:XiDef});
for quarkonia pair production it might be expressed as a function
of $y_{1},y_{2},Q^{2}$. In the light-cone gauge $A^{+}=0$ we may
rewrite the two-gluon operators in~(\ref{eq:defF},~\ref{eq:defFTilde})
as
\begin{align}
 & G^{+\mu_{\perp}\,a}\left(z_{1}\right)G_{\,\,\mu_{\perp}}^{+\,a}\left(z_{2}\right)=g_{\mu\nu}^{\perp}\left(\partial^{+}A^{\mu_{\perp},a}(z_{1})\right)\left(\partial^{+}A^{\nu_{\perp}a}(z_{2})\right),\\
 & G^{+\mu_{\perp}\,a}\left(z_{1}\right)\tilde{G}_{\,\,\mu_{\perp}}^{+\,a}\left(z_{2}\right)=G^{+\mu_{\perp}\,a}\left(z_{1}\right)\tilde{G}_{-\mu_{\perp}}^{\,a}\left(z_{2}\right)=\frac{1}{2}\varepsilon_{-\mu_{\perp}\alpha\nu}G^{+\mu_{\perp}\,a}\left(z_{1}\right)G^{\alpha\nu,\,a}\left(z_{2}\right)=\\
 & =\varepsilon_{-\mu_{\perp}+\nu_{\perp}}G^{+\mu_{\perp}\,a}\left(z_{1}\right)G^{+\nu_{\perp},\,a}\left(z_{2}\right)=\varepsilon_{\mu\nu}^{\perp}G^{+\mu_{\perp}\,a}\left(z_{1}\right)G^{+\nu_{\perp},\,a}\left(z_{2}\right)=\varepsilon_{\mu\nu}^{\perp}\left(\partial^{+}A^{\mu,a}(z_{1})\right)\left(\partial^{+}A^{\nu,\,a}(z_{2})\right).\nonumber 
\end{align}
After taking the integral over $z$ in~(\ref{eq:defF},~\ref{eq:defFTilde}),
we effectively switch to  the momentum space, where the derivatives
$\partial_{z_{1}}^{+},\,\partial_{z_{2}}^{+}$ will turn into the
factors $k_{1,2}^{+}\sim\left(x\pm\xi\right)\bar{P}^{+}$, so we may
rewrite~(\ref{eq:defF},~\ref{eq:defFTilde}) as~\cite{DVMPcc1}
\begin{align}
\frac{1}{\bar{P}^{+}}\int\frac{dz}{2\pi}\,e^{ix\bar{P}^{+}}\left.\frac{\frac{}{}}{}\left\langle P'\left|A_{\mu}^{a}\left(-\frac{z}{2}n\right)A_{\nu}^{b}\left(\frac{z}{2}n\right)\right|P\right\rangle \right|_{A^{+}=0\,{\rm gauge}} & =\frac{\delta^{ab}}{N_{c}^{2}-1}\left(\frac{-g_{\mu\nu}^{\perp}F^{g}\left(x,\xi,t\right)-\varepsilon_{\mu\nu}^{\perp}\tilde{F}^{g}\left(x,\xi,t\right)}{2\,\left(x-\xi+i0\right)\left(x+\xi-i0\right)}\right).\label{eq:defF-1}
\end{align}
We may see that it is possible to extract the coefficient functions
$C_{a}$ and $\tilde{C}_{a}$, convoluting Lorentz indices of $t$-channel
gluons in diagrams of Figures~\ref{fig:Photoproduction-A},~\ref{fig:Photoproduction-B}
with $g_{\mu\nu}^{\perp}$ and $\varepsilon_{\mu\nu}^{\perp}$ respectively,
and following~\cite{DVMPcc1} we assume that the variable $\xi$
in denominator is always replaced as $\xi\to\xi-i0$ in order to define
proper contour deformation near the poles of the amplitude. 

For evaluation of the coefficient functions, we also need to make
proper projections of the $\bar{Q}Q$ pairs onto the states with definite
color and spins. According to potential models and NRQCD, the dominant
Fock state in quarkonium is the color singlet $\bar{Q}Q$ pair in
$^{3}S_{1}^{[1]}$ state for $J/\psi$ , and $^{1}S_{0}^{[1]}$ state
for $\eta_{c}$. As discussed in~\cite{Cho:1995ce,Cho:1995vh,DVMPcc1},
the projectors on color singlet and color octet states are given respectively
by 
\begin{align}
\left(P^{[1]}\right)_{ij}=\frac{\delta_{ij}}{\sqrt{N_{c}}},\qquad\left(P_{b}^{[8]}\right)_{ij} & =\sqrt{2}\,\left(t^{b}\right)_{ij},\quad b=1,\,...,\,8.
\end{align}
The projections onto a state with definite total spin $S$ and its
projection $S_{z}$ might be found using proper Clebsch-Gordan coefficients~\cite{Cho:1995ce,Cho:1995vh,DVMPcc1},
\begin{align}
\hat{P}_{SS_{z}} & =\sum_{s_{1},\,s_{2}}\left\langle \frac{1}{2}s_{1}\frac{1}{2}s_{2}\bigg|SS_{z}\right\rangle v\left(\frac{P}{2}-q,\,s_{2}\right)\bar{u}\left(\frac{P}{2}+q,\,s_{1}\right)=\\
 & =\left\{ \begin{array}{cc}
\frac{-1}{2\sqrt{2}}\left(\frac{\hat{P}}{2}-\hat{q}-m_{Q}\right)\gamma_{5}\left(\frac{\hat{P}}{2}+\hat{q}+m_{Q}\right),\quad & S=0\\
\frac{-1}{2\sqrt{2}}\left(\frac{\hat{P}}{2}-\hat{q}-m_{Q}\right)\hat{\varepsilon}_{J/\psi}(P)\left(\frac{\hat{P}}{2}+\hat{q}+m_{Q}\right), & S=1
\end{array}\right.\nonumber 
\end{align}
where $P$ is the momentum of the produced quarkonium, $q\approx0$
is the momentum of relative motion of the quarks inside the quarkonium,
and $\varepsilon_{J/\psi}$ is the polarization vector of $J/\psi$
mesons. Combining these projectors with proper color singlet LDMEs
and disregarding momentum of the relative motion $q$, after some
algebra we may obtain effective projectors of heavy quarks onto $J/\psi$
and $\eta_{c}$ states, {\small{}
\begin{align}
\left(\hat{V}_{\eta_{c}}^{[1]}\right)_{ij} & =-\sqrt{\frac{\left\langle \mathcal{O}_{\eta_{c}}^{[1]}\right\rangle }{m_{Q}}}\,\frac{\delta_{ij}}{8N_{c}m_{Q}}\left(\frac{\hat{P}}{2}-\hat{q}-m_{Q}\right)\gamma_{5}\left(\frac{\hat{P}}{2}+\hat{q}+m_{Q}\right)\approx-\sqrt{\frac{\left\langle \mathcal{O}_{\eta_{c}}^{[1]}\right\rangle }{m_{Q}}}\,\frac{\delta_{ij}}{4N_{c}}\left(\frac{\hat{P}}{2}-m_{Q}\right)\gamma_{5}\label{eq:PEta}\\
\left(\hat{V}_{J/\psi}^{[1]}\right)_{ij} & =-\sqrt{\frac{\left\langle \mathcal{O}_{J/\psi}^{[1]}\right\rangle }{m_{Q}}}\frac{\delta_{ij}}{8N_{c}m_{Q}}\left(\frac{\hat{P}}{2}-\hat{q}-m_{Q}\right)\hat{\varepsilon}_{J/\psi}(P)\left(\frac{\hat{P}}{2}+\hat{q}+m_{Q}\right)\approx\sqrt{\frac{\left\langle \mathcal{O}_{J/\psi}^{[1]}\right\rangle }{m_{Q}}}\frac{\delta_{ij}}{4N_{c}}\hat{\varepsilon}_{J/\psi}(P)\left(\frac{\hat{P}}{2}+m_{Q}\right)\label{eq:PJPsi}
\end{align}
}where $\left\langle \mathcal{\mathcal{O}}_{M}^{[1]}\right\rangle $
are the corresponding color singlet long-distance matrix elements
for $J/\psi$ and $\eta_{c}$ mesons. These objects can be related
to the radial wave functions in potential model, and for the $S$-wave
quarkonia~\cite{DVMPcc1,Brambilla:2010cs} this relation has a form
\begin{equation}
\left\langle \mathcal{\mathcal{O}}_{M}^{[1]}\right\rangle =\frac{N_{c}}{2\pi}\left|R_{S}(0)\right|^{2}.
\end{equation}
Phenomenological estimates, for example based on analysis of the partial
decay width of $J/\psi\to e^{+}e^{-}$, suggest that $\left\langle \mathcal{\mathcal{O}}_{J/\psi}^{[1]}\left(^{3}S_{1}^{[1]}\right)\right\rangle \approx\left\langle \mathcal{\mathcal{O}}_{\eta_{c}}^{[1]}\left(^{1}S_{0}^{[1]}\right)\right\rangle \approx0.3\,{\rm GeV^{3}}$~\cite{Braaten:2002fi}. 

\begin{figure}
\includegraphics[scale=0.6]{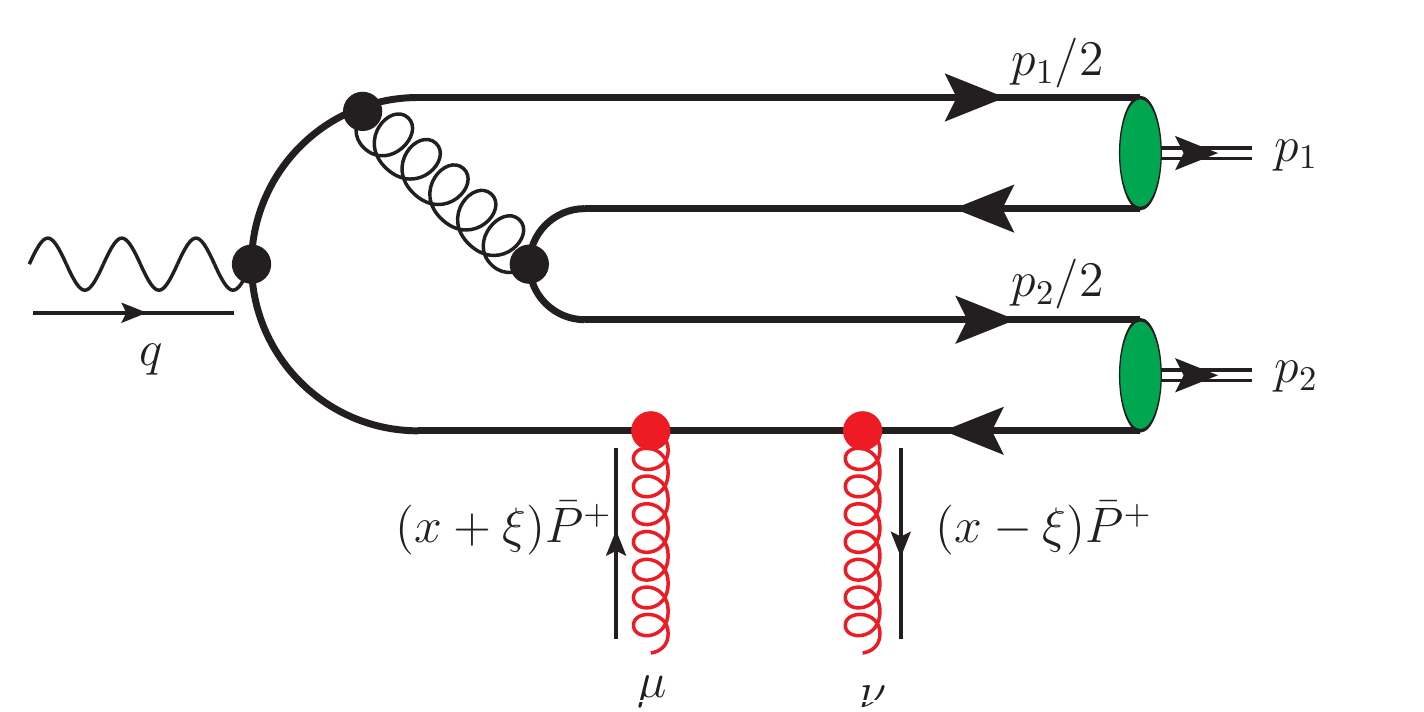}\includegraphics[scale=0.6]{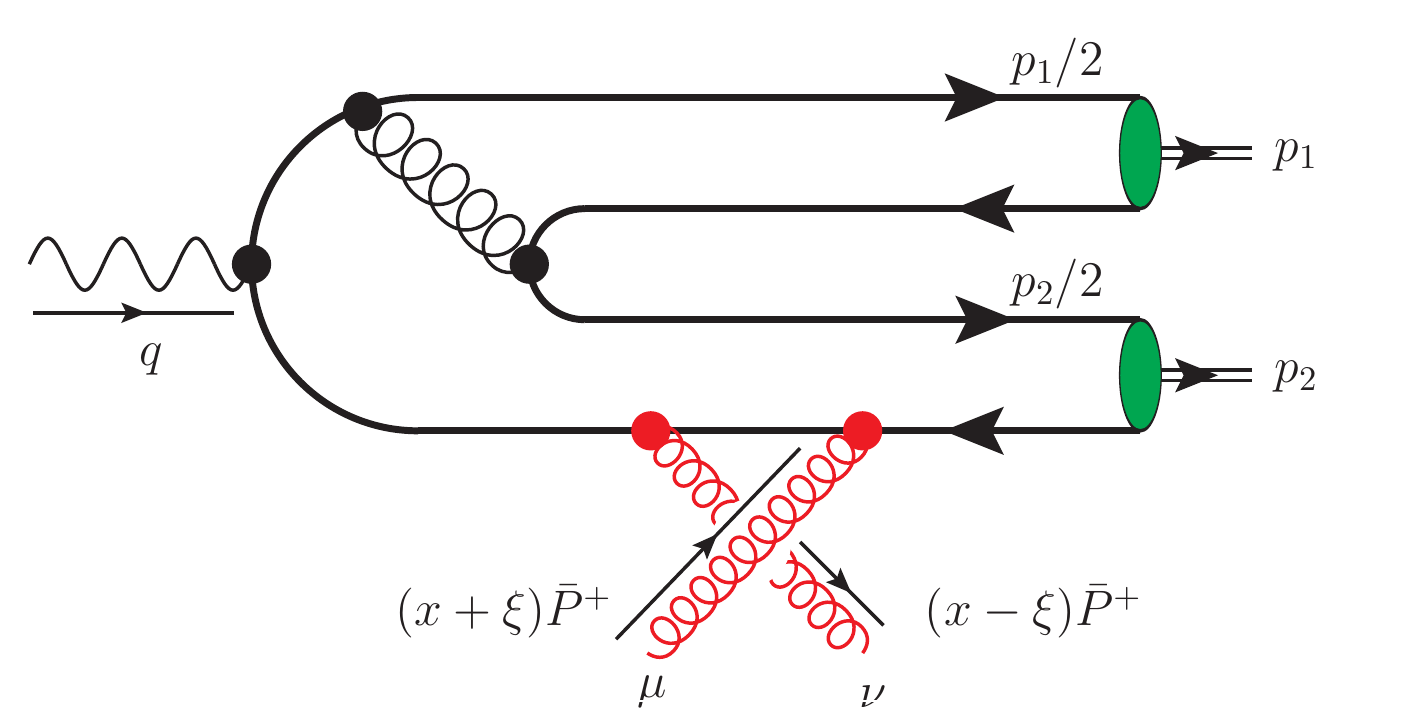}

\caption{\label{fig:Photoproduction-Permute}Schematic illustration of the
diagrams with direct and permuted $t$-channel gluons, which are related
to each other by inversion of sign in front of light-cone fraction
$x\leftrightarrow-x$, and permutation of the Lorentz indices $\mu\leftrightarrow\nu$.}
\end{figure}

In evaluation of the diagrams from Figures~\ref{fig:Photoproduction-A},~\ref{fig:Photoproduction-B}
we should take into account that each diagram should be accompanied
with another diagram with permuted final state mesons $1\leftrightarrow2$
(equivalently, diagram with inverted direction of quark lines), as
well as a diagram with permutation of $t$-channel gluons, as shown
in the Figure~\ref{fig:Photoproduction-Permute}. The latter permutation
gives contributions which differ only by change of the sign in front
of the light-cone variable $x$ and interchange of the Lorentz indices
$\mu\leftrightarrow\nu$. According to~(\ref{eq:defF-1}), we need
to contract the free Lorentz indices $\mu,\nu$ with symmetric $g_{\mu\nu}^{\perp}$
or antisymmetric $\varepsilon_{\mu\nu}^{\perp}$ in order to single
out the contributions of $F^{g}$ or $\tilde{F}^{g}$, for this reason
eventually we conclude that the coefficient functions $C_{\mathfrak{a}}$,
$\tilde{C}_{\mathfrak{a}}$ will be even or odd functions of the variable
$x$ respectively. Since we disregard internal motion of quarks inside
quarkonia, the momenta of all partons are fixed by energy-momentum
conservation and could be expressed as linear combinations of the
momenta of the quarkonia and $t$-channel gluons. Taking into account~(\ref{eq:defF-1},~\ref{eq:PEta},~\ref{eq:PJPsi}),
we may obtain for the coefficient functions

\begin{align}
C_{\mathfrak{a}}\left(x,\,\tilde{y}_{1},\,\tilde{y}_{2}\right) & =\kappa\frac{\mathcal{C}_{\mathfrak{a}}\left(x,\,\tilde{y}_{1},\,\tilde{y}_{2}\right)+\mathcal{C}_{\mathfrak{a}}\left(-x,\,\tilde{y}_{1},\,\tilde{y}_{2}\right)}{\left(x-\xi+i0\right)\left(x+\xi-i0\right)}\label{eq:CDef1}\\
\tilde{C}_{\mathfrak{a}}\left(x,\,\tilde{y}_{1},\,\tilde{y}_{2}\right) & =\kappa\frac{\tilde{\mathcal{C}}_{\mathfrak{a}}\left(x,\,\tilde{y}_{1},\,\tilde{y}_{2}\right)-\tilde{\mathcal{C}}_{\mathfrak{a}}\left(-x,\,\tilde{y}_{1},\,\tilde{y}_{2}\right)}{\left(x-\xi+i0\right)\left(x+\xi-i0\right)}\label{eq:CDef2}
\end{align}

where the constant $\kappa$ is defined as 
\begin{equation}
\kappa=\left(4\pi\alpha_{s}\right)^{2}e_{Q}\frac{\sqrt{\left\langle \mathcal{\mathcal{O}}_{J/\psi}^{[1]}\left(^{3}S_{1}^{[1]}\right)\right\rangle \left\langle \mathcal{\mathcal{O}}_{\eta_{c}}^{[1]}\left(^{1}S_{0}^{[1]}\right)\right\rangle }}{4N_{c}^{2}m_{Q}}\left(\varepsilon_{J/\psi}^{*}\cdot\varepsilon_{T}^{(\gamma)}\right),
\end{equation}
and the factors $x\pm\xi\mp i0$ in denominators of~(\ref{eq:CDef1},~\ref{eq:CDef2})
stem from~(\ref{eq:defF-1}). The contribution of each diagram from
the Figures~\ref{fig:Photoproduction-A},~\ref{fig:Photoproduction-B}
to functions $\mathcal{C}_{\mathfrak{a}}$ and $\tilde{\mathcal{C}}_{\mathfrak{a}}$
might be obtained taking Dirac and color traces over the heavy quark
loop, and contracting free Lorentz indices $\mu,\nu$ with $g_{\mu\nu}^{\perp}$
or $\varepsilon_{\mu\nu}^{\perp}$ respectively; this operation was
done using FeynCalc package for \emph{Mathematica}~\cite{FeynCalc1,FeynCalc2}.
We need to mention that gluon GPDs $H^{g},\,E^{g}$ are even functions
of variable $x$, whereas $\tilde{H}^{g},\,\tilde{E}^{g}$ are odd
functions~\cite{Diehl:2003ny}, for this reason in convolution over
$x$ both terms in numerators of~(\ref{eq:CDef1},~\ref{eq:CDef2})
give equal nonzero contributions. Numerically the dominant contribution
comes from GPD $H^{g}$, whereas contribution of $\tilde{H}^{g}$
is negligibly small. As we will see below,  the functions $\mathcal{C}_{\mathfrak{a}},\,\tilde{\mathcal{C}}_{\mathfrak{a}}$
might have other poles as a function of $x$, so the structure of
the functions~$C_{\mathfrak{a}},\,\tilde{C}_{\mathfrak{a}}$ might
be represented schematically as a sum~(\ref{eq:Monome}). 

The explicit expressions for the functions $\mathcal{C}_{\mathfrak{a}},\,\tilde{\mathcal{C}}_{\mathfrak{a}}$
depend on polarizations of the photon and are given by 
\begin{equation}
\mathcal{C}_{L}=\mathcal{O}\left(p_{a\perp}/Q,\,p_{a\perp}/m_{Q}\right)\approx0
\end{equation}
\begin{align}
\mathcal{C}_{T} & =\frac{N_{c}^{2}-1}{4N_{c}}\sum_{k=1}^{7}a_{k}-\frac{1}{4N_{c}}\sum_{k=1}^{3}b_{k}+\frac{N_{c}}{4}\sum_{k=1}^{5}c_{k}+\frac{1}{4}\sum_{k=1}^{2}d_{k}\\
\tilde{\mathcal{C}}_{T} & =\frac{N_{c}^{2}-1}{4N_{c}}\sum_{k=1}^{7}\tilde{a}_{k}-\frac{1}{4N_{c}}\sum_{k=1}^{3}\tilde{b}_{k}+\frac{N_{c}}{4}\sum_{k=1}^{5}\tilde{c}_{k}+\frac{1}{4}\sum_{k=1}^{2}\tilde{d}_{k}
\end{align}
where the contributions $a_{i},\,b_{i},\,\tilde{a}_{i},\tilde{b}_{i}$
stem from the diagrams without 3-gluon vertices in the Figure~\ref{fig:Photoproduction-A},
the terms $c_{i},\,\tilde{c}_{i}$ come from the diagrams which include
at least one three-gluon vertex, and the terms $d_{i},\tilde{d}_{i}$
stem from the diagrams in the Figure~\ref{fig:Photoproduction-B}.
Explicitly, these contributions are given by

\begin{align}
a_{1} & =4e^{\tilde{y}_{1}+\tilde{y}_{2}}Z\left(e^{\tilde{y}_{1}+\tilde{y}_{2}}Q^{2}+4\left(e^{\tilde{y}_{1}}+e^{\tilde{y}_{2}}\right)m_{Q}^{2}Z\right)(x+\xi)\times\label{eq:a1}\\
 & \times\left[m_{Q}\left(4e^{\tilde{y}_{1}}m_{Q}^{2}\left(e^{\tilde{y}_{1}}-Z\right)Z+\left(e^{2\tilde{y}_{2}}+2e^{\tilde{y}_{2}}\left(e^{\tilde{y}_{1}}-Z\right)\right)\left(e^{\tilde{y}_{1}}Q^{2}+4m_{Q}^{2}Z\right)\right)\times\frac{}{}\right.\nonumber \\
 & \left(\left(e^{\tilde{y}_{1}}Q^{2}+4m_{Q}^{2}Z\right)\left(e^{2\tilde{y}_{2}}+2e^{\tilde{y}_{2}}\left(e^{\tilde{y}_{1}}-x-Z-\xi\right)\right)+4e^{\tilde{y}_{1}}m_{Q}^{2}Z\left(e^{\tilde{y}_{1}}-x-Z-\xi\right)\right)\nonumber \\
 & \times\left.\left(1+\cosh(\tilde{y}_{1}-\tilde{y}_{2})\right)\frac{}{}\right]^{-1},\nonumber 
\end{align}
\begin{equation}
a_{2}=\frac{2e^{2\tilde{y}_{2}}\left(e^{\tilde{y}_{1}}+e^{\tilde{y}_{2}}\right)Z\left(e^{\tilde{y}_{1}+\tilde{y}_{2}}Q^{2}+4m_{Q}^{2}Z^{2}\right)}{m_{Q}\left(e^{\tilde{y}_{1}}+e^{\tilde{y}_{2}}-2Z\right)\left(e^{\tilde{y}_{1}+\tilde{y}_{2}}Q^{2}+2\left(e^{\tilde{y}_{1}}+e^{\tilde{y}_{2}}\right)m_{Q}^{2}Z\right)\left(-e^{2\tilde{y}_{2}}Q^{2}+e^{\tilde{y}_{2}}Q^{2}Z+4m_{Q}^{2}Z^{2}\right)\xi},
\end{equation}
\begin{align}
a_{3} & =8e^{\tilde{y}_{1}+\tilde{y}_{2}}Z\left(e^{\tilde{y}_{1}+2\tilde{y}_{2}}Q^{2}-8e^{\tilde{y}_{1}}m_{Q}^{2}Z^{2}-2e^{\tilde{y}_{2}}Z\left(e^{\tilde{y}_{1}}Q^{2}+2m_{Q}^{2}Z\right)\right)\times\\
 & \times\left[\frac{}{}\left(e^{\tilde{y}_{1}}+e^{\tilde{y}_{2}}\right)m_{Q}\left(e^{2\tilde{y}_{2}}Q^{2}-e^{\tilde{y}_{2}}Q^{2}Z-4m_{Q}^{2}Z^{2}\right)\right.\nonumber \\
 & \times\left.\left(\left(2e^{\tilde{y}_{2}}+e^{\tilde{y}_{1}}-2Z\right)e^{\tilde{y}_{1}+\tilde{y}_{2}}Q^{2}+4m_{Q}^{2}Z\left(e^{\tilde{y}_{1}}\left(e^{\tilde{y}_{1}}-2Z\right)+e^{2\tilde{y}_{2}}-e^{\tilde{y}_{2}}\left(Z-2e^{\tilde{y}_{1}}\right)\right)\right)\frac{}{}\right]^{-1},\nonumber 
\end{align}
\begin{align}
a_{4} & =-8e^{\tilde{y}_{1}+2\tilde{y}_{2}}Z\left(e^{\tilde{y}_{1}+\tilde{y}_{2}}Q^{2}+4m_{Q}^{2}Z^{2}\right)\times\\
 & \times\left[\frac{}{}m_{Q}\left(e^{\tilde{y}_{1}}+e^{\tilde{y}_{2}}-2Z\right)\left(e^{2\tilde{y}_{2}}Q^{2}-2e^{\tilde{y}_{2}}Q^{2}Z-4m_{Q}^{2}Z^{2}\right)\right.\nonumber \\
 & \times\left.\left(\left(2e^{\tilde{y}_{2}}+e^{\tilde{y}_{1}}-2Z\right)e^{\tilde{y}_{1}+\tilde{y}_{2}}Q^{2}+4m_{Q}^{2}Z\left(e^{\tilde{y}_{1}}\left(e^{\tilde{y}_{1}}-2Z\right)+e^{2\tilde{y}_{2}}-e^{\tilde{y}_{2}}\left(Z-2e^{\tilde{y}_{1}}\right)\right)\right)\frac{}{}\right]^{-1},\nonumber 
\end{align}

\begin{align}
a_{5} & =8e^{2(\tilde{y}_{1}+\tilde{y}_{2})}Q^{2}Z\left(e^{\tilde{y}_{1}+\tilde{y}_{2}}Q^{2}+4m_{Q}^{2}Z(-x+Z+\xi)\right)\times\\
 & \times\left[\frac{}{}m_{Q}\left(e^{\tilde{y}_{1}+\tilde{y}_{2}}Q^{2}+2\left(e^{\tilde{y}_{1}}+e^{\tilde{y}_{2}}\right)m_{Q}^{2}Z\right)\left(-e^{2\tilde{y}_{2}}Q^{2}+2e^{\tilde{y}_{2}}Q^{2}Z+4m_{Q}^{2}Z^{2}\right)\times\right.\nonumber \\
 & \times\left.\left(e^{\tilde{y}_{1}}+e^{\tilde{y}_{2}}+2x-2Z-2\xi\right)\left(-e^{2\tilde{y}_{1}}Q^{2}-2e^{\tilde{y}_{1}}Q^{2}(x-Z-\xi)+4m_{Q}^{2}Z(-x+Z+\xi)\right)\frac{}{}\right]^{-1},\nonumber 
\end{align}

\begin{align}
a_{6} & =16e^{\tilde{y}_{1}+2\tilde{y}_{2}}m_{Q}Z^{2}\left(Q^{2}e^{\tilde{y}_{2}}\left(e^{2\tilde{y}_{2}}-2e^{\tilde{y}_{1}}Z+e^{\tilde{y}_{2}}\left(e^{\tilde{y}_{1}}-2(x+Z+\xi)\right)\right)-4m_{Q}^{2}Z^{2}\left(e^{\tilde{y}_{1}}+e^{\tilde{y}_{2}}\right)\right)\times\\
 & \times\left[\frac{}{}\left(e^{\tilde{y}_{1}+\tilde{y}_{2}}Q^{2}+2\left(e^{\tilde{y}_{1}}+e^{\tilde{y}_{2}}\right)m_{Q}^{2}Z\right)\left(e^{2\tilde{y}_{2}}Q^{2}-2e^{\tilde{y}_{2}}Q^{2}Z-4m_{Q}^{2}Z^{2}\right)\left(e^{\tilde{y}_{1}}+e^{\tilde{y}_{2}}-2(x+Z+\xi)\right)\times\right.\nonumber \\
 & \times\left.\left(e^{2\tilde{y}_{2}}\left(e^{\tilde{y}_{1}}Q^{2}+4m_{Q}^{2}Z\right)+4e^{\tilde{y}_{1}}m_{Q}^{2}Z\left(e^{\tilde{y}_{1}}-x-Z-\xi\right)+2e^{\tilde{y}_{2}}\left(e^{\tilde{y}_{1}}Q^{2}+4m_{Q}^{2}Z\right)\left(e^{\tilde{y}_{1}}-x-Z-\xi\right)\right)\frac{}{}\right]^{-1},\nonumber 
\end{align}

\begin{equation}
a_{7}=\frac{4e^{2\tilde{y}_{2}}Q^{2}Z(x+\xi)}{m_{Q}\left(2e^{\tilde{y}_{2}}Q^{2}Z-e^{2\tilde{y}_{2}}Q^{2}+4m_{Q}^{2}Z^{2}\right)\left(2e^{\tilde{y}_{2}}Q^{2}(x+Z+\xi)-e^{2\tilde{y}_{2}}Q^{2}+4m_{Q}^{2}Z(x+Z+\xi)\right)\left(1+\cosh(\tilde{y}_{1}-\tilde{y}_{2})\right)},
\end{equation}

\begin{equation}
b_{1}=\frac{-8e^{3\tilde{y}_{1}+\tilde{y}_{2}}Q^{2}Z}{m_{Q}\left(e^{\tilde{y}_{1}+\tilde{y}_{2}}Q^{2}+2\left(e^{\tilde{y}_{1}}+e^{\tilde{y}_{2}}\right)m_{Q}^{2}Z\right)\left(e^{2\tilde{y}_{1}}Q^{2}-2e^{\tilde{y}_{1}}Q^{2}Z-4m_{Q}^{2}Z^{2}\right)\left(e^{\tilde{y}_{1}}+e^{\tilde{y}_{2}}+2x-2Z-2\xi\right)},
\end{equation}
\begin{equation}
b_{2}=-\frac{4e^{\tilde{y}_{1}+\tilde{y}_{2}}\left(e^{\tilde{y}_{1}}Q^{2}+2m_{Q}^{2}Z\right)}{\left(e^{\tilde{y}_{1}}+e^{\tilde{y}_{2}}\right)m_{Q}^{3}\left(e^{2\tilde{y}_{1}}Q^{2}-2e^{\tilde{y}_{1}}Q^{2}(x+Z+\xi)-4m_{Q}^{2}Z(x+Z+\xi)\right)},
\end{equation}

\begin{align}
 & b_{3}=\frac{1}{\left(e^{\tilde{y}_{1}}+e^{\tilde{y}_{2}}\right)m_{Q}}\times\\
 & \times\frac{8e^{\tilde{y}_{1}+\tilde{y}_{2}}Z\left(-e^{3\tilde{y}_{1}}Q^{2}-e^{2\tilde{y}_{1}+\tilde{y}_{2}}Q^{2}+2e^{\tilde{y}_{1}+\tilde{y}_{2}}Q^{2}Z+2e^{2\tilde{y}_{1}}Q^{2}(x+Z+\xi)+4m_{Q}^{2}Z^{2}\left(e^{\tilde{y}_{1}}+e^{\tilde{y}_{2}}\right)\right)}{\left(Q^{2}\left(e^{2\tilde{y}_{1}}-2e^{\tilde{y}_{1}}Z\right)-4m_{Q}^{2}Z^{2}\right)\left(e^{\tilde{y}_{1}}+e^{\tilde{y}_{2}}+2x-2\xi\right)\left(Q^{2}\left(e^{2\tilde{y}_{1}}-2e^{\tilde{y}_{1}}(x+Z+\xi)\right)-4m_{Q}^{2}Z(x+Z+\xi)\right)},\nonumber 
\end{align}

\begin{align}
c_{1} & =2e^{2\tilde{y}_{1}+\tilde{y}_{2}}\left[\frac{}{}e^{4\tilde{y}_{2}}(-3x+\xi)-2e^{3\tilde{y}_{2}}\left(e^{\tilde{y}_{1}}(6x-2\xi)+\xi(-5x+\xi)\right)\right.\\
 & +e^{2\tilde{y}_{1}}\left(e^{\tilde{y}_{1}}-4\xi\right)\left(2(x-\xi)\xi+e^{\tilde{y}_{1}}(-3x+\xi)\right)\nonumber \\
 & -2e^{\tilde{y}_{1}+\tilde{y}_{2}}\left(e^{2\tilde{y}_{1}}(6x-2\xi)+e^{\tilde{y}_{1}}\xi(-19x+7\xi)+2\xi\left(-2x^{2}+5x\xi+\xi^{2}\right)\right)\nonumber \\
 & \left.-2e^{2\tilde{y}_{2}}\left(e^{2\tilde{y}_{1}}(9x-3\xi)+e^{\tilde{y}_{1}}\xi(-17x+5\xi)+4\xi\left(x^{2}+x\xi-\xi^{2}\right)\right)\frac{}{}\right]\times\nonumber \\
 & \times\left[\frac{}{}\left(e^{\tilde{y}_{1}}+e^{\tilde{y}_{2}}\right)^{2}m_{Q}^{3}\left(e^{2\tilde{y}_{2}}+e^{\tilde{y}_{1}}\left(e^{\tilde{y}_{1}}-2\xi\right)+2e^{\tilde{y}_{2}}\left(e^{\tilde{y}_{1}}-2\xi\right)\right)\times\right.\nonumber \\
 & \times\left.\left(e^{2\tilde{y}_{2}}+e^{\tilde{y}_{1}}\left(e^{\tilde{y}_{1}}-4\xi\right)+2e^{\tilde{y}_{2}}\left(e^{\tilde{y}_{1}}-\xi\right)\right)(x-\xi)\left(e^{\tilde{y}_{1}}+e^{\tilde{y}_{2}}-2(x+\xi)\right)\frac{}{}\right]^{-1},\nonumber 
\end{align}

\begin{align}
c_{2} & =-2e^{\tilde{y}_{1}+\tilde{y}_{2}}\left[e^{5\tilde{y}_{1}}(x-3\xi)+e^{3\tilde{y}_{2}}\left(e^{\tilde{y}_{2}}-4\xi\right)\left(e^{\tilde{y}_{2}}-2\xi\right)(-x+\xi)\frac{}{}\right.\\
 & +2e^{3\tilde{y}_{1}}\left(e^{2\tilde{y}_{2}}(x-7\xi)-2e^{\tilde{y}_{2}}(5x-11\xi)\xi+12(x-\xi)\xi^{2}\right)+e^{4\tilde{y}_{1}}\left(e^{\tilde{y}_{2}}(3x-11\xi)+2\xi(-5x+9\xi)\right)\nonumber \\
 & +e^{\tilde{y}_{1}+2\tilde{y}_{2}}\left(4e^{\tilde{y}_{2}}(3x-\xi)\xi+e^{2\tilde{y}_{2}}(-3x+\xi)-4\xi\left(-2x^{2}+3x\xi+\xi^{2}\right)\right)\nonumber \\
 & \left.-2e^{2\tilde{y}_{1}+\tilde{y}_{2}}\left(2e^{\tilde{y}_{2}}(x-7\xi)\xi+e^{2\tilde{y}_{2}}(x+3\xi)+2\xi\left(2x^{2}-5x\xi+7\xi^{2}\right)\right)\frac{}{}\right]\times\nonumber \\
 & \times\left[\frac{}{}\left(e^{\tilde{y}_{1}}+e^{\tilde{y}_{2}}\right)m_{Q}^{3}\left(e^{2\tilde{y}_{2}}+e^{\tilde{y}_{1}}\left(e^{\tilde{y}_{1}}-2\xi\right)+2e^{\tilde{y}_{2}}\left(e^{\tilde{y}_{1}}-2\xi\right)\right)\left(e^{\tilde{y}_{1}}+e^{\tilde{y}_{2}}-4\xi\right)\times\right.\nonumber \\
 & \times\left.\left(e^{2\tilde{y}_{2}}+e^{\tilde{y}_{1}}\left(e^{\tilde{y}_{1}}-4\xi\right)+2e^{\tilde{y}_{2}}\left(e^{\tilde{y}_{1}}-\xi\right)\right)\left(e^{\tilde{y}_{1}}+e^{\tilde{y}_{2}}+2x-2\xi\right)(x-\xi)\frac{}{}\right]^{-1},\nonumber 
\end{align}

\begin{equation}
c_{3}=\frac{2e^{2\tilde{y}_{1}+\tilde{y}_{2}}\left(2e^{2\tilde{y}_{1}}+2e^{2\tilde{y}_{2}}+4e^{\tilde{y}_{1}+\tilde{y}_{2}}-2e^{\tilde{y}_{1}}(x+\xi)-e^{\tilde{y}_{2}}(x+\xi)\right)}{\left(e^{\tilde{y}_{1}}+e^{\tilde{y}_{2}}\right)^{2}m_{Q}^{3}\left(e^{\tilde{y}_{1}}+e^{\tilde{y}_{2}}-2(x+\xi)\right)\left(e^{2\tilde{y}_{1}}+e^{2\tilde{y}_{2}}+2e^{\tilde{y}_{1}+\tilde{y}_{2}}-2e^{\tilde{y}_{1}}(x+\xi)-e^{\tilde{y}_{2}}(x+\xi)\right)},
\end{equation}

\begin{align}
c_{4} & =-\frac{2e^{2(\tilde{y}_{1}+\tilde{y}_{2})}}{\left(e^{\tilde{y}_{1}}+e^{\tilde{y}_{2}}\right)^{2}m_{Q}^{3}\left(e^{\tilde{y}_{1}}+e^{\tilde{y}_{2}}-2(x+\xi)\right)}\times\\
 & \times\left[\frac{4e^{2\tilde{y}_{1}}\xi+4e^{2\tilde{y}_{2}}\xi+8e^{\tilde{y}_{1}+\tilde{y}_{2}}\xi-4e^{\tilde{y}_{1}}\xi(x+\xi)-2e^{\tilde{y}_{2}}(x+\xi)(x+3\xi)}{\left(e^{2\tilde{y}_{2}}+e^{\tilde{y}_{1}}\left(e^{\tilde{y}_{1}}-2\xi\right)+2e^{\tilde{y}_{2}}\left(e^{\tilde{y}_{1}}-2\xi\right)\right)\left(e^{2\tilde{y}_{1}}+e^{2\tilde{y}_{2}}+2e^{\tilde{y}_{1}+\tilde{y}_{2}}-e^{\tilde{y}_{1}}(x+\xi)-2e^{\tilde{y}_{2}}(x+\xi)\right)}\right.-\nonumber \\
 & -\left.\frac{(x+\xi)\left(e^{2\tilde{y}_{2}}+2e^{\tilde{y}_{2}}\left(e^{\tilde{y}_{1}}-\xi\right)+e^{\tilde{y}_{1}}\left(e^{\tilde{y}_{1}}-2(x+\xi)\right)\right)}{\left(e^{2\tilde{y}_{2}}+e^{\tilde{y}_{1}}\left(e^{\tilde{y}_{1}}-4\xi\right)+2e^{\tilde{y}_{2}}\left(e^{\tilde{y}_{1}}-\xi\right)\right)\left(e^{2\tilde{y}_{1}}+e^{2\tilde{y}_{2}}+2e^{\tilde{y}_{1}+\tilde{y}_{2}}-2e^{\tilde{y}_{1}}(x+\xi)-e^{\tilde{y}_{2}}(x+\xi)\right)}\right],\nonumber 
\end{align}
\begin{equation}
c_{5}=\frac{2e^{\tilde{y}_{1}+2\tilde{y}_{2}}\left(5\left(e^{\tilde{y}_{1}}+e^{\tilde{y}_{2}}\right)+4(x+4\xi)\right)}{\left(e^{\tilde{y}_{1}}+e^{\tilde{y}_{2}}\right)^{2}m_{Q}^{3}\left(e^{\tilde{y}_{1}}+e^{\tilde{y}_{2}}+4\xi\right)\left(e^{\tilde{y}_{1}}+e^{\tilde{y}_{2}}+2x+6\xi\right)},
\end{equation}

\begin{align}
d_{1} & =-4e^{\tilde{y}_{1}+\tilde{y}_{2}}Z\left(e^{\tilde{y}_{1}+2\tilde{y}_{2}}Q^{2}-4e^{\tilde{y}_{1}}m_{Q}^{2}Z^{2}+e^{\tilde{y}_{2}}\left(-e^{2\tilde{y}_{1}}Q^{2}+4m_{Q}^{2}Z^{2}+e^{\tilde{y}_{1}}Q^{2}(x-2Z+\xi)\right)\right)\times\\
 & \times\left[\frac{}{}m_{Q}\left(2e^{\tilde{y}_{2}}Q^{2}Z-e^{2\tilde{y}_{2}}Q^{2}+4m_{Q}^{2}Z^{2}\right)\left(e^{\tilde{y}_{1}}-x-\xi\right)\right.\times\nonumber \\
 & \times\left.\left(e^{2\tilde{y}_{2}}\left(e^{\tilde{y}_{1}}Q^{2}-4m_{Q}^{2}Z\right)-4e^{\tilde{y}_{1}}m_{Q}^{2}Z\left(e^{\tilde{y}_{1}}-x+Z-\xi\right)-2e^{\tilde{y}_{2}}\left(e^{\tilde{y}_{1}}Q^{2}-4m_{Q}^{2}Z\right)\left(e^{\tilde{y}_{1}}-x+Z-\xi\right)\right)\frac{}{}\right]^{-1},\nonumber 
\end{align}
\begin{equation}
d_{2}=-\frac{8e^{\tilde{y}_{1}+\tilde{y}_{2}}Z\left(e^{2\tilde{y}_{2}}Q^{2}-4m_{Q}^{2}Z^{2}-e^{\tilde{y}_{2}}Q^{2}(x+2Z+\xi)\right)}{m_{Q}\left(2e^{\tilde{y}_{2}}Q^{2}Z-e^{2\tilde{y}_{2}}Q^{2}+4m_{Q}^{2}Z^{2}\right)\left(e^{\tilde{y}_{1}}-x-\xi\right)\left(2e^{\tilde{y}_{2}}Q^{2}(x+Z+\xi)-e^{2\tilde{y}_{2}}Q^{2}+4m_{Q}^{2}Z(x+Z+\xi)\right)},
\end{equation}

\begin{align}
\tilde{a}_{1} & =8e^{2(\tilde{y}_{1}+\tilde{y}_{2})}Z\left(e^{\tilde{y}_{1}+\tilde{y}_{2}}Q^{2}+4e^{\tilde{y}_{1}}m_{Q}^{2}Z+4e^{\tilde{y}_{2}}m_{Q}^{2}Z\right)(x+\xi)\times\\
 & \times\left[\frac{}{}\left(e^{\tilde{y}_{1}}+e^{\tilde{y}_{2}}\right)^{2}m_{Q}\right.\times\nonumber \\
 & \times\left(e^{2\tilde{y}_{2}}\left(e^{\tilde{y}_{1}}Q^{2}+4m_{Q}^{2}Z\right)+4e^{\tilde{y}_{1}}m_{Q}^{2}Z\left(e^{\tilde{y}_{1}}-x-Z-\xi\right)+2e^{\tilde{y}_{2}}\left(e^{\tilde{y}_{1}}Q^{2}+4m_{Q}^{2}Z\right)\left(e^{\tilde{y}_{1}}-x-Z-\xi\right)\right)\nonumber \\
 & \times\left.\left(e^{2\tilde{y}_{2}}\left(e^{\tilde{y}_{1}}Q^{2}+4m_{Q}^{2}Z\right)+2e^{\tilde{y}_{2}}\left(e^{\tilde{y}_{1}}-Z\right)\left(e^{\tilde{y}_{1}}Q^{2}+4m_{Q}^{2}Z\right)+4e^{\tilde{y}_{1}}m_{Q}^{2}\left(e^{\tilde{y}_{1}}-Z\right)Z\right)\frac{}{}\right]^{-1},\nonumber 
\end{align}
\begin{equation}
\tilde{a}_{2}=\frac{2e^{2\tilde{y}_{2}}\left(e^{\tilde{y}_{1}}+e^{\tilde{y}_{2}}\right)Z\left(e^{\tilde{y}_{1}+\tilde{y}_{2}}Q^{2}+4m_{Q}^{2}Z^{2}\right)}{m_{Q}\left(e^{\tilde{y}_{1}}+e^{\tilde{y}_{2}}-2Z\right)\left(e^{\tilde{y}_{1}+\tilde{y}_{2}}Q^{2}+2m_{Q}^{2}Z\left(e^{\tilde{y}_{1}}+e^{\tilde{y}_{2}}\right)\right)\left(e^{2\tilde{y}_{2}}Q^{2}-2e^{\tilde{y}_{2}}Q^{2}Z-4m_{Q}^{2}Z^{2}\right)\xi},
\end{equation}

\begin{align}
\tilde{a}_{3} & =8e^{\tilde{y}_{1}+\tilde{y}_{2}}Z\left(e^{\tilde{y}_{1}+2\tilde{y}_{2}}Q^{2}-8e^{\tilde{y}_{1}}m_{Q}^{2}Z^{2}-2e^{\tilde{y}_{2}}Z\left(e^{\tilde{y}_{1}}Q^{2}+2m_{Q}^{2}Z\right)\right)\times\\
 & \times\left[\left(e^{\tilde{y}_{1}}+e^{\tilde{y}_{2}}\right)m_{Q}\left(e^{2\tilde{y}_{2}}Q^{2}-2e^{\tilde{y}_{2}}Q^{2}Z-4m_{Q}^{2}Z^{2}\right)\times\frac{}{}\right.\nonumber \\
 & \times\left.\left(4e^{\tilde{y}_{1}}m_{Q}^{2}\left(e^{\tilde{y}_{1}}-2Z\right)Z+2e^{2\tilde{y}_{2}}\left(e^{\tilde{y}_{1}}Q^{2}+2m_{Q}^{2}Z\right)+e^{\tilde{y}_{2}}\left(e^{2\tilde{y}_{1}}Q^{2}-2e^{\tilde{y}_{1}}\left(-4m_{Q}^{2}+Q^{2}\right)Z-4m_{Q}^{2}Z^{2}\right)\right)\frac{}{}\right]^{-1},\nonumber 
\end{align}
\begin{align}
\tilde{a}_{4} & =8e^{\tilde{y}_{1}+2\tilde{y}_{2}}Z\left(e^{\tilde{y}_{1}+\tilde{y}_{2}}Q^{2}+4m_{Q}^{2}Z^{2}\right)\times\\
 & \times\left[m_{Q}\left(e^{\tilde{y}_{1}}+e^{\tilde{y}_{2}}-2Z\right)\left(e^{2\tilde{y}_{2}}Q^{2}-2e^{\tilde{y}_{2}}Q^{2}Z-4m_{Q}^{2}Z^{2}\right)\times\frac{}{}\right.\nonumber \\
 & \times\left.\left(4e^{\tilde{y}_{1}}m_{Q}^{2}\left(e^{\tilde{y}_{1}}-2Z\right)Z+2e^{2\tilde{y}_{2}}\left(e^{\tilde{y}_{1}}Q^{2}+2m_{Q}^{2}Z\right)+e^{\tilde{y}_{2}}\left(e^{2\tilde{y}_{1}}Q^{2}-2e^{\tilde{y}_{1}}\left(-4m_{Q}^{2}+Q^{2}\right)Z-4m_{Q}^{2}Z^{2}\right)\right)\frac{}{}\right]^{-1},\nonumber 
\end{align}
\begin{align}
\tilde{a}_{5} & =8e^{2(\tilde{y}_{1}+\tilde{y}_{2})}Q^{2}Z\left(e^{\tilde{y}_{1}+\tilde{y}_{2}}Q^{2}+4m_{Q}^{2}Z(-x+Z+\xi)\right)\times\\
 & \times\left[\frac{}{}m_{Q}\left(e^{\tilde{y}_{1}+\tilde{y}_{2}}Q^{2}+2m_{Q}^{2}Z\left(e^{\tilde{y}_{1}}+e^{\tilde{y}_{2}}\right)\right)\left(-e^{2\tilde{y}_{2}}Q^{2}+2e^{\tilde{y}_{2}}Q^{2}Z+4m_{Q}^{2}Z^{2}\right)\times\right.\nonumber \\
 & \times\left.\left(e^{\tilde{y}_{1}}+e^{\tilde{y}_{2}}+2x-2Z-2\xi\right)\left(-e^{2\tilde{y}_{1}}Q^{2}-2e^{\tilde{y}_{1}}Q^{2}(x-Z-\xi)+4m_{Q}^{2}Z(-x+Z+\xi)\right)\frac{}{}\right]^{-1},\nonumber 
\end{align}

\begin{align}
\tilde{a}_{6} & =-16e^{\tilde{y}_{1}+2\tilde{y}_{2}}m_{Q}Z^{2}\left(e^{3\tilde{y}_{2}}Q^{2}-4e^{\tilde{y}_{1}}m_{Q}^{2}Z^{2}-2e^{\tilde{y}_{2}}Z\left(e^{\tilde{y}_{1}}Q^{2}+2m_{Q}^{2}Z\right)+e^{2\tilde{y}_{2}}Q^{2}\left(e^{\tilde{y}_{1}}-2(x+Z+\xi)\right)\right)\times\\
 & \times\left[\left(e^{\tilde{y}_{1}+\tilde{y}_{2}}Q^{2}+2m_{Q}^{2}Z\left(e^{\tilde{y}_{1}}+e^{\tilde{y}_{2}}\right)\right)\left(e^{2\tilde{y}_{2}}Q^{2}-2e^{\tilde{y}_{2}}Q^{2}Z-4m_{Q}^{2}Z^{2}\right)\times\frac{}{}\right.\nonumber \\
 & \times\left(e^{2\tilde{y}_{2}}\left(e^{\tilde{y}_{1}}Q^{2}+4m_{Q}^{2}Z\right)+4e^{\tilde{y}_{1}}m_{Q}^{2}Z\left(e^{\tilde{y}_{1}}-x-Z-\xi\right)+2e^{\tilde{y}_{2}}\left(e^{\tilde{y}_{1}}Q^{2}+4m_{Q}^{2}Z\right)\left(e^{\tilde{y}_{1}}-x-Z-\xi\right)\right)\nonumber \\
 & \times\left.\left(e^{\tilde{y}_{1}}+e^{\tilde{y}_{2}}-2(x+Z+\xi)\right)\frac{}{}\right]^{-1},\nonumber 
\end{align}

\begin{equation}
\tilde{a}_{7}=-\frac{8e^{\tilde{y}_{1}+3\tilde{y}_{2}}Q^{2}Z(x+\xi)}{\left(e^{\tilde{y}_{1}}+e^{\tilde{y}_{2}}\right)^{2}m_{Q}\left(e^{2\tilde{y}_{2}}Q^{2}-2e^{\tilde{y}_{2}}Q^{2}Z-4m_{Q}^{2}Z^{2}\right)\left(e^{2\tilde{y}_{2}}Q^{2}-2e^{\tilde{y}_{2}}Q^{2}(x+Z+\xi)-4m_{Q}^{2}Z(x+Z+\xi)\right)},
\end{equation}

\begin{equation}
\tilde{b}_{1}=\frac{8e^{3\tilde{y}_{1}+\tilde{y}_{2}}Q^{2}Z}{m_{Q}\left(e^{\tilde{y}_{1}+\tilde{y}_{2}}Q^{2}+2m_{Q}^{2}Z\left(e^{\tilde{y}_{1}}+e^{\tilde{y}_{2}}\right)\right)\left(e^{2\tilde{y}_{1}}Q^{2}-2e^{\tilde{y}_{1}}Q^{2}Z-4m_{Q}^{2}Z^{2}\right)\left(e^{\tilde{y}_{1}}+e^{\tilde{y}_{2}}+2x-2Z-2\xi\right)},
\end{equation}
\begin{equation}
\tilde{b}_{2}=\frac{4e^{\tilde{y}_{1}+\tilde{y}_{2}}\left(e^{\tilde{y}_{1}}Q^{2}+2m_{Q}^{2}Z\right)}{\left(e^{\tilde{y}_{1}}+e^{\tilde{y}_{2}}\right)m_{Q}^{3}\left(e^{2\tilde{y}_{1}}Q^{2}-2e^{\tilde{y}_{1}}Q^{2}(x+Z+\xi)-4m_{Q}^{2}Z(x+Z+\xi)\right)},
\end{equation}

\begin{equation}
\tilde{b}_{3}=\frac{8e^{\tilde{y}_{1}+\tilde{y}_{2}}Z\left(e^{3\tilde{y}_{1}}Q^{2}+e^{2\tilde{y}_{1}+\tilde{y}_{2}}Q^{2}-2e^{\tilde{y}_{1}+\tilde{y}_{2}}Q^{2}Z-4\left(e^{\tilde{y}_{1}}+e^{\tilde{y}_{2}}\right)m_{Q}^{2}Z^{2}-2e^{2\tilde{y}_{1}}Q^{2}(x+Z+\xi)\right)\left(e^{\tilde{y}_{1}}+e^{\tilde{y}_{2}}\right)^{-1}}{m_{Q}\left(e^{2\tilde{y}_{1}}Q^{2}-2e^{\tilde{y}_{1}}Q^{2}Z-4m_{Q}^{2}Z^{2}\right)\left(e^{\tilde{y}_{1}}+e^{\tilde{y}_{2}}+2x-2\xi\right)\left(Q^{2}\left(e^{2\tilde{y}_{1}}-2e^{\tilde{y}_{1}}(x+Z+\xi)\right)-4m_{Q}^{2}Z(x+Z+\xi)\right)},
\end{equation}

\begin{align*}
\tilde{c}_{1} & =-2e^{2\tilde{y}_{1}+\tilde{y}_{2}}\left[e^{4\tilde{y}_{2}}+e^{2\tilde{y}_{1}}\left(e^{\tilde{y}_{1}}-4\xi\right)\left(e^{\tilde{y}_{1}}-2\xi\right)+4e^{3\tilde{y}_{2}}\left(e^{\tilde{y}_{1}}-\xi\right)+\frac{}{}\right.\\
 & \left.\frac{}{}+2e^{2\tilde{y}_{2}}\left(3e^{2\tilde{y}_{1}}-7e^{\tilde{y}_{1}}\xi-2\xi(2x+\xi)\right)+4e^{\tilde{y}_{1}+\tilde{y}_{2}}\left(e^{2\tilde{y}_{1}}-4e^{\tilde{y}_{1}}\xi+\xi(2x+5\xi)\right)\right]\times\\
 & \times\left[\frac{}{}\left(e^{\tilde{y}_{1}}+e^{\tilde{y}_{2}}\right)^{2}m_{Q}^{3}\left(e^{2\tilde{y}_{2}}+e^{\tilde{y}_{1}}\left(e^{\tilde{y}_{1}}-2\xi\right)+2e^{\tilde{y}_{2}}\left(e^{\tilde{y}_{1}}-2\xi\right)\right)\times\right.\\
 & \times\left.\left(e^{2\tilde{y}_{2}}+e^{\tilde{y}_{1}}\left(e^{\tilde{y}_{1}}-4\xi\right)+2e^{\tilde{y}_{2}}\left(e^{\tilde{y}_{1}}-\xi\right)\right)\left(e^{\tilde{y}_{1}}+e^{\tilde{y}_{2}}-2(x+\xi)\right)\frac{}{}\right]^{-1},
\end{align*}
\begin{align}
\tilde{c}_{2} & =-2e^{\tilde{y}_{1}+\tilde{y}_{2}}\left[\frac{}{}e^{5\tilde{y}_{1}}+e^{4\tilde{y}_{1}}\left(5e^{\tilde{y}_{2}}-6\xi\right)+e^{3\tilde{y}_{2}}\left(e^{\tilde{y}_{2}}-4\xi\right)\left(e^{\tilde{y}_{2}}-2\xi\right)\right.\\
 & +2e^{3\tilde{y}_{1}}\left(5e^{2\tilde{y}_{2}}+e^{\tilde{y}_{2}}(4x-9\xi)+4\xi^{2}\right)\nonumber \\
 & +e^{\tilde{y}_{1}+2\tilde{y}_{2}}\left(5e^{2\tilde{y}_{2}}+2e^{\tilde{y}_{2}}(4x-9\xi)+8\xi(-3x+\xi)\right)\nonumber \\
 & \left.+2e^{2\tilde{y}_{1}+\tilde{y}_{2}}\left(5e^{2\tilde{y}_{2}}+4e^{\tilde{y}_{2}}(2x-3\xi)+6\xi(-2x+\xi)\right)\frac{}{}\right]\times\nonumber \\
 & \times\left[\frac{}{}\left(e^{\tilde{y}_{1}}+e^{\tilde{y}_{2}}\right)m_{Q}^{3}\left(e^{2\tilde{y}_{2}}+e^{\tilde{y}_{1}}\left(e^{\tilde{y}_{1}}-2\xi\right)+2e^{\tilde{y}_{2}}\left(e^{\tilde{y}_{1}}-2\xi\right)\right)\times\right.\nonumber \\
 & \times\left.\left(e^{2\tilde{y}_{2}}+e^{\tilde{y}_{1}}\left(e^{\tilde{y}_{1}}-4\xi\right)+2e^{\tilde{y}_{2}}\left(e^{\tilde{y}_{1}}-\xi\right)\right)\left(e^{\tilde{y}_{1}}+e^{\tilde{y}_{2}}-4\xi\right)\left(e^{\tilde{y}_{1}}+e^{\tilde{y}_{2}}+2x-2\xi\right)\frac{}{}\right]^{-1},\nonumber 
\end{align}
\begin{align}
\tilde{c}_{3} & =4e^{\tilde{y}_{1}+\tilde{y}_{2}}\left[e^{5\tilde{y}_{1}}+e^{4\tilde{y}_{1}}\left(5e^{\tilde{y}_{2}}+x-5\xi\right)+e^{\tilde{y}_{1}+2\tilde{y}_{2}}\left(5e^{\tilde{y}_{2}}+2(x-7\xi)\right)\left(e^{\tilde{y}_{2}}-2\xi\right)\frac{}{}\right.\\
 & +e^{3\tilde{y}_{2}}\left(e^{\tilde{y}_{2}}+x-3\xi\right)\left(e^{\tilde{y}_{2}}-2\xi\right)+2e^{2\tilde{y}_{1}+\tilde{y}_{2}}\left(5e^{2\tilde{y}_{2}}+e^{\tilde{y}_{2}}(x-19\xi)-2(x-8\xi)\xi\right)\nonumber \\
 & \left.\frac{}{}+2e^{3\tilde{y}_{1}}\left(5e^{2\tilde{y}_{2}}+e^{\tilde{y}_{2}}(x-12\xi)+\xi(-x+3\xi)\right)\right]\times\nonumber \\
 & \times\left[\frac{}{}\left(e^{\tilde{y}_{1}}+e^{\tilde{y}_{2}}\right)m_{Q}^{3}\left(e^{2\tilde{y}_{2}}+e^{\tilde{y}_{1}}\left(e^{\tilde{y}_{1}}-2\xi\right)+2e^{\tilde{y}_{2}}\left(e^{\tilde{y}_{1}}-2\xi\right)\right)\left(e^{\tilde{y}_{1}}+e^{\tilde{y}_{2}}-4\xi\right)\times\right.\nonumber \\
 & \times\left.\left(e^{2\tilde{y}_{2}}+e^{\tilde{y}_{1}}\left(e^{\tilde{y}_{1}}-4\xi\right)+2e^{\tilde{y}_{2}}\left(e^{\tilde{y}_{1}}-\xi\right)\right)\left(e^{\tilde{y}_{1}}+e^{\tilde{y}_{2}}+2x-2\xi\right)\frac{}{}\right]^{-1},\nonumber 
\end{align}
\begin{equation}
\tilde{c}_{4}=\frac{2e^{2\tilde{y}_{1}+\tilde{y}_{2}}\left(2e^{2\tilde{y}_{1}}+2e^{2\tilde{y}_{2}}+4e^{\tilde{y}_{1}+\tilde{y}_{2}}-2e^{\tilde{y}_{1}}(x+\xi)-e^{\tilde{y}_{2}}(x+\xi)\right)}{\left(e^{\tilde{y}_{1}}+e^{\tilde{y}_{2}}\right)^{2}m_{Q}^{3}\left(e^{\tilde{y}_{1}}+e^{\tilde{y}_{2}}-2(x+\xi)\right)\left(e^{2\tilde{y}_{1}}+e^{2\tilde{y}_{2}}+2e^{\tilde{y}_{1}+\tilde{y}_{2}}-2e^{\tilde{y}_{1}}(x+\xi)-e^{\tilde{y}_{2}}(x+\xi)\right)},
\end{equation}
\begin{align}
\tilde{c}_{5}=- & \frac{2e^{2(\tilde{y}_{1}+\tilde{y}_{2})}}{\left(e^{\tilde{y}_{1}}+e^{\tilde{y}_{2}}\right)^{2}m_{Q}^{3}\left(e^{\tilde{y}_{1}}+e^{\tilde{y}_{2}}-2(x+\xi)\right)}\times\\
 & \times\left(\frac{-4e^{2\tilde{y}_{1}}\xi-4e^{2\tilde{y}_{2}}\xi-8e^{\tilde{y}_{1}+\tilde{y}_{2}}\xi+4e^{\tilde{y}_{1}}\xi(x+\xi)+2e^{\tilde{y}_{2}}(x+\xi)(x+3\xi)}{\left(e^{2\tilde{y}_{2}}+e^{\tilde{y}_{1}}\left(e^{\tilde{y}_{1}}-2\xi\right)+2e^{\tilde{y}_{2}}\left(e^{\tilde{y}_{1}}-2\xi\right)\right)\left(e^{2\tilde{y}_{1}}+e^{2\tilde{y}_{2}}+2e^{\tilde{y}_{1}+\tilde{y}_{2}}-e^{\tilde{y}_{1}}(x+\xi)-2e^{\tilde{y}_{2}}(x+\xi)\right)}\right.+\nonumber \\
 & +\left.\frac{(x+\xi)\left(e^{2\tilde{y}_{2}}+2e^{\tilde{y}_{2}}\left(e^{\tilde{y}_{1}}-\xi\right)+e^{\tilde{y}_{1}}\left(e^{\tilde{y}_{1}}-2(x+\xi)\right)\right)}{\left(e^{2\tilde{y}_{2}}+e^{\tilde{y}_{1}}\left(e^{\tilde{y}_{1}}-4\xi\right)+2e^{\tilde{y}_{2}}\left(e^{\tilde{y}_{1}}-\xi\right)\right)\left(e^{2\tilde{y}_{1}}+e^{2\tilde{y}_{2}}+2e^{\tilde{y}_{1}+\tilde{y}_{2}}-2e^{\tilde{y}_{1}}(x+\xi)-e^{\tilde{y}_{2}}(x+\xi)\right)}\right),\nonumber 
\end{align}
\begin{equation}
\tilde{c}_{6}=-\frac{2ie^{\tilde{y}_{1}+2\tilde{y}_{2}}\left(5\left(e^{\tilde{y}_{1}}+e^{\tilde{y}_{2}}\right)+4(x+4\xi)\right)}{\left(e^{\tilde{y}_{1}}+e^{\tilde{y}_{2}}\right)^{2}m_{Q}^{3}\left(e^{\tilde{y}_{1}}+e^{\tilde{y}_{2}}+4\xi\right)\left(e^{\tilde{y}_{1}}+e^{\tilde{y}_{2}}+2x+6\xi\right)},
\end{equation}

\begin{align}
\tilde{d}_{1} & =4e^{2\tilde{y}_{2}}Z\left(e^{\tilde{y}_{1}+2\tilde{y}_{2}}Q^{2}-4e^{\tilde{y}_{1}}m_{Q}^{2}Z^{2}+e^{\tilde{y}_{2}}\left(-e^{2\tilde{y}_{1}}Q^{2}+4m_{Q}^{2}Z^{2}+e^{\tilde{y}_{1}}Q^{2}(x-2Z+\xi)\right)\right)\times\\
 & \times\left[\frac{}{}m_{Q}\left(-e^{2\tilde{y}_{2}}Q^{2}+2e^{\tilde{y}_{2}}Q^{2}Z+4m_{Q}^{2}Z^{2}\right)\left(e^{\tilde{y}_{2}}-x-\xi\right)\times\right.\nonumber \\
 & \times\left.\left(e^{2\tilde{y}_{2}}\left(e^{\tilde{y}_{1}}Q^{2}-4m_{Q}^{2}Z\right)-4e^{\tilde{y}_{1}}m_{Q}^{2}Z\left(e^{\tilde{y}_{1}}-x+Z-\xi\right)-2e^{\tilde{y}_{2}}\left(e^{\tilde{y}_{1}}Q^{2}-4m_{Q}^{2}Z\right)\left(e^{\tilde{y}_{1}}-x+Z-\xi\right)\right)\frac{}{}\right]^{-1},\nonumber 
\end{align}
\begin{equation}
\tilde{d}_{2}=\frac{8e^{2\tilde{y}_{2}}Z\left(e^{2\tilde{y}_{2}}Q^{2}-4m_{Q}^{2}Z^{2}-e^{\tilde{y}_{2}}Q^{2}(x+2Z+\xi)\right)}{m_{Q}\left(-e^{2\tilde{y}_{2}}Q^{2}+2e^{\tilde{y}_{2}}Q^{2}Z+4m_{Q}^{2}Z^{2}\right)\left(e^{\tilde{y}_{2}}-x-\xi\right)\left(-e^{2\tilde{y}_{2}}Q^{2}+2e^{\tilde{y}_{2}}Q^{2}(x+Z+\xi)+4m_{Q}^{2}Z(x+Z+\xi)\right)}.\label{eq:d2Tilde}
\end{equation}
We may see that all the contributions, as a function of $x$, include
poles; for this reason all the integrals which include convolution
of these coefficient functions with GPDs should be understood in the
principal value sense, taking into account the above-mentioned $\xi\to\xi-i0$
prescription~\cite{DVMPcc1} for contour deformation near the poles.
A special point of concern are the contributions $c_{1}$,$c_{2}$,
which stem from the three-gluon diagrams 11-14 in Figure~(\ref{fig:Photoproduction-A})
and contain singularities $\sim\left(x-\xi\right)^{-1}$. These singularities
apparently overlap with similar singularities in~(\ref{eq:CDef1}),
leading to  the second-order poles. The integral in the vicinity of
such singularities is defined via integration by parts~\cite{Baranov:2010zzb},
\begin{align}
\int_{-1}^{1}dx\frac{H^{g}\left(x,\,\xi\right)}{\left(x\mp\xi\pm i0\right)^{2}} & =-\int_{-1}^{1}dxH^{g}\left(x,\,\xi\right)\frac{d}{dx}\left(\frac{1}{x\mp\xi\pm i0}\right)=-\left.\frac{H^{g}\left(x,\,\xi\right)}{x\mp\xi\pm i0}\right|_{-1}^{1}+\int_{-1}^{1}dx\,\frac{\partial_{x}H^{g}\left(x,\,\xi\right)}{x\mp\xi\pm i0}
\end{align}
and exists only if the derivative $\partial_{x}H^{g}\left(x,\,\xi\right)$
is a continuous function near the points $x=\pm\xi$. Fortunately,
in the process under consideration such second-order poles cancel,
since near the point $x\approx\xi$ we have for residues 
\begin{equation}
\underset{x=\xi}{{\rm Res}}\,c_{1}=-\underset{x=\xi}{{\rm Res}}\,c_{2}.
\end{equation}
 A careful analysis demonstrates that such singularities occur only
in the $z_{1}=z_{2}=1/2$ approximation. Beyond that limit, the two
poles are separated from each other by a distance $\pm\left(\frac{1}{4z_{a}}-z_{a}\right)e^{\tilde{y}_{a}}$
or an equivalent expression, which might be found by the replacement
$z_{a}\to1-z_{a}$. 

Finally, we need to mention that in the limit $Q=0$ it is possible
to express the coefficients~(\ref{eq:a1}-\ref{eq:d2Tilde}) in a
compact form, as a function of skewedness variable $\xi$ and rapidity
difference $\Delta y=y_{1}-y_{2}$. Since photoproduction gives the
dominant contribution to the cross-section and might present special
interest for future phenomenological studies, below we provide explicit
expressions for this case: 

\begin{equation}
a_{5}=a_{7}=b_{1}=\tilde{a}_{5}=\tilde{a}_{7}=\tilde{b}_{1}=0
\end{equation}
\begin{equation}
a_{1}=-\frac{2e^{2\Delta y}(\xi+1)(\xi+x)}{m_{Q}^{3}\left(e^{\Delta y}+1\right)^{2}\left(2e^{\Delta y}(\xi+1)+4\xi+3\right)\left(\xi\left(e^{\Delta y}(\xi+1)+2\xi+1\right)-\left(e^{\Delta y}+2\right)(\xi+1)x\right)}
\end{equation}
\begin{equation}
a_{2}=\frac{1}{m_{Q}^{3}\left(e^{\Delta y}+1\right)^{2}\left(4\xi^{2}+7\xi+3\right)}
\end{equation}
\begin{equation}
a_{3}=\frac{2e^{\Delta y}\left(2e^{\Delta y}+1\right)}{m_{Q}^{3}\left(e^{\Delta y}+1\right)^{2}\left(e^{\Delta y}(4\xi+3)+2(\xi+1)\right)}
\end{equation}
\begin{equation}
a_{4}=\frac{2e^{\Delta y}}{m_{Q}^{3}\left(e^{\Delta y}+1\right)^{2}(4\xi+3)\left(e^{\Delta y}(4\xi+3)+2(\xi+1)\right)}
\end{equation}
\begin{equation}
a_{6}=\frac{2e^{\Delta y}\xi^{2}}{m_{Q}^{3}\left(e^{\Delta y}+1\right)^{2}(\xi(2\xi+1)-2(\xi+1)x)\left(\xi\left(e^{\Delta y}(\xi+1)+2\xi+1\right)-\left(e^{\Delta y}+2\right)(\xi+1)x\right)}
\end{equation}
\begin{equation}
b_{2}=\frac{\xi}{m_{Q}^{3}(1+\cosh(\Delta y))\left(-\xi^{2}+\xi x+x\right)}
\end{equation}
\begin{equation}
b_{3}=\frac{2e^{\Delta y}\xi^{2}}{m_{Q}^{3}\left(e^{\Delta y}+1\right)^{2}\left(\xi^{2}-(\xi+1)x\right)(\xi(2\xi+1)-2(\xi+1)x)}
\end{equation}
\begin{align}
c_{1} & =\frac{2e^{2\Delta y}}{\left(e^{\Delta y}+1\right)^{3}m_{Q}^{3}\left(e^{\Delta y}(2\xi+1)+4\xi+3\right)\left(e^{\Delta y}(4\xi+3)+2\xi+1\right)(x-\xi)\left(\xi+2\left(\xi^{2}+\xi x+x\right)\right)}\times\\
 & \times\left[\xi^{2}\left(-e^{2\Delta y}(2\xi+1)(4\xi+3)+2e^{\Delta y}(2\xi(\xi+4)+5)-2\xi(4\xi+7)-7\right)-8\left(e^{\Delta y}-1\right)(\xi+1)^{2}x^{2}\right.\nonumber \\
 & \left.\quad+2e^{\Delta y}\xi x\left(\left(8\xi^{2}+4\xi-1\right)\cosh(\Delta y)-2(\xi+1)\sinh(\Delta y)+2\xi(5\xi+4)+1\right)\right]\nonumber 
\end{align}
\begin{align}
 & c_{2}=\frac{\text{sech}^{2}\left(\frac{\Delta y}{2}\right)}{2m_{Q}^{3}(4\xi+3)(x-\xi)(2(\xi+1)x-\xi(2\xi+1))\left((\xi+1)\tanh\left(\frac{\Delta y}{2}\right)-3\xi-2\right)\left((\xi+1)\tanh\left(\frac{\Delta y}{2}\right)+3\xi+2\right)}\\
 & \times\left[\frac{\xi\left((4\xi+3)(\sinh(\Delta y)(2\xi(2\xi+1)-(4\xi+3)x)+\cosh(\Delta y)(\xi(2\xi+1)-2(\xi+1)x))+\xi+2(\xi+1)\left(2\xi^{2}+\xi x+x\right)\right)}{\cosh(\Delta y)+1}\right.\nonumber \\
 & \left.{\color{white}.}\quad-8(\xi+1)^{2}\sinh^{4}\left(\frac{\Delta y}{2}\right)\text{csch}^{3}(\Delta y)\left(\xi^{2}-2x^{2}\right)\frac{}{}\right]\nonumber 
\end{align}
\begin{equation}
c_{3}=\frac{2e^{2\Delta y}\xi\left(\left(2e^{\Delta y}+1\right)\left(\xi^{2}+\xi x+x\right)-\xi\right)}{\left(e^{\Delta y}+1\right)^{3}m_{Q}^{3}\left(\xi+2\left(\xi^{2}+\xi x+x\right)\right)\left(e^{\Delta y}\left(\xi+2\left(\xi^{2}+\xi x+x\right)\right)+\xi^{2}+\xi x+x\right)}
\end{equation}
\begin{align}
c_{4} & =\frac{\left(e^{\Delta y}+1\right)(\xi+1)\text{sech}^{4}\left(\frac{\Delta y}{2}\right)}{8m_{Q}^{3}\left(e^{\Delta y}(2\xi+1)+4\xi+3\right)\left(e^{\Delta y}(4\xi+3)+2\xi+1\right)\left(\xi+2\left(\xi^{2}+\xi x+x\right)\right)\left(\left(e^{\Delta y}+2\right)\left(\xi^{2}+\xi x+x\right)+\xi\right)}\times\\
 & \times\left[e^{2\Delta y}\left(\xi^{3}\left(60\xi^{3}+78\xi^{2}+32\xi+5\right)+2(\xi+1)^{2}x^{3}+2\xi(\xi+1)(3\xi+4)(10\xi+7)x^{2}\right)\right.\nonumber \\
 & {\color{white}.}\quad+e^{2\Delta y}\xi^{2}(2\xi(3\xi(20\xi+43)+83)+31)x+e^{\Delta y}\xi^{2}(2\xi(\xi(24\xi+41)+14)-3)x\nonumber \\
 & {\color{white}.}\quad+e^{\Delta y}\left(\xi^{3}(\xi(12\xi(2\xi+1)-5)-2)-2(\xi+1)^{2}x^{3}+\xi(\xi+1)(4\xi(6\xi+11)+17)x^{2}\right)\nonumber \\
 & {\color{white}.}\quad+(2\xi+1)\left(-\xi^{3}(2\xi(\xi+4)+3)+2(\xi+1)^{2}x^{3}+2\xi(\xi+1)^{2}x^{2}-\xi^{2}(2\xi(\xi+4)+7)x\right)\nonumber \\
 & \left.{\color{white}.}\quad-e^{3\Delta y}\left(\xi^{2}+\xi x+x\right)(-\xi(14\xi+11)+2\xi x+x)\left(\xi+2\left(\xi^{2}+\xi x+x\right)\right)\right]\left(\xi e^{\Delta y}+\left(\xi^{2}+\xi x+x\right)\left(1+2e^{\Delta y}\right)\right)^{-1}\nonumber \\
c_{5} & =\frac{4e^{\Delta y}(\xi(16\xi+21)+4(\xi+1)x)}{\left(e^{\Delta y}+1\right)^{3}m_{Q}^{3}(4\xi+5)(\xi(6\xi+7)+2(\xi+1)x)}
\end{align}
\begin{align}
d_{1} & =\frac{4\xi^{2}\sinh\left(\frac{\Delta y}{2}\right)\left(\cosh\left(\frac{\Delta y}{2}\right)\left(\xi+2\left(\xi^{2}+\xi x+x\right)\right)-\xi\sinh\left(\frac{\Delta y}{2}\right)\right)^{-1}}{m_{Q}^{3}\left(-3\sinh(\Delta y)\left(3\xi^{2}+\xi(x+2)+x\right)+\cosh(\Delta y)\left(3\xi^{2}+\xi(x+4)+x\right)+3\xi^{2}+\xi x+x\right)}\\
d_{2} & =-\frac{2e^{\Delta y}\xi^{2}}{m_{Q}^{3}\left(e^{\Delta y}+1\right)\left(-\xi^{2}+\xi x+x\right)\left(\left(e^{\Delta y}+1\right)\left(\xi^{2}+\xi x+x\right)+\xi\right)}
\end{align}

\begin{equation}
\tilde{a}_{1}=\frac{2e^{2\Delta y}(\xi+1)(\xi+x)}{m_{Q}^{3}\left(e^{\Delta y}+1\right)^{2}\left(2e^{\Delta y}(\xi+1)+4\xi+3\right)\left(\xi\left(e^{\Delta y}(\xi+1)+2\xi+1\right)-\left(e^{\Delta y}+2\right)(\xi+1)x\right)}
\end{equation}
\begin{equation}
\tilde{a}_{2}=-\frac{1}{m_{Q}^{3}\left(e^{\Delta y}+1\right)^{2}\left(4\xi^{2}+7\xi+3\right)}
\end{equation}
\begin{equation}
\tilde{a}_{3}=\frac{2e^{\Delta y}\left(2e^{\Delta y}+1\right)}{m_{Q}^{3}\left(e^{\Delta y}+1\right)^{2}\left(e^{\Delta y}(4\xi+3)+2(\xi+1)\right)}
\end{equation}
\begin{equation}
\tilde{a}_{4}=-\frac{2e^{\Delta y}}{m_{Q}^{3}\left(e^{\Delta y}+1\right)^{2}(4\xi+3)\left(e^{\Delta y}(4\xi+3)+2(\xi+1)\right)}
\end{equation}
\begin{equation}
\tilde{a}_{6}=-\frac{\xi^{2}}{m_{Q}^{3}(\cosh(\Delta y)+1)(\xi(2\xi+1)-2(\xi+1)x)\left(\xi\left(e^{\Delta y}(\xi+1)+2\xi+1\right)-\left(e^{\Delta y}+2\right)(\xi+1)x\right)}
\end{equation}
\begin{equation}
\tilde{b}_{2}=-\frac{\xi}{m_{Q}^{3}(\cosh(\Delta y)+1)\left(-\xi^{2}+\xi x+x\right)}
\end{equation}
\begin{equation}
\tilde{b}_{3}=-\frac{\xi^{2}}{m_{Q}^{3}(\cosh(\Delta y)+1)\left(-\xi^{2}+\xi x+x\right)(2(\xi+1)x-\xi(2\xi+1))}
\end{equation}
\begin{equation}
\tilde{c}_{1}=\frac{2e^{2\Delta y}\left(\xi\left(e^{2\Delta y}(2\xi+1)(4\xi+3)+2e^{\Delta y}(5\xi(2\xi+3)+6)-4\xi(\xi+3)-7\right)+8\left(e^{\Delta y}-1\right)(\xi+1)^{2}x\right)}{m_{Q}^{3}\left(e^{\Delta y}+1\right)^{3}\left(e^{\Delta y}(2\xi+1)+4\xi+3\right)\left(e^{\Delta y}(4\xi+3)+2\xi+1\right)\left(\xi+2\left(\xi^{2}+\xi x+x\right)\right)}
\end{equation}
\begin{align}
\tilde{c}_{2} & =-\frac{2e^{\Delta y}}{m_{Q}^{3}\left(e^{\Delta y}+1\right)^{3}(4\xi+3)\left(e^{\Delta y}(2\xi+1)+4\xi+3\right)\left(e^{\Delta y}(4\xi+3)+2\xi+1\right)(\xi(2\xi+1)-2(\xi+1)x)}\times\\
 & \times\left[e^{3\Delta y}\xi(2\xi+1)(4\xi+3)-e^{\Delta y}\left(-4(2\xi+1)\xi^{2}+\xi+8(\xi+1)(3\xi+2)x\right)\right.+\nonumber \\
 & \left.\qquad+e^{2\Delta y}\left(3\xi(2\xi+1)^{2}-8(\xi+1)(3\xi+2)x\right)+\xi(2\xi+1)(4\xi+3)\right]\nonumber 
\end{align}
\begin{align}
\tilde{c}_{3} & =-\frac{\text{sech}^{4}\left(\frac{\Delta y}{2}\right)}{2m_{Q}^{3}(4\xi+3)(\xi(2\xi+1)-2(\xi+1)x)\left((\xi+1)\tanh\left(\frac{\Delta y}{2}\right)-3\xi-2\right)\left((\xi+1)\tanh\left(\frac{\Delta y}{2}\right)+3\xi+2\right)}\times\\
 & \times\left(-\xi(\xi+1)^{2}\tanh\left(\frac{\Delta y}{2}\right)+(2\xi+1)\cosh(\Delta y)\left(-3\xi^{2}+\xi(x-2)+x\right)-\xi(3\xi+2)(4\xi+3)+(\xi+1)^{2}x\right)
\end{align}
\begin{equation}
\tilde{c}_{4}=\frac{2e^{2\Delta y}\xi\left(\left(2e^{\Delta y}+1\right)\left(\xi^{2}+\xi x+x\right)-\xi\right)}{m_{Q}^{3}\left(e^{\Delta y}+1\right)^{3}\left(\xi+2\left(\xi^{2}+\xi x+x\right)\right)\left(e^{\Delta y}\left(\xi+2\left(\xi^{2}+\xi x+x\right)\right)+\xi^{2}+\xi x+x\right)}
\end{equation}
\begin{align}
\tilde{c}_{5} & =-\frac{\left(e^{\Delta y}+1\right)(\xi+1)\text{sech}^{4}\left(\frac{\Delta y}{2}\right)\left(\xi e^{\Delta y}+\left(\xi^{2}+\xi x+x\right)\left(1+2e^{\Delta y}\right)\right)^{-1}}{8m_{Q}^{3}\left(e^{\Delta y}(2\xi+1)+4\xi+3\right)\left(e^{\Delta y}(4\xi+3)+2\xi+1\right)\left(\xi+2\left(\xi^{2}+\xi x+x\right)\right)\left(\left(e^{\Delta y}+2\right)\left(\xi^{2}+\xi x+x\right)+\xi\right)}\times\\
 & \times\left[\frac{}{}e^{2\Delta y}\left(\xi^{3}\left(60\xi^{3}+78\xi^{2}+32\xi+5\right)+2(\xi+1)^{2}x^{3}+2\xi(\xi+1)(3\xi+4)(10\xi+7)x^{2}\right)\right.\nonumber \\
 & {\color{white}.}\quad+e^{\Delta y}\left(\xi^{3}(\xi(12\xi(2\xi+1)-5)-2)-2(\xi+1)^{2}x^{3}+\xi(\xi+1)(4\xi(6\xi+11)+17)x^{2}\right)\nonumber \\
 & {\color{white}.}\quad+x\xi^{2}e^{2\Delta y}(2\xi(3\xi(20\xi+43)+83)+31)+x\xi^{2}e^{\Delta y}(2\xi(\xi(24\xi+41)+14)-3)\\
 & {\color{white}.}\quad+(2\xi+1)\left(-\xi^{3}(2\xi(\xi+4)+3)+2(\xi+1)^{2}x^{3}+2\xi(\xi+1)^{2}x^{2}-\xi^{2}(2\xi(\xi+4)+7)x\right)\nonumber \\
 & {\color{white}.}\quad\left.-e^{3\Delta y}\left(\xi^{2}+\xi x+x\right)(-\xi(14\xi+11)+2\xi x+x)\left(\xi+2\left(\xi^{2}+\xi x+x\right)\right)\frac{}{}\right]\nonumber 
\end{align}
\begin{equation}
\tilde{c}_{6}=-\frac{4e^{\Delta y}(\xi(16\xi+21)+4(\xi+1)x)}{m_{Q}^{3}\left(e^{\Delta y}+1\right)^{3}(4\xi+5)(\xi(6\xi+7)+2(\xi+1)x)}
\end{equation}
\begin{equation}
\tilde{d}_{1}=\frac{2\left(e^{\Delta y}-1\right)\xi^{2}\left(e^{\Delta y}(\xi+1)(\xi+x)+\xi^{2}+\xi x+x\right)^{-1}}{m_{Q}^{3}\left(\xi\left(-3e^{\Delta y}\xi+e^{2\Delta y}(3\xi+1)-6\xi-5\right)+\left(e^{\Delta y}-2\right)\left(e^{\Delta y}+1\right)(\xi+1)x\right)}
\end{equation}
\begin{equation}
\tilde{d}_{2}=\frac{2\xi^{2}}{m_{Q}^{3}\left(e^{\Delta y}+1\right)\left(-\xi^{2}+\xi x+x\right)\left(e^{\Delta y}(\xi+1)(\xi+x)+\xi^{2}+\xi x+x\right)}
\end{equation}

\end{document}